%% file: Sayers_ApJ_pressure_evolution.tex
\documentclass[twocolumn]{aastex631}

\usepackage{float,amsmath}

\newcommand{\msun}{M$_{\sun}$}
\newcommand{\mfive}{M$_{500}$}
\newcommand{\thetafive}{$\theta_{500}$}

\newcommand{\rfive}{R$_{500}$}
\newcommand{\rtwo}{R$_{200}$}
\newcommand{\pfive}{P$_{500}$}
\newcommand{\chandra}{{\it Chandra}}
\newcommand{\planck}{{\it Planck}}
\newcommand{\rosat}{{\it ROSAT}}
\newcommand{\lowz}{low-$z$}
\newcommand{\midz}{mid-$z$}
\newcommand{\three}{{\sc The Three Hundred}}

\shorttitle{Galaxy Cluster Pressure Profiles}
\shortauthors{Sayers et al.}

\graphicspath{{./}{figures/}}

\begin{document}

\title{The Evolution and Mass Dependence of Galaxy Cluster Pressure Profiles
  at 0.05~$\le z \le$~0.60 and $4 \times 10^{14}$~\msun~$\le \textrm{M}_{500} \le 30 \times 10^{14}$~\msun}

\correspondingauthor{Jack Sayers}
\email{jack@caltech.edu}

\author[0000-0002-8213-3784]{Jack Sayers}
\affiliation{California Institute of Technology, 1200 East California Boulevard, Pasadena, California 91125, USA}

\author[0000-0002-8031-1217]{Adam B. Mantz}
\affiliation{Kavli Institute for Particle Astrophysics and Cosmology, Stanford University, 452 Lomita Mall, Stanford, California 94305, USA}

\author[0000-0003-4175-002X]{Elena Rasia}
\affiliation{INAF - Osservatorio Astronomico di Trieste, via Tiepolo 11, I-34143 Trieste, Italy}

\author[0000-0003-0667-5941]{Steven W. Allen}
\affiliation{Kavli Institute for Particle Astrophysics and Cosmology, Stanford University, 452 Lomita Mall, Stanford, California 94305, USA}
\affiliation{Department of Physics, Stanford University, 382 Via Pueblo Mall, Stanford, California 94305, USA}
\affiliation{SLAC National Accelerator Laboratory, 2575 Sand Hill Road, Menlo Park, CA 94025, USA}

\author[0000-0002-2113-4863]{Weiguang Cui}
\affiliation{Institute for Astronomy, University of Edinburgh, Royal Observatory, Edinburgh EH9 3HJ, United Kingdom}
\author[0000-0002-1098-7174]{Sunil R. Golwala}
\affiliation{California Institute of Technology, 1200 East California Boulevard, Pasadena, California 91125, USA}

\author[0000-0003-2985-9962]{R. Glenn Morris}
\affiliation{Kavli Institute for Particle Astrophysics and Cosmology, Stanford University, 452 Lomita Mall, Stanford, California 94305, USA}
\affiliation{SLAC National Accelerator Laboratory, 2575 Sand Hill Road, Menlo Park, CA 94025, USA}

\author[0000-0001-8872-4991]{Jenny T. Wan}
\affiliation{California Institute of Technology, 1200 East California Boulevard, Pasadena, California 91125, USA}

\begin{abstract}

  We have combined X-ray observations from \chandra\ with Sunyaev-Zel'dovich (SZ) effect data from \planck\
  and Bolocam to measure intra-cluster medium pressure profiles from 0.03\rfive~$\le$~R~$\le$~5\rfive\
  for a sample of 21 \lowz\ galaxy clusters with a median redshift $\langle z \rangle = 0.08$ and a 
  median mass $\langle \textrm{M}_{500} \rangle = 6.1 \times 10^{14}$~\msun\ and a sample of
  19 \midz\ galaxy clusters with $\langle z \rangle = 0.50$ and
  $\langle \textrm{M}_{500} \rangle = 10.6 \times 10^{14}$~\msun.
  The mean scaled pressure in the \lowz\ sample
  is lower at small radii and higher at large radii, a trend that is accurately reproduced in
  similarly selected samples from \three\ simulations. This difference appears to be primarily due to
  dynamical state at small radii, evolution at intermediate radii, and a combination of evolution
  and mass dependence at large radii. Furthermore, the overall flattening of the mean
  scaled pressure profile in the \lowz\
  sample compared to the \midz\ sample is consistent with expectations due to
  differences in mass accretion rate and the fractional impact of feedback mechanisms.
  In agreement with previous studies, the fractional scatter about the mean
  scaled pressure profile reaches a minimum of $\simeq 20$ per cent near 0.5\rfive.
  This scatter is consistent between the \lowz\ and \midz\ samples at all radii,
  suggesting it is not strongly impacted by sample selection, and this general behavior is
  reproduced in \three\ simulations.
  Finally, analytic functions that approximately describe the mass and redshift trends in mean
  pressure profile shape are provided.
  
\end{abstract}


\section{Introduction} \label{sec:intro}

Galaxy clusters have played an important role in constraining cosmological models
and in probing the physics related to
large scale structure formation \citep{Voit2005, Allen2011,Kravtsov2012}.
Central to many of these studies are the thermodynamic properties
of the intra-cluster medium (ICM), which are often characterized
by azimuthally averaged radial profiles \citep{Vikhlinin2005,Cavagnolo2009}.
Because galaxy cluster formation is dominated by gravitational physics, these
ICM radial profiles were long ago predicted to
follow simple scaling relations based on mass
and redshift, with minimal cluster-to-cluster scatter after accounting for
these scalings \citep{Kaiser1986,Kaiser1991}.

In particular, ICM pressure is directly proportional to the Sunyaev-Zel'dovich (SZ)
effect brightness and provides a measure of the total thermal energy of the ICM
\citep{Sunyaev1972, Carlstrom2002}. To describe the mean scaled
pressure profile, \citet{Nagai2007} proposed the use of a generalized
\citet[][hereafter gNFW]{Navarro1997} parameterization, which has since become
the standard for most analyses that probe beyond the inner regions of the
ICM that have traditionally been measured with X-rays. While many early studies relied
entirely or in part on simulations to constrain the parameters
of this pressure model \citep{Nagai2007,Arnaud2010,Battaglia2012}, improvements
in SZ effect data, often combined with X-ray observations, quickly
opened the possibility of obtaining constraints derived entirely
from observational data
\citep{Plagge2010, Bonamente2012, Planck2013_V, Sayers2013_pressure}.
All of these initial studies broadly confirmed the simple theoretical prediction
of self similarity (i.e., a universal scaled pressure profile
with minimal cluster-to-cluster scatter).

However, baryonic processes, in particular active galactic nucleus (AGN) feedback, the
thermalization of newly accreted material, and merger-induced
disruptions of the ICM core, can produce deviations from
self similarity. Perhaps the clearest example is the dichotomy
in central ICM pressure between dynamically relaxed and disturbed
galaxy clusters \citep[which are sometimes distinguished by the presence
  or lack of a cool-core, see e.g.,][]{Arnaud2010,McDonald2014}.
A range of simulations have made predictions for how these
processes will impact pressure profiles. For instance,
enhanced mass accretion rates, which occur for galaxy clusters
with larger masses and/or higher redshifts, will tend to
steepen the pressure profiles at large radii due to halo contraction during
accretion \citep{Diemer2014,Diemer2022}, 
further penetration of the shocks required for thermalization
\citep{Battaglia2012-1, Lau2015, Planelles2017},
and the initial accretion shock occuring at smaller radii
\citep{Aung2021}.
As another example, the fractional amount of gas ejected
from the central ICM due to AGN outbursts is larger
in lower mass systems \citep{Battaglia2010, McCarthy2011,
  Battaglia2012-1,LeBrun2015, Barnes2017, Truong2018, Henden2020}.

Quantitative predictions for the mass and redshift dependence
of these deviations from self similarity have been provided
based on gNFW fits to some of these simulations
\citep{Battaglia2012, LeBrun2015, Planelles2017}, and generally
point to flatter pressure profiles in lower mass and/or lower
redshift galaxy clusters, particularly at larger radii.
With the recent availability of high
quality observational data covering well-defined samples
spanning a range of masses/redshifts, it has become possible
to search for such trends. For instance,
\citet{McDonald2014}, \citet{McDonald2017}, and \citet{Ghirardini2021-highz}, using
X-ray observations of a large
sample of SZ selected galaxy clusters extending to $z = 1.8$,
find significant deviations from self similarity in the
evolution of scaled pressure profiles near the core.
This appears to be due to AGN feedback maintaining this region
in a relatively fixed state while the bulk of
the ICM evolves in a self-similar manner outside of it
\citep[although we note that][based on an independent analysis
  of the same data, find no such trend in ICM pressure near the core]{Sanders2018}.

In contrast to many simulations, none of these observational
studies find evidence
for evolution in the scaled pressure profiles outside of the core
region. Furthermore, the observational studies of \citet{Sayers2016},
\citet{Bourdin2017}, and \citet{Bartalucci2017}
also found a similar lack of evolution, although selection effects
may have played a role in these works.
Fewer observational studies have focused on trends in scaled pressure
with mass, but we note that \citet{Sayers2016}
did find the expected level of flattening in the profiles
of lower mass systems.

In sum, while several physical processes have been identified
in simulations as potential causes for deviations from self similarity
in ICM pressure profiles, there is a relative lack of consensus in the
quantitative level of the trends with mass and redshift that should result from
these processes. This lack of consensus exists between the various
simulations, between different observational results, and when comparing
the simulations to observations. To better address this situation, recent
observational studies
have demonstrated the power of combining data from different facilities
to probe a wide range of radial scales \citep{Planck2013_V, Romero2017,
  Bartalucci2017, Bourdin2017, Ghirardini2019, Pointecouteau2021}, and
such data are now becoming available for large samples with well defined
selections. In addition, mock observations of similarly selected samples
from large volume simulations are increasingly being used for comparison
in order to mitigate systematics related to analysis methods and
population statistics \citep{Gianfagna2021, Paliwal2022}.

In this work, we make use of \chandra\ X-ray observations, along with
\planck\ and Bolocam SZ effect data, to measure ICM pressure profiles
over a wide range in radius for a sample of 40 
galaxy clusters selected primarily based on X-ray luminosity.
This sample was specifically chosen to search
for trends as a function of mass and redshift in the highest mass
objects between $z \simeq 0.5$ and the present day. To facilitate
interpretation of these observed profiles, we perform an identical analysis
of simulated clusters similarly selected from \three\ simulations.
The galaxy cluster sample and suite of observational data are described
in Section~\ref{sec:data}, and our analysis techniques to generate
pressure profiles are presented in Section~\ref{sec:deprojection}.
Section~\ref{sec:mean} details our method to compute mean scaled pressure profiles,
and the intrinsic cluster-to-cluster scatter about this mean, for
both the observational and simulated data. Simple parametric models
that approximately capture the shape of these mean profiles are
given in Section~\ref{sec:gNFW}, and a brief discussion and summary
of the results is presented in Section~\ref{sec:discussion}.
Throughout this work we assume a flat $\Lambda$CDM cosmology with
$\Omega_{\textrm{m}} = 0.3$, $\Omega_{\Lambda} = 0.7$, and $h = 0.7$
to convert between observed and physical quantities.

\section{Observational Data} \label{sec:data}

\subsection{Galaxy Cluster Sample}

In order to study the evolution and mass dependence of galaxy cluster pressure profiles we have
compiled two samples selected primarily based on X-ray luminosity. The ``\lowz'' sample
contains 21 objects at $0.05 \le z \le 0.10$ from the {\it ROSAT} Brightest Cluster Sample
\citep[BCS,][]{Ebeling1998} and the {\it ROSAT}-ESO Flux Limited X-ray \citep[REFLEX,][]{Bohringer2004}
catalogs with a 0.1--2.4~keV luminosity $\ge 2.5 \times 10^{44}$~erg s$^{-1}$ (see Table~\ref{tab:sample}).
The ``\midz'' sample contains 19 objects at $0.40 \le z \le 0.60$ from the Massive Cluster
Survey \citep[MACS,][]{Ebeling2001,Ebeling2007,Ebeling2010,Mann2012,Mantz2015}.
While we sought to include all of the galaxy clusters matching the redshift and luminosity
cuts given above, the following objects were removed from our study:
the merging pair Abell 0399/Abell 0401 from the \lowz\ sample due to the significant
contamination from their overlapping SZ effect signals; Abell 2597
from the \lowz\ sample because we are not able to adequately measure its pressure deprojection,
likely due to the low SNR of the {\it Planck} SZ effect data; MACS~J0717.5 from
the \midz\ sample due to its large kinematic SZ effect signal
\citep{Mroczkowski2012,Sayers2013,Adam2017}; 23 additional objects from the
MACS catalog that lack the required ground-based SZ effect data from Bolocam to
be included in the \midz\ sample.\footnote{
  While in operation, the Bolocam instrument team aimed to observe the entire published MACS sample,
  selecting targets primarily based on visibility at the time of the observations.
  Therefore, in general, the 20 MACS objects observed by Bolocam in the mid-$z$ sample
  were not selected based on characteristics such as X-ray morphology or dynamical state,
  and they should thus provide a reasonable representation of the full population in the MACS catalog.
  In addition, we note that many of the 23 missing
  objects were not known in the literature at the time of the final Bolocam
  observations in 2011 October.}

\begin{deluxetable*}{cccccccccc}
\tablenum{1}
\tablecaption{Galaxy Cluster Sample}
\tablewidth{0pt}
\tablehead{
  \colhead{Name} & \colhead{R. A.} & \colhead{Decl.} & \colhead{$z$} &
  \colhead{\mfive} & \colhead{\thetafive} & \colhead{{\it CXO/RASS}} & \colhead{\planck} &
  \colhead{Bolocam} & \colhead{Morph.} \\
  \colhead{} & \colhead{} & \colhead{} & \colhead{} &
  \colhead{($10^{14}$\msun)} & \colhead{(\arcmin)} & \colhead{(ksec)} & \colhead{(SNR)} &
  \colhead{(SNR)}}
\startdata
  \sidehead{\lowz\ Sample}
  Abell 0754 & 09\fh09\fm20\fs06 & $-$09\arcdeg40\arcmin57\arcsec.0 & 0.054 & \phn5.3$\pm$0.6 & 20.2$\pm$0.8 & 123.7/16.6 & 36.2 & --- & Dist. \\
  Abell 0085 & 00\fh41\fm50\fs47 & $-$09\arcdeg18\arcmin11\arcsec.8 & 0.056 & \phn6.4$\pm$0.6 & 18.2$\pm$0.6 & 169.6/13.6 & 22.6 & --- & Dist. \\
  Abell 3667 & 20\fh12\fm31\fs47 & $-$56\arcdeg50\arcmin31\arcsec.8 & 0.056 & \phn8.2$\pm$1.2 & 19.8$\pm$1.0 & 388.2/\phn6.3 & 25.8 & --- & Dist. \\
  Abell 2256 & 17\fh04\fm08\fs89 & $+$78\arcdeg39\arcmin00\arcsec.1 & 0.058 & \phn9.5$\pm$1.0 & 20.8$\pm$0.7 & \phn16.6/16.3 & 39.8 & --- &   \\
  Abell 3158 & 03\fh42\fm52\fs91 & $-$53\arcdeg37\arcmin47\arcsec.5 & 0.059 & \phn4.6$\pm$0.6 & 16.3$\pm$0.6 & \phn57.7/\phn2.9 & 19.8 & --- &   \\
  Abell 3266 & 04\fh31\fm13\fs47 & $-$61\arcdeg27\arcmin16\arcsec.2 & 0.059 & \phn9.8$\pm$1.2 & 21.1$\pm$0.8 & \phn34.8/19.8 & 40.0 & --- &   \\
  Abell 1795 & 13\fh48\fm52\fs59 & $+$26\arcdeg35\arcmin28\arcsec.8 & 0.063 & \phn5.3$\pm$0.5 & 17.1$\pm$0.6 & 847.0/\phn0.0 & 18.3 & --- &   \\
  Abell 2065 & 15\fh22\fm29\fs24 & $+$27\arcdeg42\arcmin25\arcsec.5 & 0.072 & \phn4.8$\pm$0.5 & 14.3$\pm$0.5 & \phn25.0/\phn0.0 & 12.3 & --- & Dist. \\
  Abell 3112 & 03\fh17\fm57\fs55 & $-$44\arcdeg14\arcmin18\arcsec.1 & 0.075 & \phn4.5$\pm$0.6 & 12.4$\pm$0.5 & 114.1\phn7.0 & 10.5 & --- &   \\
  Abell 2029 & 15\fh10\fm56\fs00 & $+$05\arcdeg44\arcmin40\arcsec.3 & 0.078 & \phn9.3$\pm$1.0 & 15.7$\pm$0.5 & 118.9/\phn8.7 & 23.2 & --- & Rel. \\
  Abell 2255 & 17\fh12\fm47\fs57 & $+$64\arcdeg03\arcmin47\arcsec.4 & 0.081 & \phn6.2$\pm$0.8 & 13.7$\pm$0.5 & \phn36.7/11.8 & 26.2 & --- &   \\
  Abell 1650 & 12\fh58\fm41\fs49 & $-$01\arcdeg45\arcmin44\arcsec.1 & 0.084 & \phn4.6$\pm$0.5 & 12.5$\pm$0.4 & 203.6/\phn0.0 & 11.2 & --- &   \\
  Abell 1651 & 12\fh59\fm22\fs30 & $-$04\arcdeg11\arcmin46\arcsec.8 & 0.084 & \phn6.0$\pm$0.7 & 13.6$\pm$0.5 & \phn\phn9.1/\phn7.4 & 14.5 & --- &   \\
  Abell 2420 & 22\fh10\fm18\fs91 & $-$12\arcdeg10\arcmin24\arcsec.6 & 0.085 & \phn4.6$\pm$0.5 & 11.3$\pm$0.4 & \phn\phn7.8/\phn0.0 & 12.4 & --- &   \\
  Abell 0478 & 04\fh13\fm24\fs88 & $+$10\arcdeg27\arcmin54\arcsec.1 & 0.088 & \phn9.2$\pm$1.1 & 14.1$\pm$0.6 & 129.4/21.9 & 15.8 & --- & Rel. \\
  Abell 2142 & 15\fh58\fm20\fs52 & $+$27\arcdeg13\arcmin48\arcsec.9 & 0.089 & 11.3$\pm$1.1 & 15.1$\pm$0.5 & 192.8/15.1 & 28.4 & --- &   \\
  Abell 3695 & 20\fh34\fm48\fs96 & $-$35\arcdeg49\arcmin32\arcsec.5 & 0.089 & \phn5.8$\pm$0.9 & 12.1$\pm$0.6 & \phn\phn9.9/\phn0.0 & 11.9 & --- &   \\
  Abell 3921 & 22\fh49\fm57\fs36 & $-$64\arcdeg25\arcmin45\arcsec.7 & 0.094 & \phn5.0$\pm$0.5 & 11.5$\pm$0.4 & \phn25.8/\phn0.0 & 13.3 & --- &   \\
  Abell 2244 & 17\fh02\fm42\fs53 & $+$34\arcdeg03\arcmin39\arcsec.4 & 0.097 & \phn8.2$\pm$1.1 & 12.3$\pm$0.5 & \phn58.7/\phn2.9 & 12.4 & --- &   \\
  Abell 2426 & 22\fh14\fm33\fs66 & $-$10\arcdeg22\arcmin09\arcsec.9 & 0.098 & \phn3.7$\pm$0.4 & \phn9.5$\pm$0.4 & \phn\phn9.4/\phn0.0 & \phn7.3 & --- &   \\
  Abell 3827 & 22\fh01\fm53\fs36 & $-$59\arcdeg56\arcmin46\arcsec.8 & 0.098 & \phn6.1$\pm$0.7 & 11.2$\pm$0.4 & \phn45.6/\phn0.0 & 19.6 & --- &   \\
  \sidehead{\midz\ Sample}
  MACS J0416.1 & 04\fh16\fm08\fs80 & $-$24\arcdeg04\arcmin13\arcsec.0 & 0.395 & \phn9.1$\pm$2.0 & \phn3.9$\pm$0.5 & 285.6/\phn0.0 & \phn4.7 & \phn8.5 &   \\
  MACS J2211.7 & 22\fh11\fm45\fs90 & $-$03\arcdeg49\arcmin42\arcsec.0 & 0.396 & 18.1$\pm$2.5 & \phn5.0$\pm$0.2 & \phn13.4/\phn0.0 & 11.8 & 14.7 &   \\
  MACS J0429.6 & 04\fh29\fm36\fs00 & $-$02\arcdeg53\arcmin05\arcsec.0 & 0.399 & \phn5.8$\pm$0.8 & \phn3.4$\pm$0.2 & \phn19.3/\phn0.0 & \phn2.9 & \phn8.9 & Rel. \\
  MACS J0451.9 & 04\fh51\fm54\fs70 & $+$00\arcdeg06\arcmin18\arcsec.0 & 0.430 & \phn6.3$\pm$1.1 & \phn3.3$\pm$0.2 & \phn\phn9.7/\phn0.0 & \phn5.8 & \phn8.1 &   \\
  MACS J1206.2 & 12\fh06\fm12\fs30 & $-$08\arcdeg48\arcmin05\arcsec.0 & 0.439 & 19.2$\pm$3.0 & \phn4.7$\pm$0.2 & \phn19.9/\phn0.0 & 13.3 & 21.7 &   \\
  MACS J0417.5 & 04\fh17\fm34\fs30 & $-$11\arcdeg54\arcmin27\arcsec.0 & 0.443 & 22.1$\pm$2.7 & \phn4.9$\pm$0.2 & \phn81.5/\phn0.0 & 13.3 & 22.7 &   \\
  MACS J0329.6 & 03\fh29\fm41\fs50 & $-$02\arcdeg11\arcmin46\arcsec.0 & 0.450 & \phn7.9$\pm$1.3 & \phn3.4$\pm$0.2 & \phn22.2/\phn0.0 & \phn3.1 & 12.1 &   \\
  MACS J1347.5 & 13\fh47\fm30\fs80 & $-$11\arcdeg45\arcmin08\arcsec.0 & 0.451 & 21.7$\pm$3.0 & \phn4.8$\pm$0.2 & 206.5/\phn0.0 & 11.2 & 36.6 & Rel. \\
  MACS J1311.0 & 13\fh11\fm01\fs70 & $-$03\arcdeg10\arcmin39\arcsec.0 & 0.494 & \phn3.9$\pm$0.5 & \phn2.6$\pm$0.1 & \phn88.7/\phn0.0 & \phn2.0 & \phn9.6 & Rel. \\
  MACS J2214.9 & 22\fh14\fm57\fs29 & $-$14\arcdeg00\arcmin11\arcsec.0 & 0.503 & 13.2$\pm$2.3 & \phn3.8$\pm$0.2 & \phn26.0/\phn0.0 & \phn8.3 & 12.6 &   \\
  MACS J0257.1 & 02\fh57\fm09\fs10 & $-$23\arcdeg26\arcmin03\arcsec.0 & 0.505 & \phn8.5$\pm$1.3 & \phn3.2$\pm$0.2 & \phn35.2/\phn0.0 & \phn5.4 & 10.1 &   \\
  MACS J0911.2 & 09\fh11\fm10\fs90 & $+$17\arcdeg46\arcmin31\arcsec.0 & 0.505 & \phn9.0$\pm$1.2 & \phn3.3$\pm$0.2 & \phn36.6/\phn0.0 & \phn5.1 & \phn4.8 & Dist. \\
  MACS J0454.1 & 04\fh54\fm11\fs40 & $-$03\arcdeg00\arcmin50\arcsec.0 & 0.538 & 11.5$\pm$1.5 & \phn3.4$\pm$0.2 & \phn46.5/\phn0.0 & \phn7.1 & 24.3 &   \\
  MACS J1423.8 & 14\fh23\fm47\fs90 & $+$24\arcdeg04\arcmin43\arcsec.0 & 0.543 & \phn6.6$\pm$0.9 & \phn2.9$\pm$0.1 & 123.7/\phn0.0 & \phn0.8 & \phn9.4 & Rel. \\
  MACS J1149.6 & 11\fh49\fm35\fs40 & $+$22\arcdeg24\arcmin04\arcsec.0 & 0.544 & 18.7$\pm$3.0 & \phn4.0$\pm$0.2 & 248.3/\phn0.0 & 11.3 & 17.4 & Dist. \\
  MACS J0018.5 & 00\fh18\fm33\fs40 & $+$16\arcdeg26\arcmin13\arcsec.0 & 0.546 & 16.5$\pm$2.5 & \phn3.8$\pm$0.2 & \phn61.0/\phn0.0 & \phn8.6 & 15.7 &   \\
  MACS J0025.4 & 00\fh25\fm29\fs90 & $-$12\arcdeg22\arcmin44\arcsec.0 & 0.584 & \phn7.6$\pm$0.9 & \phn2.8$\pm$0.1 & 145.7/\phn0.0 & \phn1.5 & 12.3 &   \\
  MACS J0647.8 & 06\fh47\fm49\fs70 & $+$70\arcdeg14\arcmin55\arcsec.0 & 0.591 & 10.9$\pm$1.6 & \phn3.2$\pm$0.2 & \phn33.1/\phn0.0 & \phn5.8 & 14.4 &   \\
  MACS J2129.4 & 21\fh29\fm25\fs70 & $-$07\arcdeg41\arcmin31\arcsec.0 & 0.589 & 10.6$\pm$1.4 & \phn3.1$\pm$0.2 & \phn26.1/\phn0.0 & \phn2.8 & 15.2 & Dist. \\
\enddata
\tablecomments{Name, center coordinates, redshift, mass, radius, data quality, and morphological classification of the sample.}
\label{tab:sample}
\end{deluxetable*}

Based on the symmetry-peakiness-alignment (SPA) criteria of \citet{Mantz2015} we classify two
of the \lowz\ and four of the \midz\ galaxy clusters as highly relaxed, and from the
similarly defined symmetry-alignment criteria of \citet{Sayers2016} we classify four of the \lowz\
and three of the \midz\ galaxy clusters as highly disturbed. Thus, there is a slight
difference in population statistics between the two samples, with the \midz\
objects being more relaxed on average. As a result, there is the potential
for differences in the ensemble average scaled pressure profiles between these samples due to dynamical
state rather than the mass and redshift trends we seek to measure, and we address
this issue in Section~\ref{sec:three_comparison}. 

For each cluster in our sample we estimate the total mass within
an aperture enclosing 500 times the critical density of the universe (\mfive)
using the ICM density profiles derived in Section~\ref{sec:chandra}
in combination with an assumed gas mass fraction of $f_{\textrm{gas}}(\textrm{R} \le \textrm{R}_{500}) = 0.125$
\citep[][see Table~\ref{tab:sample}]{Mantz2016_scaling}.
Specifically, we solve the equation
\begin{equation}
  \textrm{M}_{500} = \frac{\textrm{M}_{\textrm{gas}}(\textrm{R} \le \textrm{R}_{500})}
         {f_{\textrm{gas}}(\textrm{R} \le \textrm{R}_{500})} =
         \frac{4 \pi}{3} 500 \rho_{\textrm{cr}}(z) \textrm{R}^3_{500},
\end{equation}
where $\textrm{M}_{\textrm{gas}}$ is the enclosed gas mass within \rfive\
and $\rho_{\textrm{cr}}(z)$ is the critical density of the universe.
Given the redshift ranges of the two samples, the \lowz\ galaxy clusters are drawn
from a much smaller volume. As a result, their masses are lower on average compared to
the \midz\ sample (e.g., the median masses
are $6.1 \times 10^{14}$~\msun\ and $10.6 \times 10^{14}$~\msun, respectively).
However, there is significant overlap, with masses
that range from 3.7--$11.3 \times 10^{14}$~\msun\ for the \lowz\ sample and
from 3.9--$22.1 \times 10^{14}$~\msun\ for the \midz\ sample. Therefore,
even though the mass distributions are not identical,
the samples do provide good leverage for separating redshift evolution
from mass dependence.

\chandra\ X-ray and \planck\ SZ effect data are available for all of the
objects in the \lowz\ and \midz\ samples, and Bolocam SZ effect data
are available for all of the objects in the \midz\ sample. Our initial
reduction of these data followed the procedure detailed
in \citet{Wan2021}, which we briefly summarize below.

\subsection{\chandra}
\label{sec:chandra}

Our reduction and cleaning procedure is based on the techniques described in
\citet{Mantz2014,Mantz2015} using \texttt{CIAO} version 4.9 and \texttt{CALDB}
version 4.7.4. From these reduced data, we obtain deprojected ICM density
and temperature profiles based on the method described in \citet{Mantz2014,Mantz2016}.
Specifically, the ICM is modeled as a set of spherical shells, and the
density, temperature, and metallicity are simultaneously fitted for in
all of the shells, fully accounting for covariances between
the fitted values. The assumed centers for the fits, given in Table~\ref{tab:sample},
were determined by an iterative median procedure, as described
by \citet{Mantz2015}.
For both the \lowz\ and \midz\ samples the profiles generally cover the radial
range $0.01\textrm{R}_{500} \lesssim \textrm{R} \lesssim 0.6$\rfive.

\subsection{\rosat}

For most of the \lowz\ galaxy clusters, we make use of \rosat\ PSPC data in
addition to \chandra. Given its larger field of view, these \rosat\
data allow us to probe larger angular scales in these nearby objects.
The reduction of these data are described in detail in \citet{Mantz2010}
and \citet{Mantz2016_scaling}. In particular, the \rosat\ gas
density profiles were rescaled to forge agreement
with \chandra\ at overlapping radii far enough from the galaxy cluster
centers to be minimally impacted by the \rosat\
point spread function.

\subsection{\planck}

We use the publicly available R.2.00 all-sky \planck\ $y$-maps generated using the \texttt{MILCA}
algorithm \citep{Planck2016_XXII}, utilizing the procedures detailed in \citet{Sayers2016}.
While these data have been processed to remove
unwanted astrophysical contamination from primary CMB anisotropies, radio emission,
and other relevant sources, the SZ effect signal
from other objects aligned in projection with the target of interest still
remain in the $y$-maps. Additionally, residual galactic dust emission is also
present in some regions of the sky. To mitigate the impact of these unwanted signals,
we search for any map pixels with a signal to noise ratio $\textrm{SNR} \ge 5$ in the radial range
$2\textrm{R}_{500} \le \textrm{R} \le 6$\rfive. Any data within 2 full-width at half maxima (FWHM)
of the point spread function (PSF), equal to 20\arcmin,
of such pixels within this radial range are removed from our analysis. In addition,
we remove any data within 2 PSF FWHM of objects listed in the meta-catalog
of X-ray detected clusters of galaxies \citep[MCXC,][]{Piffaretti2011}. Due to
the large projected angular sizes of the \lowz\ objects at $\textrm{R} \le 6$\rfive,
we typically identify up to 2 such regions to remove from each galaxy cluster,
mainly based on the positions of MCXC sources. No such regions are found
in any of the \midz\ galaxy clusters, likely due to their small projected
sizes and generally high galactic latitudes. For the \lowz\ sample the
\planck\ data provide fidelity to radial scales between
$0.3\textrm{R}_{500} \lesssim \textrm{R} \lesssim 6$\rfive, and for the \midz\ sample
the \planck\ data provide fidelity to radial scales between
$1.5\textrm{R}_{500} \lesssim \textrm{R} \lesssim 6$\rfive.

\subsection{Bolocam}

Our analysis is based on the publicly available ``filtered'' SZ images
from Bolocam \citep{Sayers2013_pressure}, using the updated flux
calibration described in \citet{Sayers2019}. We estimate this calibration
to be accurate to 1.6~per~cent relative to \planck\
(which is accurate to 0.1~per~cent). No Bolocam data are available
for the \lowz\ sample. For the \midz\ sample
Bolocam data provide fidelity to radial scales between
$0.2\textrm{R}_{500} \lesssim \textrm{R} \lesssim 2$\rfive.

In aggregate, we thus have good radial overlap between all three datasets
used in this analysis, with X-ray observations generally providing constraints
within $\simeq 0.6$\rfive\ and SZ data generally providing constraints outside
of $\simeq 0.2$\rfive. 

\section{Joint X-ray/SZ Pressure Profile Deprojections}
\label{sec:deprojection}

As with the data reduction detailed above, we follow the techniques
described in \citet{Wan2021} to obtain a single ICM pressure profile fit
to each galaxy cluster under the assumption of spherical symmetry.
In brief, the radial pressure profile is described by a set of 11
pressure values P$_i$ located at logarithmically spaced radii R$_i$. An identical set of
R$_i$, in scaled coordinates relative to \rfive\ and spaced between 0.028\rfive\
and 4.6\rfive, is used for every galaxy cluster. Furthermore, the P$_i$ values
for every object are scaled relative to
\begin{equation}
  \textrm{P}_{500} = (3.68 \times 10^{-3}) \times \left( \frac{\textrm{M}_{500}}{10^{15} \: \textrm{M}_{\sun}} \right)^{2/3}
    \times E(z)^{8/3}
  \label{eqn:p500}
\end{equation}
where \pfive\ is in units of keV cm$^{-3}$ and
$E(z)^2 = \Omega_{\textrm{\tiny{M}}}(1+z)^3 + \Omega_{\Lambda}$ \citep{Nagai2007,Arnaud2010}.
To generate a continuous
model of the pressure, which is needed for comparison with the observed data,
the profile is assumed to follow a power law with
a constant exponent between the P$_i$. Beyond 6\rfive, the pressure
is set equal to zero. To allow for a calibration offset in
the X-ray data we also include a parameter $\mathcal{R}$,
which is a constant multiplicative factor applied to the P$_i$ prior
to comparison with the X-ray data.

\begin{figure}
  \centering \includegraphics[width=\columnwidth]{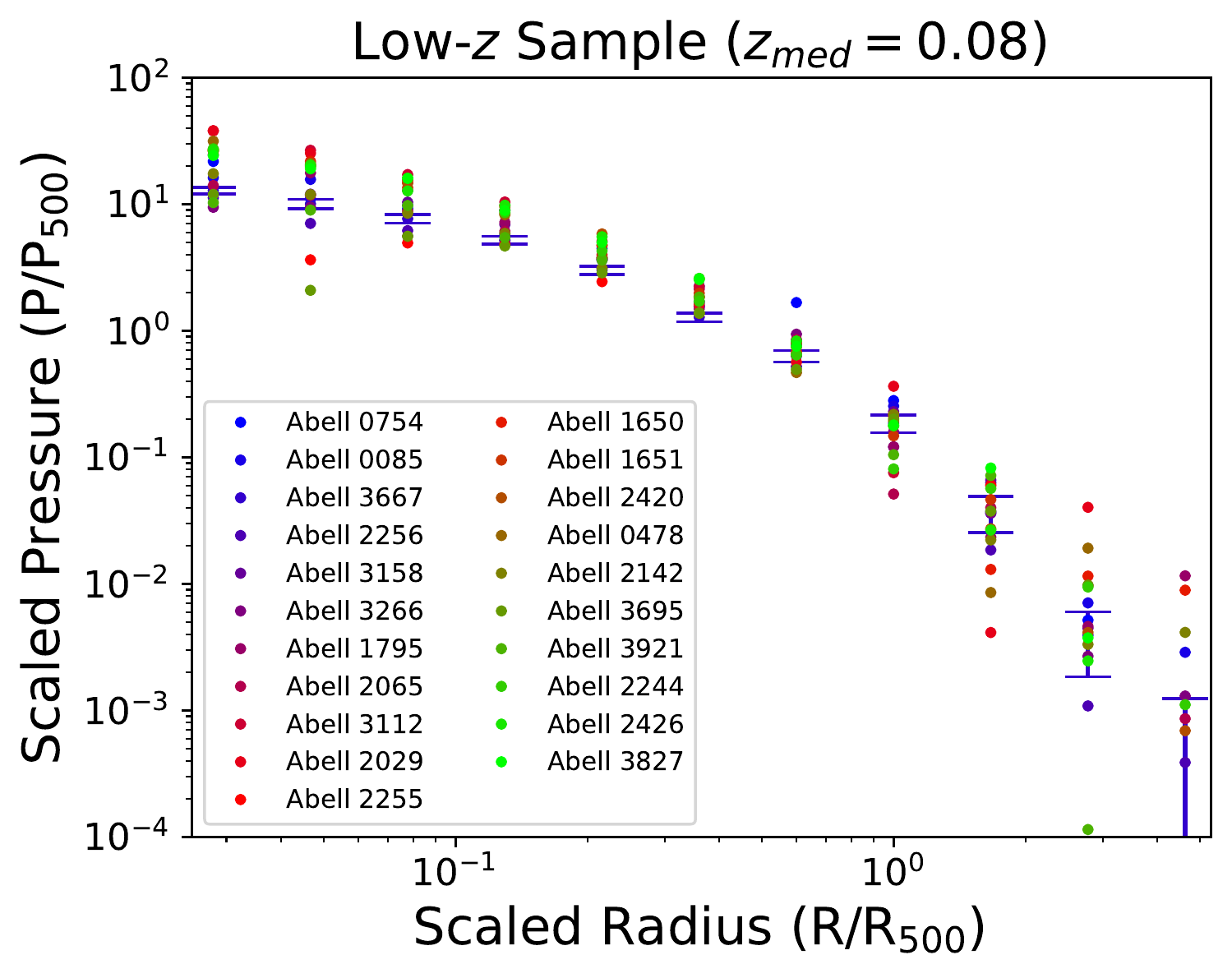}
    \includegraphics[width=\columnwidth]{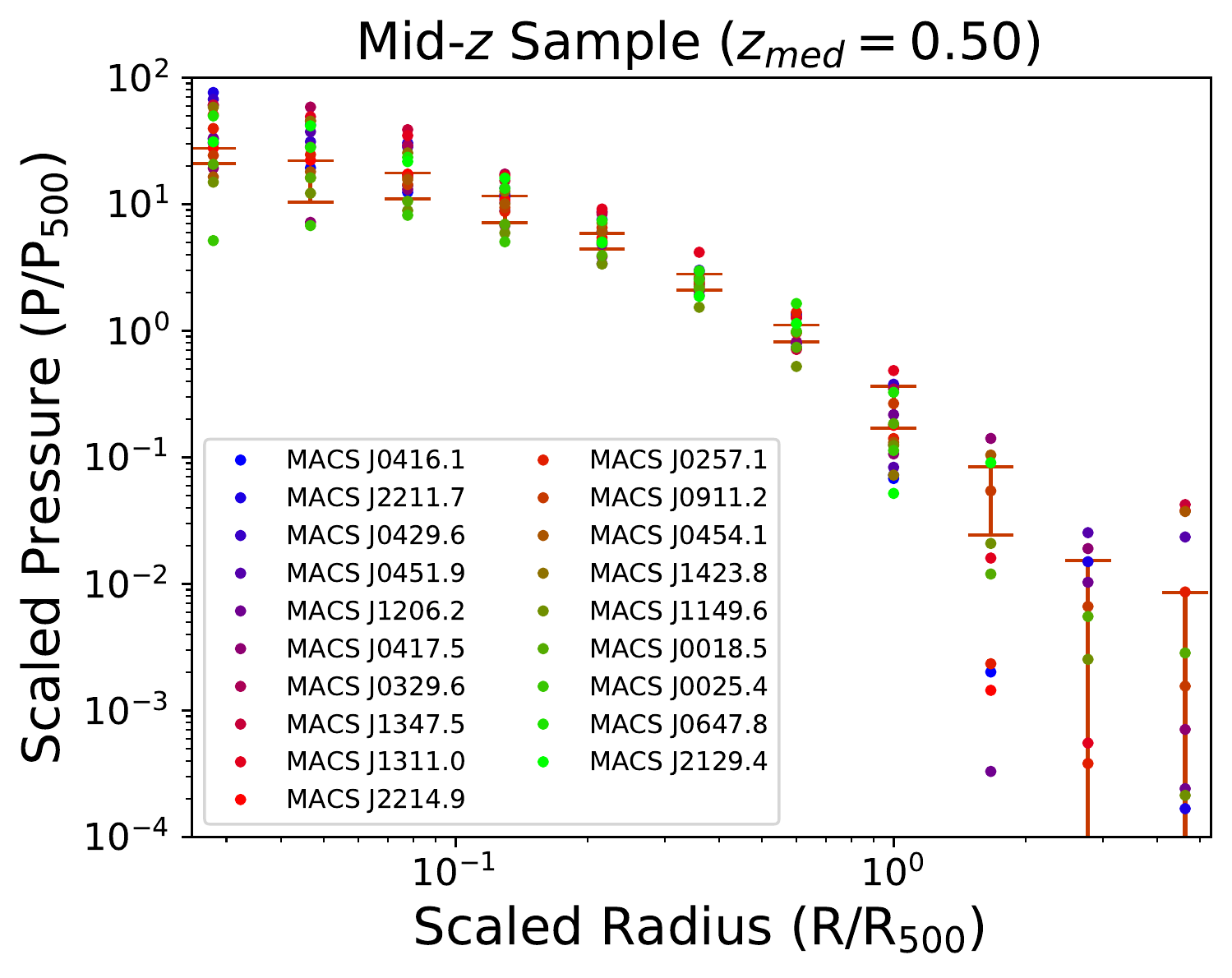}
    \caption{Joint X-ray/SZ effect scaled pressure deprojections for all 40 galaxy clusters
      in our sample, with the \lowz\ objects plotted on the top and the \midz\
      objects plotted on the bottom. The dots indicate the best-fit value at each
      radius for each galaxy cluster. The square root of the diagonal elements of the
      covariance matrix are shown as error bars for a single galaxy cluster in each plot
    to provide an indication of the typical measurement uncertainty.}
  \label{fig:individual}
\end{figure}

We simultaneously fit the X-ray and SZ effect data using a generalized
least squares algorithm to obtain the best-fit values of the P$_i$
and $\mathcal{R}$ \citep{Markwardt2009}.
In the case of the X-ray data, the deprojected
ICM density and temperature profiles are compared directly with the
pressure profile model. To compare with the SZ data,
the three-dimensional pressure
model is projected along one dimension to obtain a two-dimensional
$y$-map. As part of this projection, relativistic corrections to the
SZ effect are included based on the X-ray temperature profile using
\texttt{SZpack} \citep{Chluba2012,Chluba2013}. For \planck, this
projected model is convolved with the PSF shape, which is assumed
to be Gaussian with a FWHM of 10\arcmin. An analogous PSF convolution
is also used for Bolocam, based on a Gaussian with a FWHM of 58\arcsec.
In addition, the filtering effects of the Bolocam data processing,
which are described by a two-dimensional transfer function included
in the public data release, are applied to the projected model.
The deprojections for all 40 galaxy clusters in our sample are shown
in Figure~\ref{fig:individual}, with numerical values provided
in Table~\ref{tab:deproj_individual} of the Appendix.

In performing these fits, we assume the noise is accurately modeled by a diagonal
covariance matrix, primarily because it is not possible to quantitatively
estimate the off-diagonal elements of the Bolocam covariance matrix
\citep[for more details, see][]{Wan2021}.
To test for biases due to this assumption, we generate a set of mock datasets
by inserting a galaxy cluster with a known pressure profile into random
noise realizations of the various datasets. For Bolocam, 1000 such noise
realizations are included with the public data release, while for \planck\
they were generated from a combination of the homogeneous and inhomogeneous
noise spectra associated with those data. Samples from the Markov
chain Monte Carlo employed to analyze the X-ray data were used to create
noise realizations associated with the \chandra\ and \rosat\ data.
For each of these samples, the values of \mfive\ and \rfive\ were
also extracted and used to scale the $\textrm{P}_i$ and $\textrm{R}_i$
values, thus ensuring that uncertainties on our mass estimates
are properly accounted for in our overall error budget.

A total of 100 mocks were then analyzed in an identical
manner to the observed data, including the assumption of a diagonal
covariance matrix, and the fitted pressure profiles were compared with
the known input profiles. We find no evidence for a radial trend
in the ratio of fitted to input pressure, and we also find no evidence
for an on-average non-zero difference between the two values.
The standard deviation of the ratio between
the fitted and true pressure value is 0.022 for the \lowz\ sample
and 0.033 for the \midz\ sample. These numbers represent the fluctuations
in the recovered pressure profiles due to our fitting procedure. Compared
to the measurement noise and calibration uncertainties on the pressure
values, these systematic variations due to fitting are a factor
of approximately 5--10 smaller. Since these fluctuations are negligible
to our overall error budget, and they also do not produce a measurable
bias that might appear in ensemble averages, we do not account for them
in our analysis.

\begin{figure}
  \centering \includegraphics[width=\columnwidth]{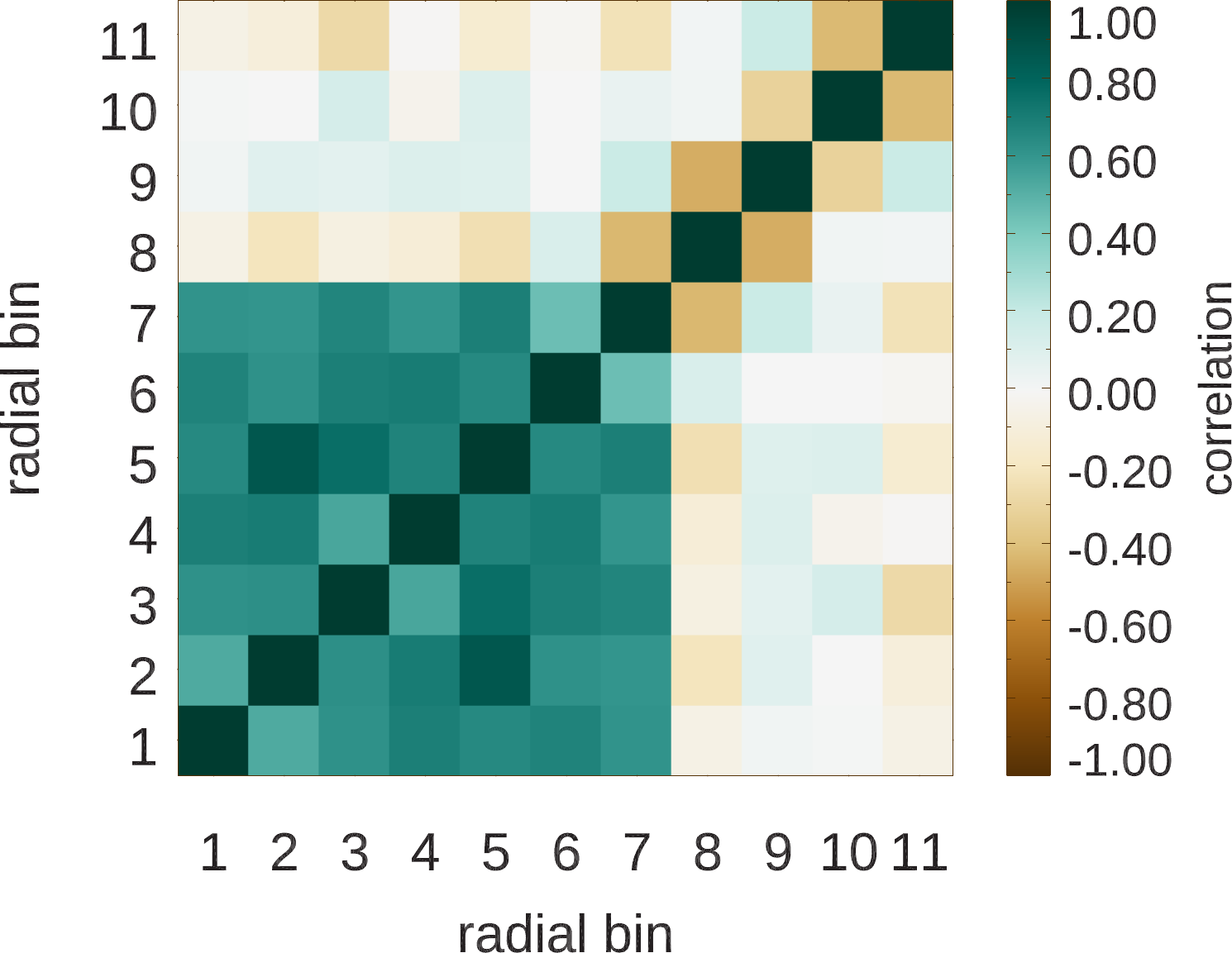}
  \caption{Correlation values of the noise covariance matrix associated
    with the joint X-ray/SZ effect deprojection of Abell 1650. A similar
    structure is observed in the covariance matrices of the other galaxy
  clusters.}
  \label{fig:corr}
\end{figure}

In addition to using our fits to the mocks to search for biases, we also
use them to characterize the noise covariance matrix associated with the
joint X-ray/SZ effect pressure deprojection values P$_i$. In general, the off-diagonal
elements of this covariance matrix are significant. For the inner radial
bins, which are largely constrained by the X-ray data, the P$_i$ are strongly
positively correlated due to the constant multiplicative factor
$\mathcal{R}$ included in the fits to account for uncertainties in the X-ray
temperature calibration. In the outer radial bins, which are largely
constrained by the SZ effect data, there is typically a large
negative correlation between adjacent P$_i$. This arises from the course
angular resolution of the data, whereby the observed maps are largely
unchanged when an upward fluctuation in one pressure bin is compensated
by downward fluctuations in the adjacent bins. A typical example of
the correlation values of the covariance matrix is given in Figure~\ref{fig:corr}.

Because of the typically large correlations in the noise covariance matrices for
each galaxy cluster, the matrices contain
small eigenvalues that can lead to unstable fits \citep[e.g.][]{Michael1994, Yoon2012}.
In order to use these individual pressure profiles to obtain
the ensemble fits described in Sections~\ref{sec:mean} and \ref{sec:gNFW},
we thus need to address this problem.
Several solutions to obtain more stable fits from
such covariance matrices have been proposed, including setting
all of the off-diagonal elements to zero
\citep{Bae2010}. For this analysis, we choose to employ the cutoff method,
also known as the singular value decomposition method, in order to obtain
covariance matrices that produce more stable fits
\citep{Bhattacharya1999, Gamiz2011, Yoon2012}. Specifically, we compute
the eigenvalues of the covariance matrices for all 40 galaxy clusters. For each
object, we then search for sufficiently small eigenvalues, quantified as
those smaller than $10^{-4}$ of the maximum eigenvalue. The eigenmodes associated
with these small values are removed, and the covariance matrix is reformed
based on the remaining eigenmodes and eigenvalues. This trimming occurs
in a total of 5 galaxy clusters, and involves 1--2 eigenmodes per object.
Thus, only a small fraction of our sample is impacted by this procedure.

\section{Ensemble Mean Scaled Pressure Profiles}
\label{sec:mean}

\subsection{Analysis Procedure}

Using the set of individual pressure deprojections described in Section~\ref{sec:deprojection},
which have all been computed in units of scaled pressure relative to \pfive\
at a set of common scaled radii relative to \rfive,
we constrain the ensemble mean profile for each of the two subsamples (\lowz\ and \midz).
To determine the mean profile, we use the publicly available \texttt{LRGS}
software package from \cite{Mantz2016_LRGS}, which also returns the covariance
matrix of the intrinsic scatter about the mean profile (i.e.,
$\Sigma_{\textrm{int}}(\textrm{R}_i,\textrm{R}_i)$). We perform these fits
in a two-step process.

First, an initial estimate of the mean profile is obtained directly from
the scaled pressure deprojections. Using this initial estimate,
we compute a single multiplicative normalization
factor N$_{l}$ for the deprojection of each individual galaxy cluster $l$.
Specifically, we first compute
\begin{equation}
  \mathcal{P}_i^l = \mathrm{P}_i^l/\langle \mathrm{P}_i^l \rangle 
\end{equation}
where $\langle \mathrm{P}_i^l \rangle$ is the mean profile.
Next, we compute
\begin{equation}
  \textrm{N}_l = [\vec{\mathcal{P}^l}]^{\mathrm{T}} [C^l]^{-1} \vec{\mathcal{P}^l}
\end{equation}
where $C^l$ is the weighted covariance matrix associated with $\vec{\mathrm{P}^l}$,
and the vector represents the set of radial bins $i$.
Then, we fit the N$_{l}$ as a function of
$\tilde{\textrm{M}}_l = \textrm{M}_{500,l}/ \langle \textrm{M}_{500} \rangle$,
where $\langle \textrm{M}_{500} \rangle$ is the mean mass of the given
subsample and the assumed functional form is
N$_{l} = \tilde{\textrm{M}}_l^{\alpha_{\textrm{\tiny{M}}}}$. We find consistent
values for the power law exponent in both the \lowz\ and \midz\ subsamples,
with $\alpha_{\textrm{\tiny{M}}} = -0.30 \pm 0.19$ and $-0.14 \pm 0.18$, respectively.
These values indicate that our data favor, at low significance, a
slightly shallower dependence of \pfive\ with mass compared to
Equation~\ref{eqn:p500} (i.e., closer to \mfive$^{1/2}$ rather than
\mfive$^{2/3}$). 
Given the narrow redshift range of each subsample, we do not attempt to
fit an analogous relation as a function of $E(z)$.

\begin{figure}
  \centering \includegraphics[width=\columnwidth]{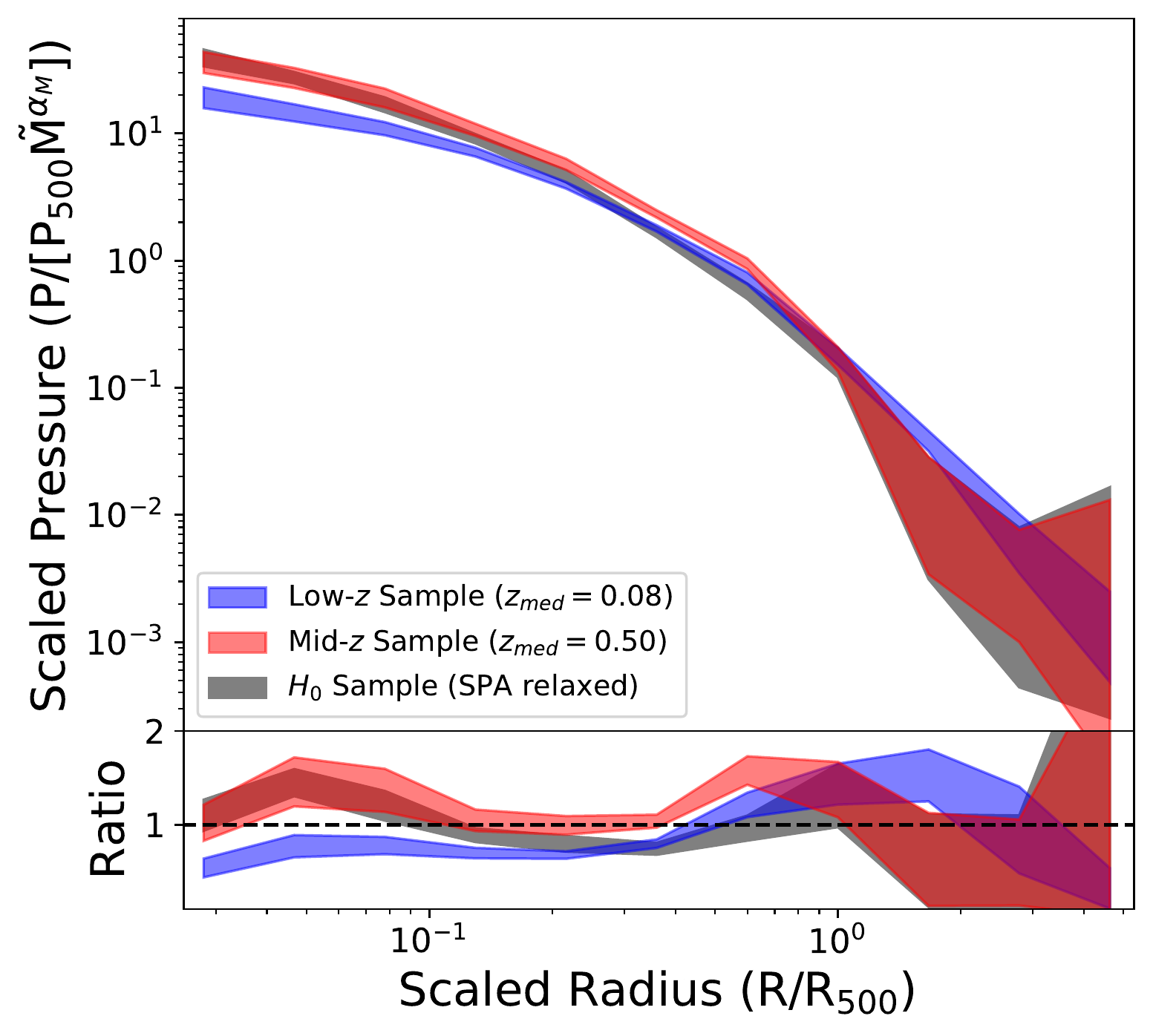}
    \includegraphics[width=\columnwidth]{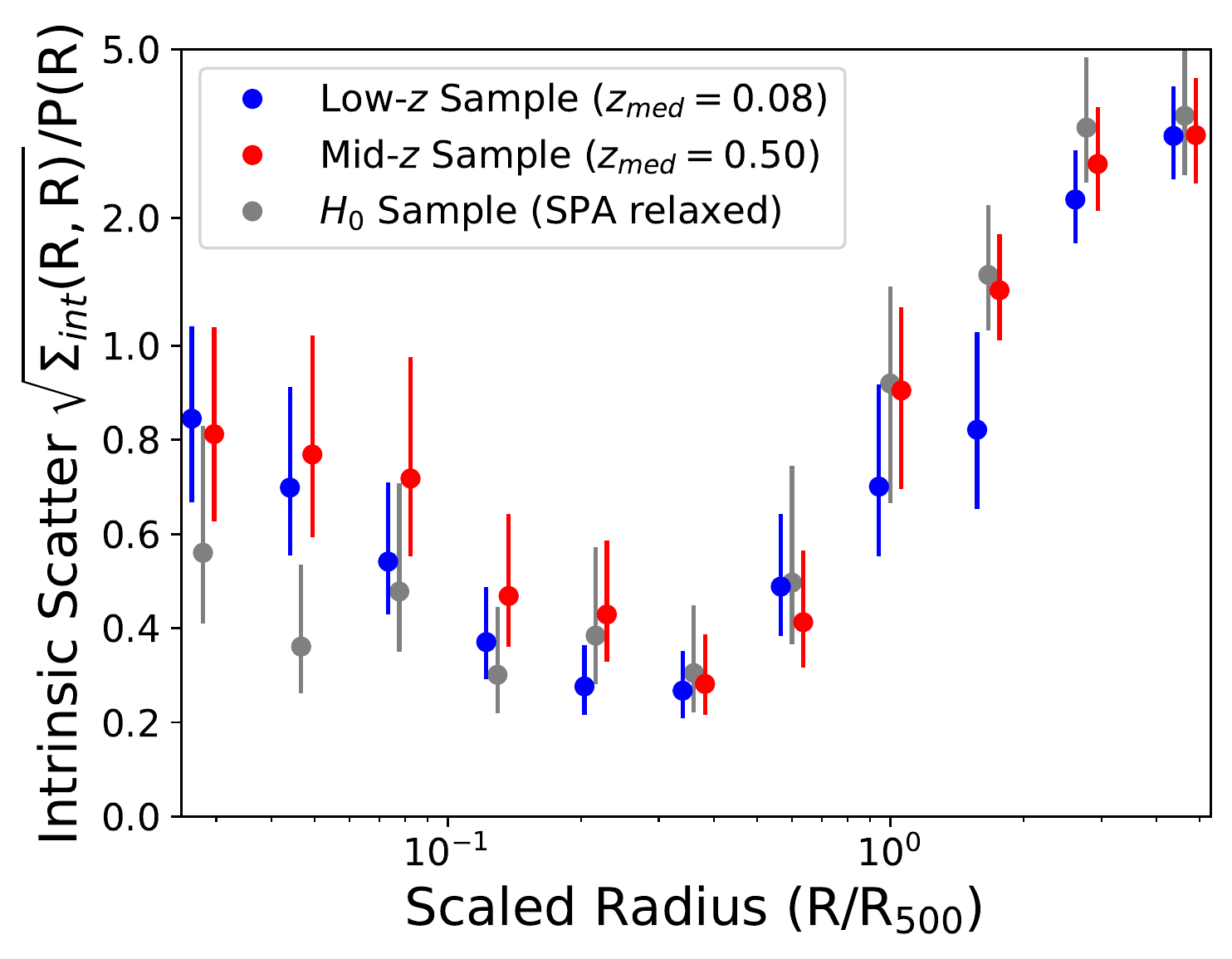}
  \caption{The top panel shows the
    ensemble mean scaled pressure profile for the \lowz, \midz, and $H_0$ samples, with
    the ratio of each profile to the average of the three profiles shown
    in the lower part of the panel.
    The shaded region indicates the 68~per cent confidence region.
    As noted in the text, the factor of $\tilde{\mathrm{M}}^{\alpha_{\textrm{\tiny{M}}}}$
    has almost no impact on the mean scaled pressure profiles.
    The bottom panel shows the fractional intrinsic scatter about the
    mean profile, with the \lowz\ and \midz\ points slightly offset
    in radius for clarity.}
  \label{fig:mean_results}  
\end{figure}

Second, using the fitted value of $\alpha_{\textrm{\tiny{M}}}$ for each subsample,
we rescale the pressure values of each individual deprojection.
We then refit for the ensemble mean profile, along with the intrinsic scatter
about that mean profile. Compared to the initial fit (with $\alpha_{\textrm{\tiny{M}}} = 0$),
there is a slightly smaller intrinsic scatter but essentially no change to the
ensemble mean profile. A summary of the results is given in Figure~\ref{fig:mean_results},
with numerical values provided in Tables~\ref{tab:deproj_lowz} and \ref{tab:deproj_midz}
in the Appendix.
For the bins at R$_i \le 1.7$\rfive\ the individual P$_i$ are generally measured at high
significance, with a SNR of at least 3 when estimated from the diagonal elements of
the covariance matrix. The same is true for the ensemble mean profiles, where the
SNR is comparable to the individual cluster values  due to the large intrinsic scatter.
In the two outermost bins, the SNR is typically lower, corresponding to 
$\simeq 2$ for the \lowz\ sample and to $\simeq 1$ for the \midz\ sample for
both the individual profiles and the mean profiles.
We note that the \planck\
data for the \midz\ sample are in general noisier than for the \lowz\ sample, which
is the primary reason for the difference in uncertainties on the mean
profiles at the largest radii. The scaled pressure profiles
for both samples monotonically steepen with increasing radius, corresponding
to an effective power law exponent of $\sim -0.5$ at the smallest radii and
$\sim -3$ to $\sim -5$ at the largest radii (see Section~\ref{sec:three_comparison}).
Finally, in both the \lowz\ and \midz\ samples the fractional intrinsic scatter decreases
from near unity in the smallest radial bin to a minimum of $\sim 0.2$ near
0.5\rfive. Beyond that radius the fractional intrinsic scatter increases,
reaching a value of $\gtrsim 2$ in the largest radial bin.

\subsection{Comparison of the \lowz, \midz, and $H_0$ Samples}
\label{sS:redshift}

From Figure~\ref{fig:mean_results}, it is clear that there is a systematic difference in the
mean scaled pressure profile shape between the \lowz\ and \midz\ samples. Specifically, at radii
smaller than approximately 0.6\rfive, the mean scaled pressure is significantly higher
in the \midz\ sample compared to the \lowz\ sample. The trend reverses near 2\rfive,
with a higher mean scaled pressure in the \lowz\ sample, although at modest statistical
significance. The intrinsic scatter about the mean profile is consistent at all
radii for the two samples.

We have also repeated the same
analysis using the highly relaxed sample of galaxy clusters from
\citet[hereafter the $H_0$ sample]{Wan2021}
to search for any obvious trends with dynamical state.
The $H_0$ sample contains 14 objects distributed approximately uniformly
between $0.08 \le z \le 0.54$ with a median
redshift of $z=0.33$. The masses range from $4.7 \le \textrm{M}_{500} \le 17.4 \times 10^{14}$~\msun\
with a median mass of $8.5 \times 10^{14}$~\msun. Thus, based on both mass and redshift,
the $H_0$ sample is almost perfectly intermediate to the \lowz\ and \midz\
samples.

At the smallest and largest radii, corresponding to $\textrm{R} \lesssim 0.2$\rfive\ and
$\textrm{R} \gtrsim 1$\rfive, we find that the mean scaled pressure profiles
of the $H_0$ and \midz\ samples are in good agreement.
At intermediate radii,
the mean scaled pressure profile of the \lowz\ sample more closely traces that of the
$H_0$ sample. Thus, rather than falling in between the mean scaled pressure profiles
of the \lowz\ and \midz\ samples as would be expected purely due to
differences in redshift and mass, the mean profile from the $H_0$ sample
is instead much more sharply peaked at small radii. We attribute this
to: 1) the uniform presence of cool cores in the $H_0$ sample, which are known to
enhance the central pressure \citep[e.g.,][]{Arnaud2010,Sayers2013}, and 2)
the center determined from the procedure in Section~\ref{sec:chandra} is likely
to be coincident with the X-ray peak for cool-core systems, thus retaining
a higher central pressure for those galaxy clusters.

We find that the fractional intrinsic scatter about the mean scaled pressure profile
is somewhat smaller in the $H_0$ sample for the innermost radial bins, although at
low statistical significance. Such a difference is expected, again due to the
uniformity of cool cores within this sample which
reduces the relative spread among the population in the central regions.

\subsection{Comparison with \three\ Simulations}
\label{sec:three_comparison}

To better understand if the observed difference between the mean scaled pressure profiles
in the \lowz\ and \midz\ samples is due to evolution, mass scaling, and/or
dynamical state,
we also performed our analysis on a set of scaled pressure profiles obtained
from \three\ simulations \citep{Cui2018}. A range of redshift snapshots for
324 galaxy clusters are available in these simulations,
and we have selected the two that most closely
match the median redshifts of our \lowz\ and \midz\ samples, which
are at $z=0.07$ and $z=0.46$. Within each snapshot, we would ideally like
to apply a simple and quantitative selection based on the
X-ray luminosity in order to obtain a large statistically representative
set of galaxy clusters matching the properties of the observed samples.
However, such a selection is unlikely to yield the best set of
matched galaxy clusters.
First, the \midz\ sample contains numerous additional selections
beyond a simple luminosity threshold that are not straightforward
to include, such as the timing between when subsamples of MACS clusters were known
in the literature and when Bolocam conducted its observations.
Furthermore, our observed samples were selected from the full sky,
and include the most massive and thus rare objects.
While the underlying dark-matter only simulation used to obtain
\three\ sample is large, with a comoving side length of 1 $h^{-1}$ Gpc,
it is unclear if a selection based on X-ray luminosity from this
volume will provide the best individual matches to the observed \midz\ systems.

\begin{figure}
  \centering \includegraphics[width=\columnwidth]{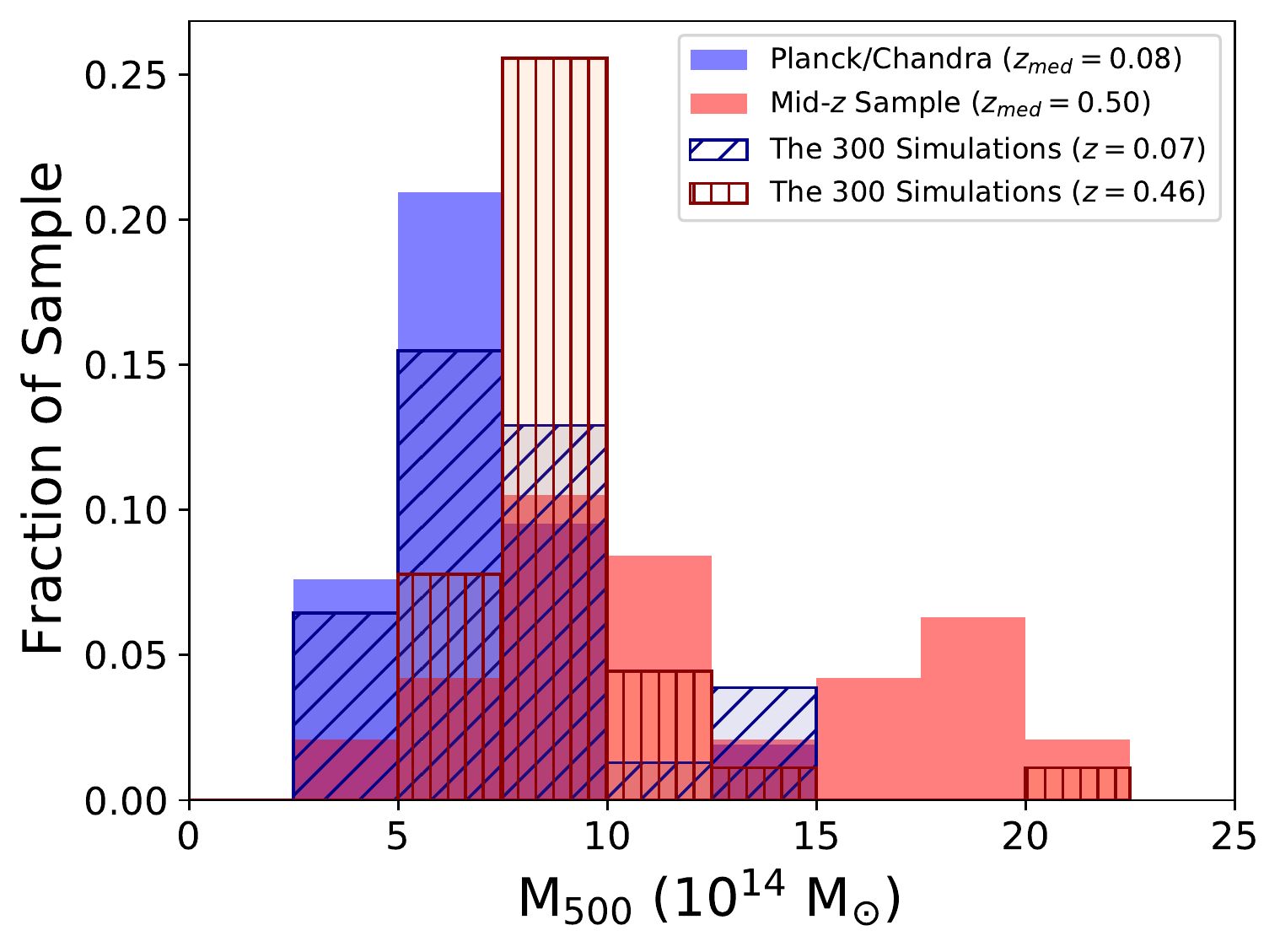}
  \caption{Mass distributions of the observed samples and the matched
    samples selected from \three\ simulations.}
  \label{fig:mass}
\end{figure}

Instead, within each snapshot, we select the three simulated
galaxy clusters most similar in mass to each observed galaxy cluster, removing
duplicates. Three matches was chosen as a reasonable compromise between obtaining
sufficiently large samples (which motivates a larger number of matches)
while still retaining a good representation of the
population statistics (which motivates a smaller number of matches).
From these candidate samples, we then remove all of the objects that
contain two or more distinct sub-clusters, indicating they are in a pre-merger
phase (i.e., analogous to the removal of Abell 399/401 from the observational sample).
Next, we determine which objects are
relaxed or disturbed based on the criteria defined by \citet{Cui2018}, and
randomly remove galaxy clusters until the fraction of both types approximately
match the observed samples. Following these cuts, we were left with a samples
of 31 $z=0.07$ and 36 $z=0.46$ simulated galaxy clusters.
With the exception of the mass distribution of the simulated sample
at $z=0.46$, where there were not enough high mass objects to match the
observed \midz\ sample, these simulated samples are well matched to the
properties of the observed samples (i.e., redshift, mass, and dynamical state,
see Figure~\ref{fig:mass} and Table~\ref{tab:othersamples}).

\begin{deluxetable}{cccccc}
\tablenum{2}
\tablecaption{Observed and Simulated Samples}
\tablewidth{0pt}
\tablehead{
  \colhead{Sample} & N$_{\textrm{clus}}$ & \colhead{N$_{\textrm{r}}$} & \colhead{N$_{\textrm{d}}$} & \colhead{$\langle$\mfive $\rangle$} & \colhead{$\langle$$z$$\rangle$}}
\startdata
This Work (\lowz) & 21 & \phn2 & \phn4 & \phn6.1\phantom{$^{\textrm{a}}$} & 0.08 \\
\citet{Arnaud2010} & 31 & 10 & 12 & \phn2.6\phantom{$^{\textrm{a}}$} & 0.12 \\
\citet{Bourdin2017} & 61 & 22 & 39 & \phn7.2\phantom{$^{\textrm{a}}$} & 0.15 \\
\citet{Ghirardini2019} & 12 & \phn4 & \phn8 & \phn5.7\phantom{$^{\textrm{a}}$} & 0.06 \\
The300 (\lowz\ Matched) & 31 & \phn3 & \phn6 & \phn6.7\phantom{$^{\textrm{a}}$} & 0.07 \\ \hline 
This Work (\midz) & 19 & 4 & 3 & 10.6\phantom{$^{\textrm{a}}$} & 0.50 \\
\citet{McDonald2014} & 40 & 19 & 21 & \phn5.5\phantom{$^{\textrm{a}}$} & 0.46 \\
\citet{Bourdin2017} & 23 & --- & --- & \phn7.9\phantom{$^{\textrm{a}}$} & 0.56 \\
\citet{Ghirardini2017} & 15 & \phn8 & \phn7 & \phn9.7$^{\textrm{a}}$  & 0.45 \\
The300 (\midz\ Matched) & 36 & \phn8 & \phn6 & \phn8.2\phantom{$^{\textrm{a}}$} & 0.46 \\ 
\enddata
\tablecomments{Studies of ensemble mean scaled pressure profiles from galaxy cluster samples
  with similar masses and redshifts to those in our \lowz\ (top) and \midz\ (bottom) samples,
  along with the samples selected from \three\ simulations.
  The total number of galaxy clusters, the number of relaxed and disturbed galaxy clusters,
  median mass, and median redshift of each sample is
  listed. For studies from the literature, unless otherwise noted by the authors,
  we consider the objects listed as cool cores to
  be relaxed and the non cool cores to be disturbed. \\ $^{\textrm{a}}$Since no masses were provided
  in \citet{Ghirardini2017},
  we compute them from the values of M$_{\textrm{gas,500}}$ given
  in \citet{Amodeo2016} assuming a gas mass fraction of 0.125 \citep{Mantz2016_scaling}}
\label{tab:othersamples}
\end{deluxetable}

\begin{figure}
  \centering \includegraphics[width=\columnwidth]{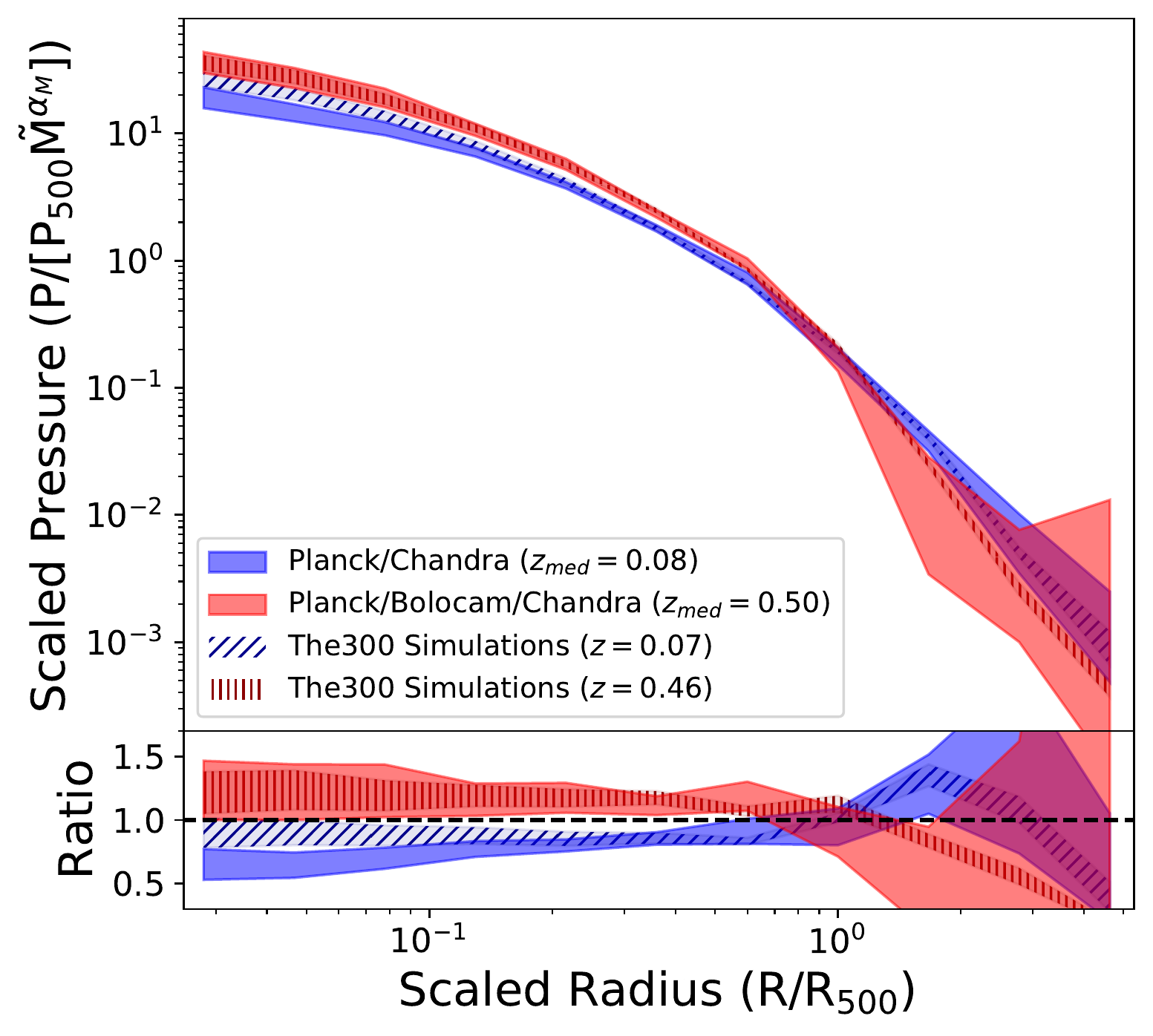}
    \includegraphics[width=\columnwidth]{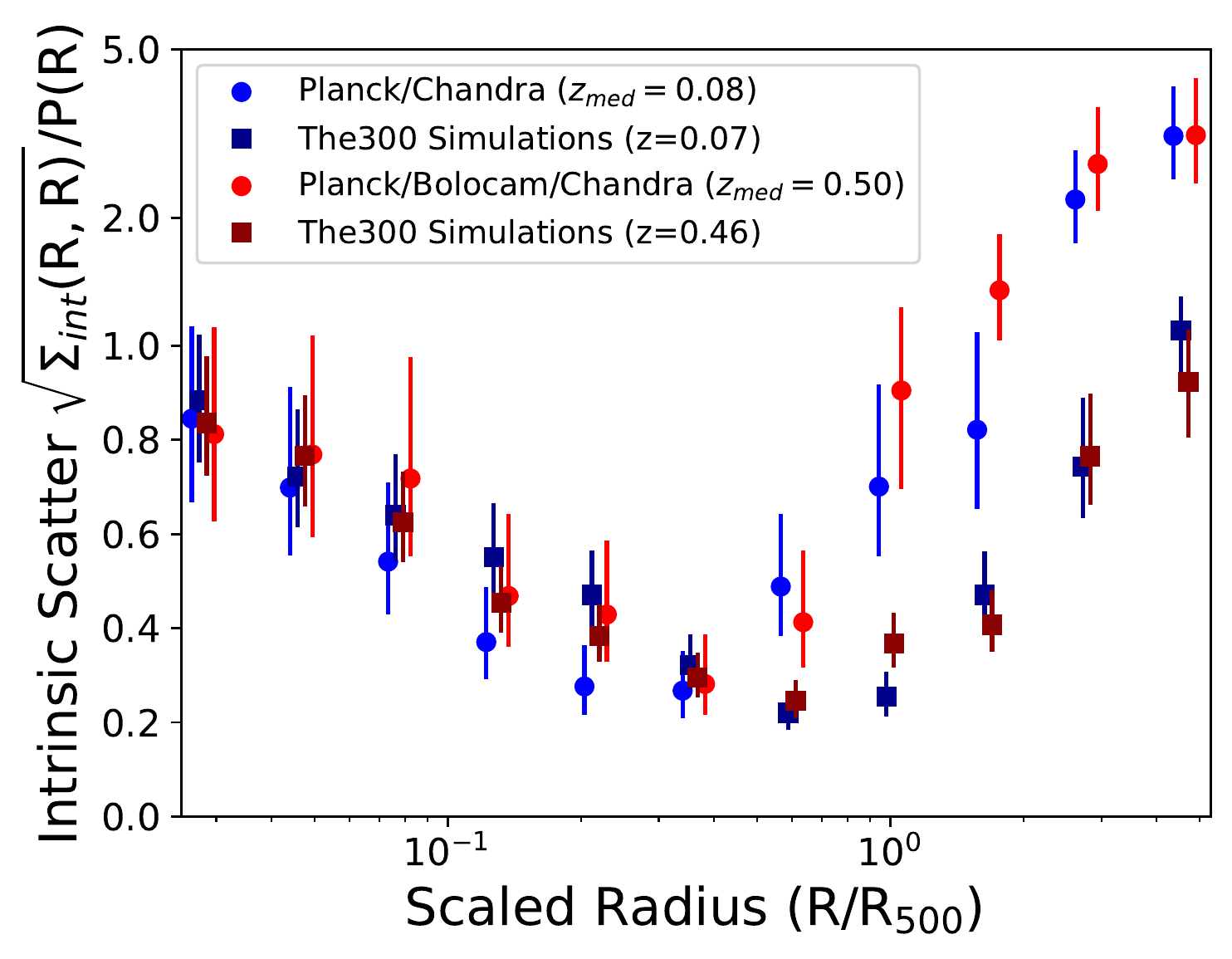}
    \caption{Same as Figure~\ref{fig:mean_results}, showing results from the
      matched samples selected from \three\ simulations.}
  \label{fig:mean_sim_results}  
\end{figure}

\begin{figure}
  \centering
  \includegraphics[width=\columnwidth]{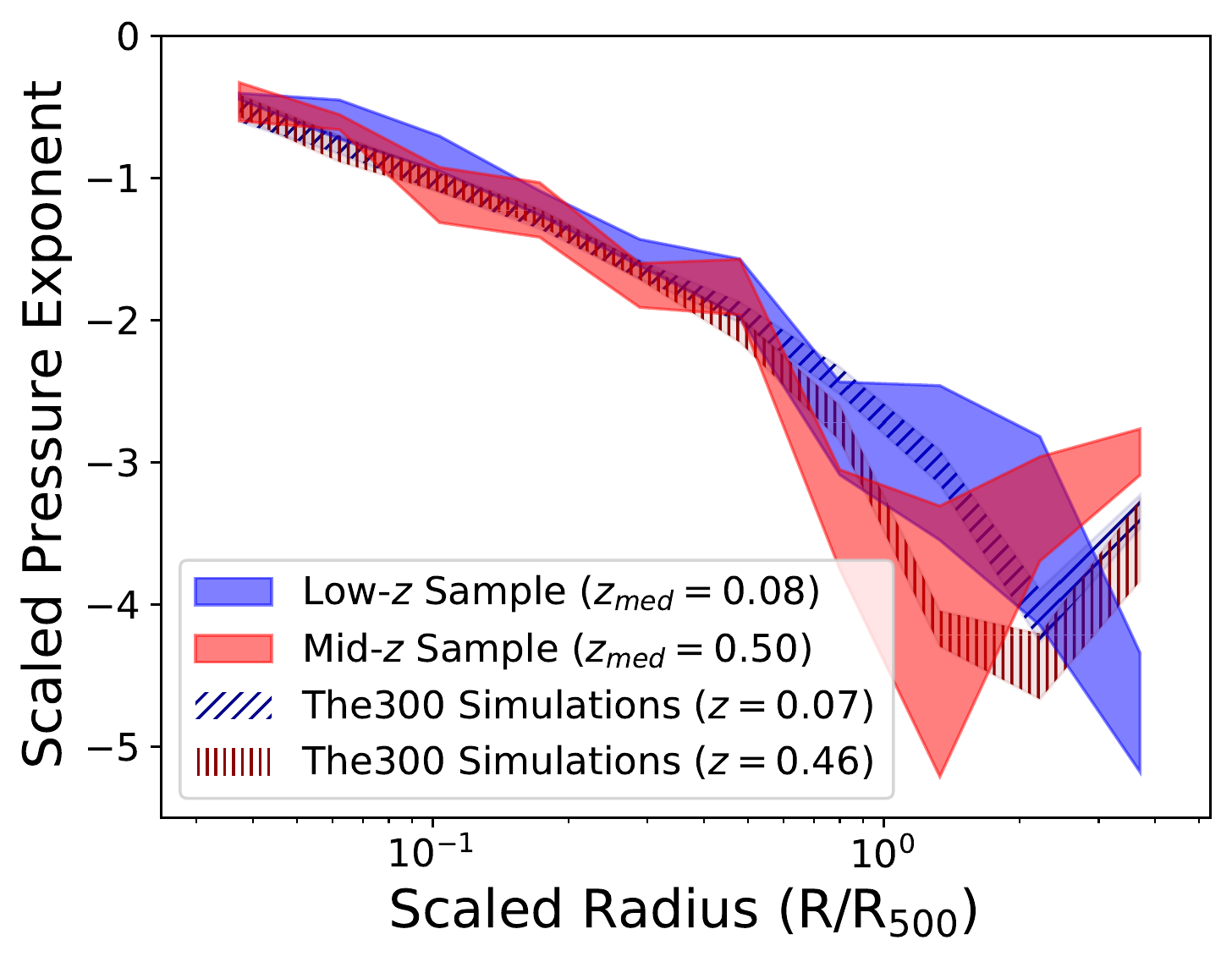}
  \caption{Effective power law exponent of the ensemble mean pressure versus radius.
    Labels are the same as in Figure~\ref{fig:mean_sim_results}.}
  \label{fig:slope}
\end{figure}

To determine the pressure profiles of the simulated galaxy clusters, we use
the center positions obtained from the Amiga Halo Finder algorithm
\citep[AHF,][]{Knollmann2009}. These centers correspond to the location of the maximum
in total density, and are thus likely to be more similar to the observed centers
obtained from the procedure in Section~\ref{sec:chandra} than the location of the ICM peak.
From the initial fits, we find $\alpha_{\textrm{\tiny{M}}} = -0.39 \pm 0.15$ and
$-0.26 \pm 0.16$, for the $z=0.07$ and $z=0.46$ samples, in good agreement
with the values obtained from the observational data.
The shape of the mean scaled pressure profiles recovered from the matched \three\ samples
are shown in the top panel of Figure~\ref{fig:mean_sim_results}, and numerical values
are provided in Tables~\ref{tab:deproj_lowz_sim} and \ref{tab:deproj_midz_sim} of the
Appendix. Overall, the agreement with the
observed data is excellent, with the possible exception of the core region of the \lowz\ sample,
indicating that the simulations accurately reproduce
the measured differences between the \lowz\ and \midz\ samples.
Furthermore, in regards to the uncertainties on the ensemble mean profiles in
Figure~\ref{fig:mean_sim_results}, we note that the intrinsic scatter is larger than the measurement
uncertainties for the observed data at all radii other than the outermost bins. It thus dominates the
overall uncertainty on the mean profile, resulting in similar constraints for
both the observed data and the (perfectly known) simulated data.

In Figure~\ref{fig:slope} we show the effective power law exponent of the
scaled pressure versus radius for the observed and matched simulated samples,
which are generally in excellent agreement. In the central region the exponent
is near $-0.5$, steepening to approximately $-4$ near the maximum radii
included in our analysis.
At both redshifts, \three\ simulations suggest a significant steepening
of the profile between the radial bins at 1.7\rfive\ and 2.8\rfive\,
with a minimum exponent of less than $-4$.
The statistical significance of this minimum is approximately
$5\sigma$ for the \lowz\ sample and $2\sigma$ for the \midz\ sample.
The observed data also indicate a minimum exponent of less than $-4$ at large radii,
between the 2.8\rfive\ and 4.6\rfive\ radial bins for the \lowz\ sample and between
the 1.0\rfive\ and 1.7\rfive\ radial bins for the \midz\ sample. The statistical
significance of these minima is lower than for the simulated samples,
approximately $2\sigma$ for the \lowz\ sample and $1\sigma$ for the \midz\
sample.

Recent results have used SZ data to similarly probe large radius features in the
pressure profile, including \citet[][for a single massive cluster at \lowz]{Hurier2019},
\citet[][for a sample of 10 group-size objects in the local universe]{Pratt2021},
and \citet[][for a sample of 516 clusters detected in the South Pole Telescope
SPT-SZ survey]{Anbajagane2022}.
The latter study, along with \citet{Baxter2021}, also included simulated clusters
from \three.

In mass, redshift, and sample size, the \citet{Anbajagane2022} result
is the only one directly comparable to our analysis, and so we consider that study
in more detail. From \three\ simulations, they find a minimum exponent
near \rtwo\ (which is $\sim 1.5$\rfive), corresponding to slightly smaller radii
than the minimum in our analysis. They also
find a minimum exponent in the observed data at a similar radius, in good agreement
with the location of the minimum we find in the observed \midz\ sample but at
smaller radii than in the observed \lowz\ sample.
The overall agreement between our analyses is thus quite good, particularly given
the differences in analysis technique (our work considers deprojected pressure
profiles and the effect of intrinsic scatter while \citealt{Anbajagane2022}
considers projected pressure profiles) and overall sample characteristics
(the \citealt{Anbajagane2022} sample is most similar in mass to our \lowz\
sample and most similar in redshift to our \midz\ sample).
In addition, the coarse radial binning of our analysis prevents an accurate
determination of the location of the minimum exponent, which may also
be contributing to some of the differences in our results.
We further note that, while the underlying physical cause of this minimum exponent is not
well understood, \citet{Anbajagane2022} suggest it may be due to shock-induced
thermal non-equilibrium between electrons and ions \citep[see, e.g.][]{Avestruz2015}.

\begin{figure*}
  \centering
  \includegraphics[width=0.45\textwidth]{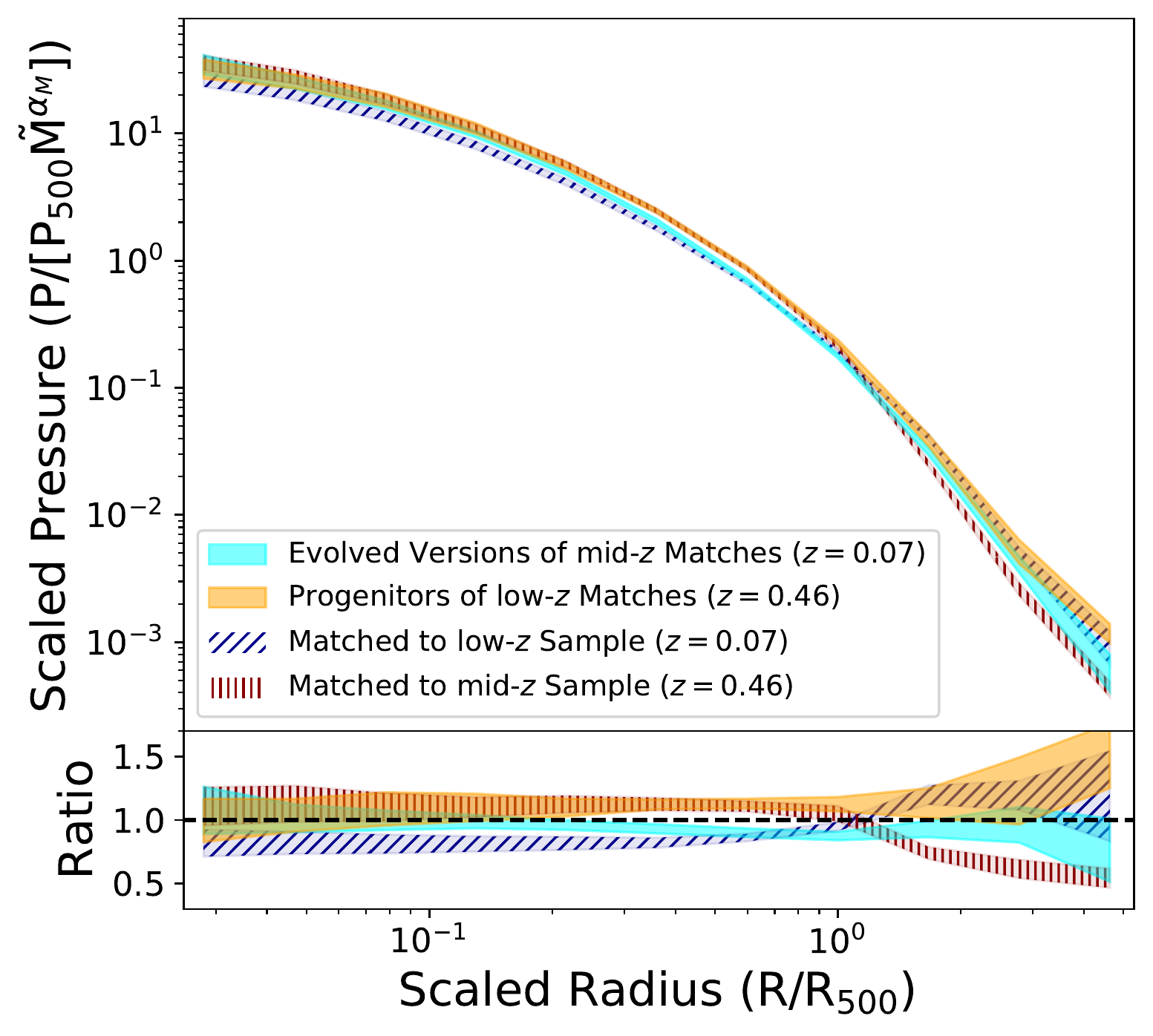}
  \hspace{0.05\textwidth}
  \includegraphics[width=0.45\textwidth]{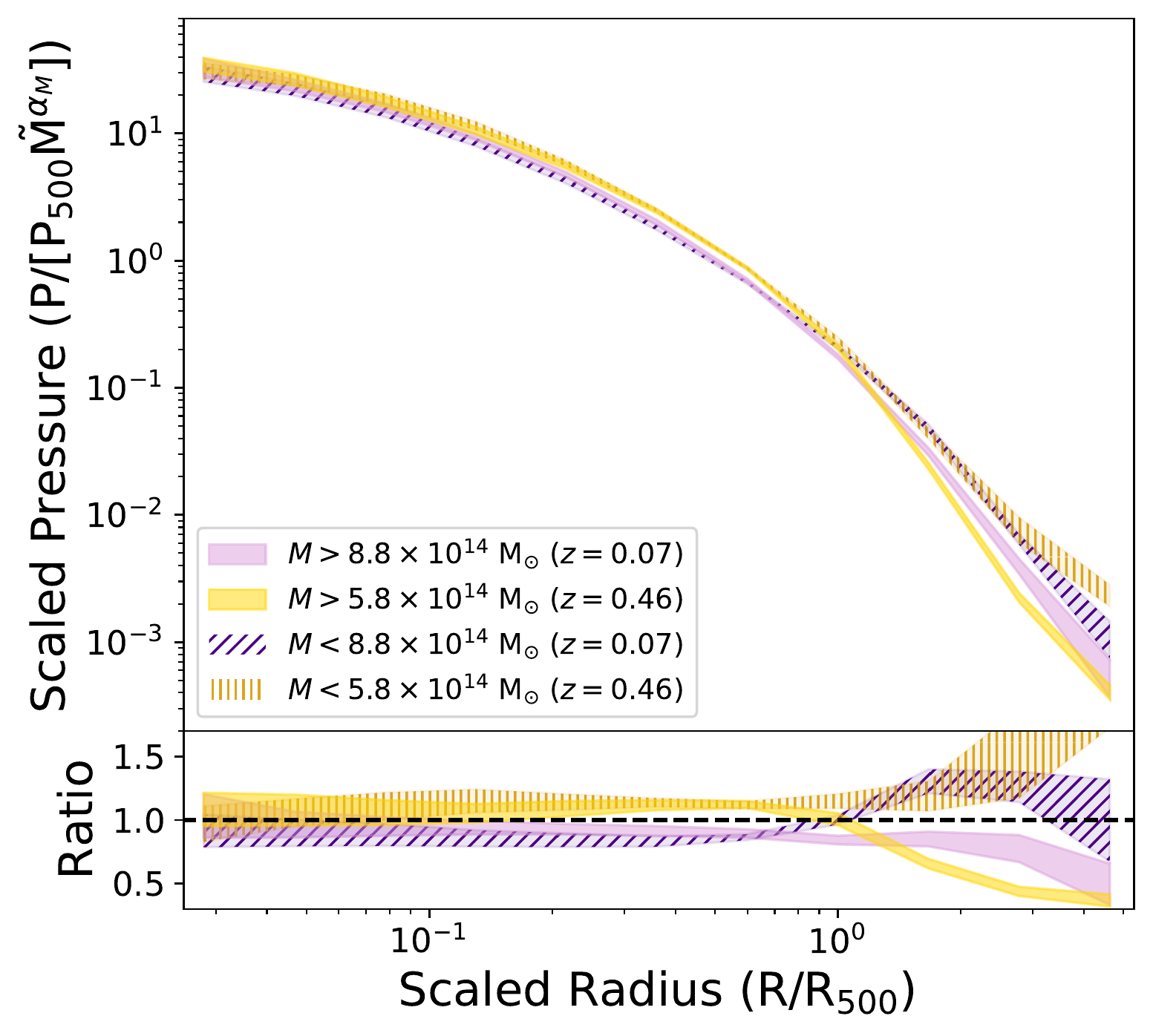}
  \includegraphics[width=0.45\textwidth]{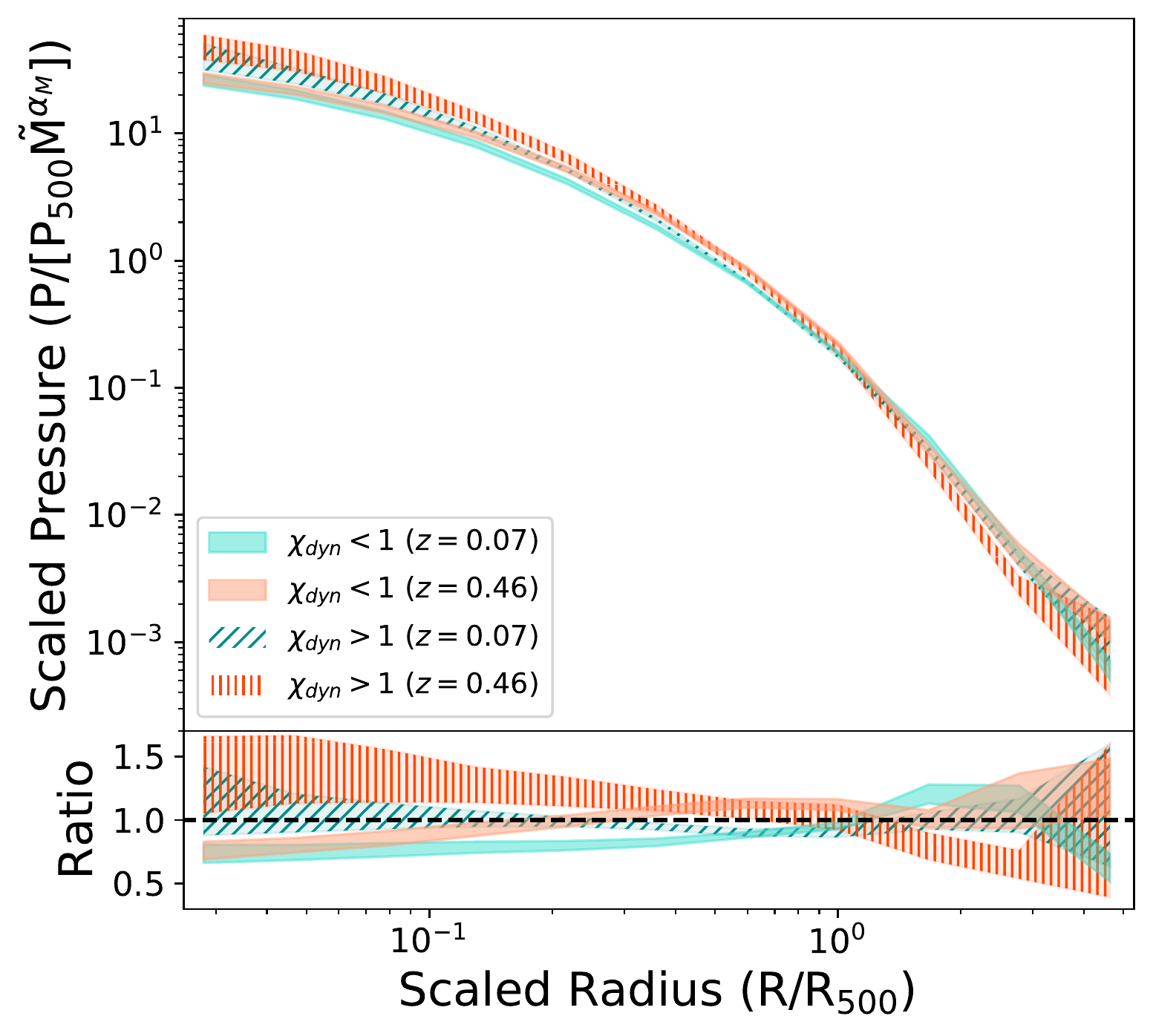}  
  \caption{Upper left: Mean scaled pressure profiles from \three\ for the sets of 31 and 36 galaxy
    clusters selected to match the \lowz\ and \midz\ observational samples. The same sets of
    objects are also shown at different redshifts to isolate the impact of evolution.
    Upper right: Mean scaled pressure profiles from \three\ for samples selected to be above or
    below the median mass at $z=0.07$ and $z=0.46$ to isolate
    the impact of mass.
    Bottom: Mean scaled pressure profiles from \three\ for the set of more dynamically relaxed
    ($\chi_{\textrm{dyn}} > 1$) and less dynamically relaxed
    ($\chi_{\textrm{dyn}} < 1$) objects at each redshift to isolate the impact of dynamical state.}
  \label{fig:the300_evolved}
\end{figure*}

Given the good agreement between the observed and simulated mean scaled pressure profiles,
we defined three additional samples from \three\ to better understand
the underlying cause of the difference between the \lowz\ and \midz\ samples.
First, in order to best isolate the impact of evolution between the observed samples,
we repeated our analysis
for the $z=0.07$ counterparts of the matched $z=0.46$ galaxy clusters (i.e., the
same set of objects were extracted from the simulations at two different redshifts).
A similar analysis was also performed on the $z=0.46$ counterparts of the
matched $z = 0.07$ galaxy clusters. The results are shown in the top left panel of
Figure~\ref{fig:the300_evolved}, and indicate: 1) there is a clear trend of decreasing
scaled pressure with decreasing redshift at intermediate radii, with the \lowz\ and
\midz\ profiles separated by 2--3 times the width of 68 per cent confidence regions;
2) there are hints of a similar trend at small radii near the core, although
all of the profiles overlap within the 68 per cent confidence regions;
3) at large radii there
is evidence of increasing scaled pressure with decreasing redshift in the \midz-matched
galaxy clusters, which are separated by approximately 2 times the width of the 68~per cent
confidence regions, but there is no such trend in the \lowz-matched objects.

Next, to best isolate the impact of mass, we considered the full set of \three\
galaxy clusters at both $z=0.07$ and $z=0.46$ selected either as a match to
the \lowz\ or \midz\ samples
(i.e., 67 objects). We then split
this set of galaxy clusters into two 
samples based on the overall median mass at each snapshot, which is equal to
$8.8 \times 10^{14}$~\msun\ at $z=0.07$ and $5.8 \times 10^{14}$~\msun\ at $z=0.46$. The results are shown
in the top right panel of Figure~\ref{fig:the300_evolved},
and indicate: 1) there is no significant trend in the mean scaled pressure profile
shape based on mass within \rfive, where the profiles at a given redshift snapshot overlap within
the 68 per cent confidence regions at all radii; 2) there is strong trend of increasing scaled pressure
with decreasing mass outside of that radius, with the profiles separated by approximately
2--3 times the width of the 68 per cent confidence region at $z=0.07$ and by even more than
that amount at $z=0.46$;
3) the trend of decreasing scaled pressure with decreasing redshift at intermediate
radii noted in the previous paragraph is again reproduced here with a separation
of approximately 3 times the width of the 68 per cent region.

Finally, to best isolate the impact of dynamical state, we again split the
full set of \three\ galaxy clusters at $z=0.07$ and $z=0.46$ based on the value
of the dynamical state parameter $\chi_{\textrm{dyn}}$ defined by
\citet{DeLuca2021}. In brief, this parameter is computed as a weighted
quadratic sum of the ratio between the thermal and potential energy $\eta$,
the offset between the galaxy cluster center and the center of mass $\Delta_{\textrm{r}}$,
and the mass fraction of all the sub-halos in the galaxy cluster $f_{\textrm{s}}$,
all of which were evaluated within R$_{200}$.
The set of objects with $\chi_{\textrm{dyn}} > 1$ were
defined to be the more dynamically relaxed samples, and the set of objects with
$\chi_{\textrm{dyn}} <  1$ were defined to be the less dynamically relaxed samples.
The results are shown in the bottom panel of Figure~\ref{fig:the300_evolved},
and indicate: 1) a higher mean scaled pressure at small radii in the more dynamically
relaxed sample, with the profiles separated by
approximately 2 times the width of the 68 per cent confidence regions;
2) at larger radii there is less difference in the profiles based on dynamical
state, although the more relaxed samples have slightly lower pressure
in both redshift snapshots;
3) again, the trend of decreasing scaled pressure with decreasing redshift
at intermediate radii is evident, with the profiles separated by approximately
twice the width of the 68 per cent confidence regions.

The lower
panel of Figure~\ref{fig:mean_sim_results} compares the fractional intrinsic scatter
about the mean scaled pressure profiles for the observations and \three.
While the simulations show the same basic radial trend as the
observed data (i.e., a minimum scatter near 0.5\rfive), the
fractional intrinsic scatter in \three\ samples is lower than the observed data at
large radii. This could, for example, indicate
that the observed data have a wider range of accretion histories,
although we note the potential for systematics.
One such possibility is the presence of unmodeled noise in the observed data at
large angular scales, for example due to galactic dust emission or primary
CMB anisotropies, which is then interpreted as additional scatter
in our fits. Astrophysical systematics in the observed data are another
possibility. For instance, the presence of unrelated structures along the LOS
that were not flagged by our analysis procedure would also enhance
the intrinsic scatter recovered in our fits. However, given that the \lowz\ sample
is far more sensitive to such projection effects due to their larger
angular extent, we would expect this
systematic to produce a difference between the \lowz\ and \midz\ samples,
which is not observed.

\subsection{Comparison to Previous Observations}

\begin{figure}
  \centering
  \includegraphics[width=\columnwidth]{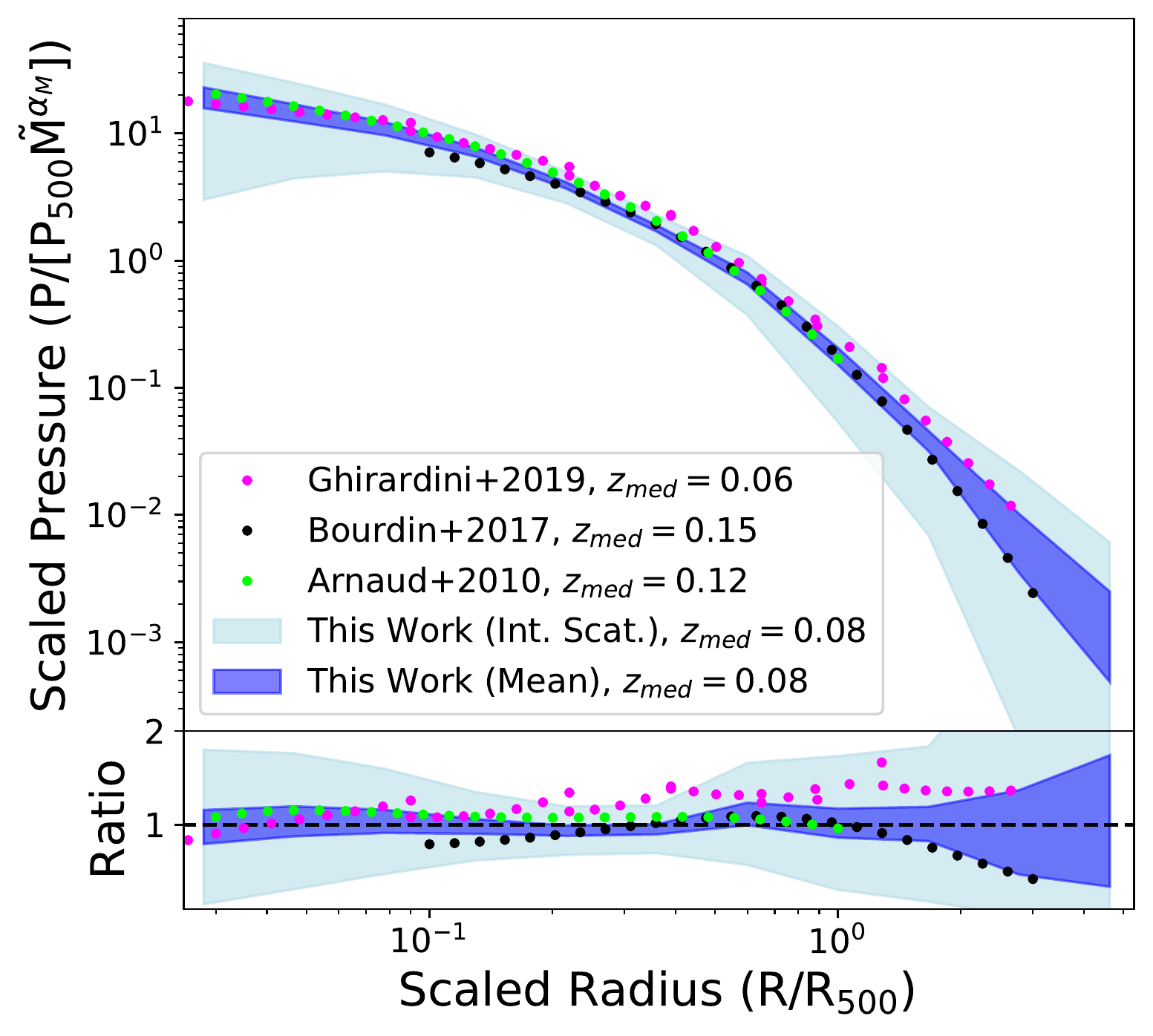}
  \includegraphics[width=\columnwidth]{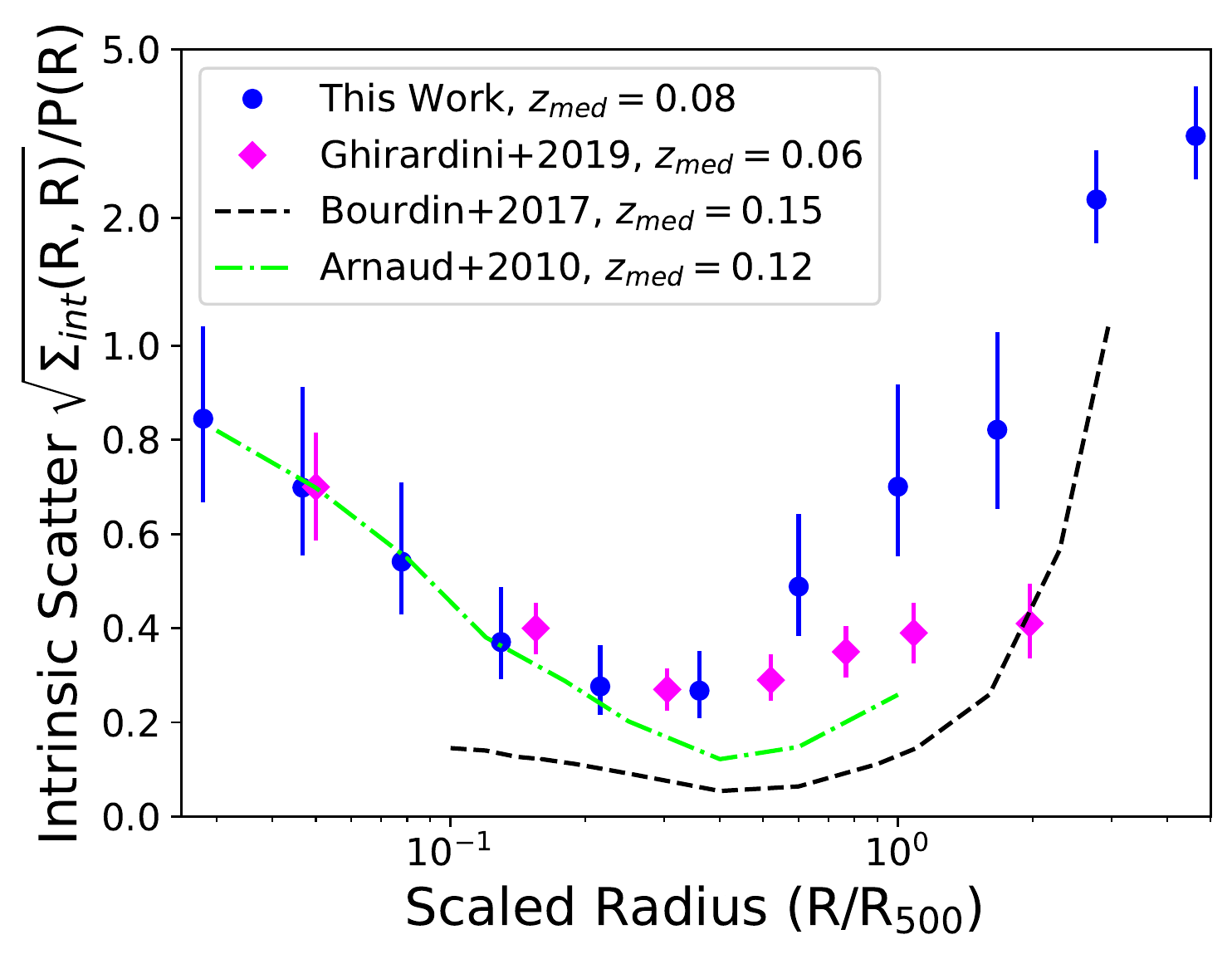}
  \caption{Ensemble mean scaled pressure profiles and intrinsic scatter about the mean profile
    for observational samples near $z=0.10$.}
  \label{fig:lowz_obs}  
\end{figure}

\begin{figure}
  \centering
  \includegraphics[width=\columnwidth]{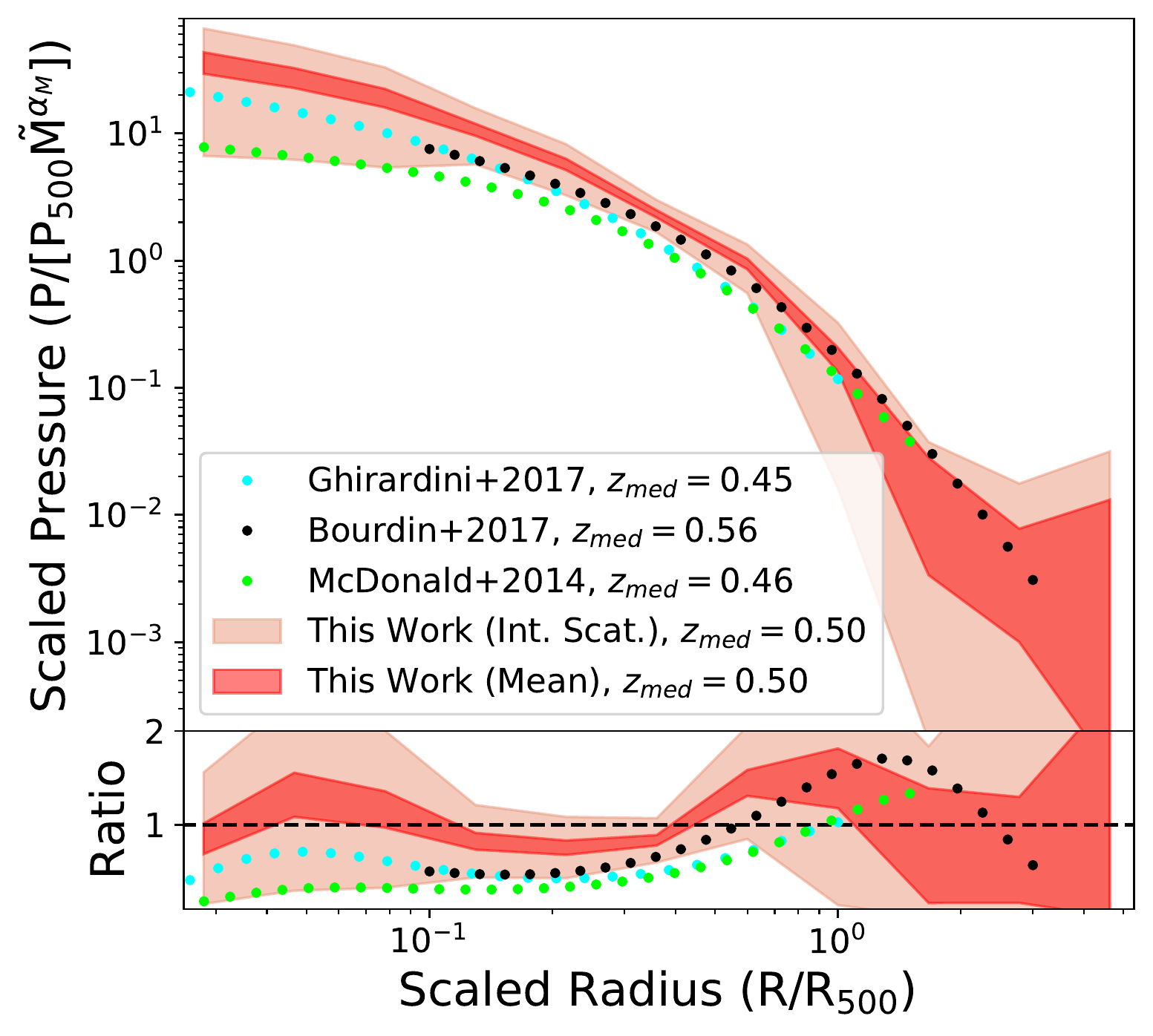}
  \includegraphics[width=\columnwidth]{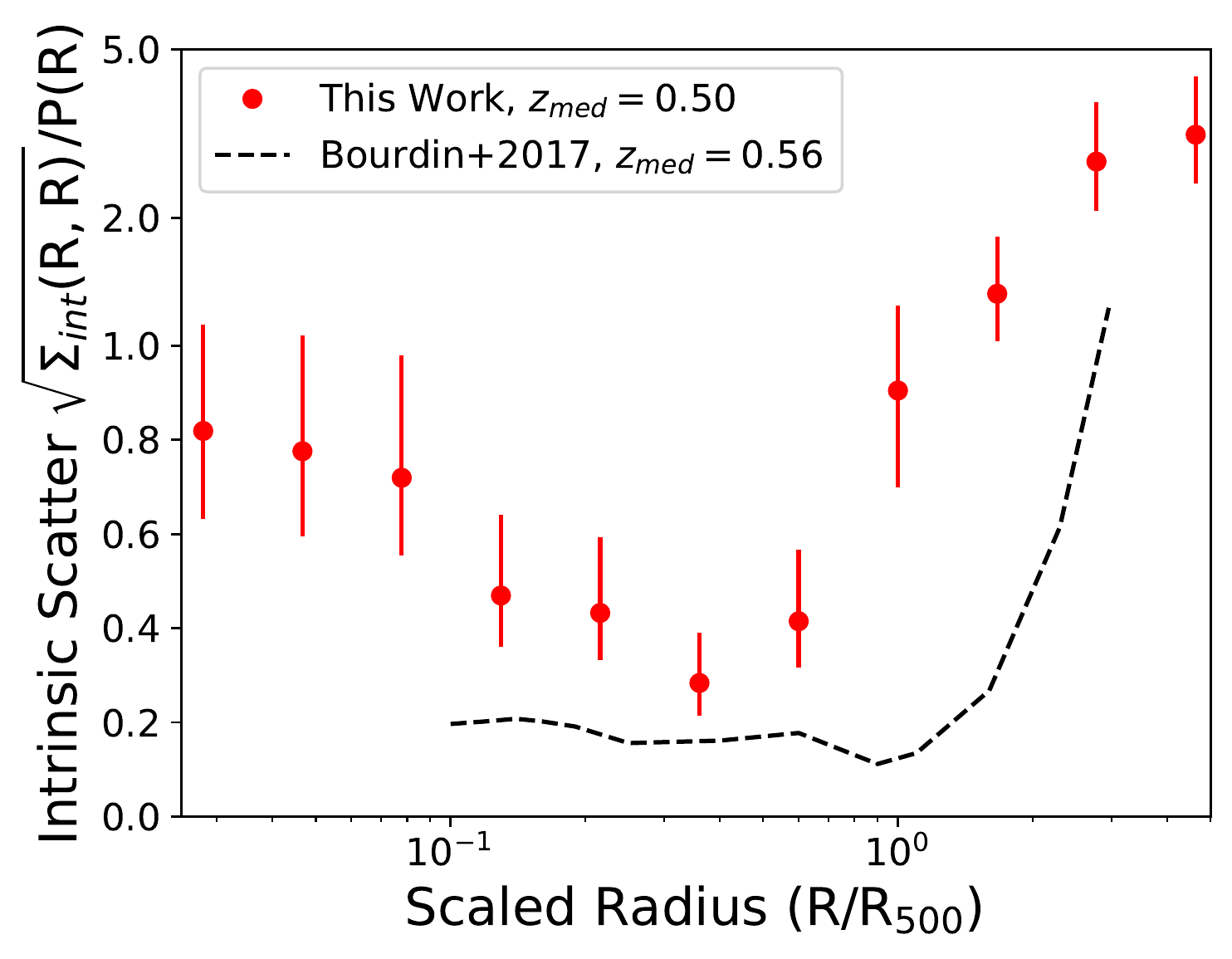}
  \caption{The same as Figure~\ref{fig:lowz_obs}, but for observational samples
    near $z=0.50$.}
  \label{fig:highz_obs}  
\end{figure}

Three previous analyses have measured ensemble mean scaled pressure profiles and the
associated intrinsic scatter for galaxy cluster samples with similar mass
and redshift distributions to our \lowz\ sample
\citep[, see Table~\ref{tab:othersamples}]{Arnaud2010, Bourdin2017, Ghirardini2019}.
As shown in the top panel of Figure~\ref{fig:lowz_obs},
both the overall shape and the normalization of the mean profiles
found in these works are in excellent agreement with our analysis. The level
of consistency is remarkable, particularly given the range of analysis techniques
and observational datasets. There is also generally good agreement in the
fractional scatter about the mean profile at small and intermediate radii
(lower panel of Figure~\ref{fig:lowz_obs}),
although our observed results are somewhat higher at large radii
and \citet{Bourdin2017} find
a lower scatter at all radii.

There are also three previous analyses that have constrained ensemble mean scaled pressure
profiles using samples well matched in mass and redshift to our \midz\ sample
\citep[, see Figure~\ref{fig:highz_obs}]{McDonald2014, Bourdin2017, Ghirardini2017}.
While the agreement is fairly
good near \rfive, there is a noticeable divergence at smaller radii. Our work finds the highest mean
scaled pressure at those radii, while \citet{McDonald2014} obtains much lower values
and the other two studies are approximately half way between.
The trend approximately correlates with the mean mass of the samples, and so
that is a possible explanation for the difference.
Given that the differences are most pronounced near the core, the centering
algorithms used in these works may also play a role \citep[e.g.,][used a centroid determined from an annulus
  at 250--500~kpc, which could result in a lower central scaled pressure]{McDonald2014}.
Alternatively, these discrepancies could be due to the distribution of dynamical states within
the samples. While no obvious differences exist based on the relative fractions of
relaxed and disturbed systems listed in Table~\ref{tab:othersamples}, we note that
each analysis used different criteria for identifying such systems.
For instance, when a uniform set of criteria are used,
there is an observed difference in cool core fraction between SZ- and
X-ray-selected samples \citep{Rossetti2017,Andrade-Santos2017,Lovisari2017}.
The \citet{McDonald2014} sample is purely SZ-selected, which could thus result
in a lower cool core fraction and correspondingly lower inner scaled pressure. In contrast,
our \midz\ sample is X-ray-selected from MACS, which is known to have a significant
bias towards cool core systems \citep{Rossetti2017}. The samples of \citet{Bourdin2017} and
\citet{Ghirardini2017} have less defined selections that are likely somewhat intermediate.
Therefore, it is possible that at least some of the difference between these observed
profiles is due to dynamical state.
Only \citet{Bourdin2017} measured the fractional intrinsic scatter as part of their
analysis. As with the lower $z$
samples, they again find a similar overall trend with radius as our analysis, but
with significantly smaller values.

\section{gNFW Fits}
\label{sec:gNFW}

Particularly for recent analyses involving SZ effect data, the 
gNFW parameterization proposed by \citet{Nagai2007} has become standard
in describing ICM pressure profiles. Specifically,
\begin{equation}
  \textrm{P}(x) = \textrm{P}_0 x^{-\gamma} (1 + x^{\alpha})^{-(\beta-\gamma)/\alpha},
\end{equation}
where $\textrm{P}_0$ is the pressure normalization; $\gamma$, $\alpha$, and $\beta$
primarily describe the power law exponent at small, intermediate, and large radii;
and $x = c_{500} (\textrm{R}/\textrm{R}_{500})$ is the scaled radial coordinate
with concentration $c_{500}$. As originally noted by \citet{Nagai2007}, degeneracies
between the five parameters in the model often prevent meaningful constraints
when all are varied \citep[see also][]{Battaglia2012}.
Thus, to describe the individual galaxy cluster pressure profiles obtained from our deprojection
procedure in Section~\ref{sec:deprojection}, we first considered gNFW 
fits that varied every possible permutation of $\textrm{P}_0$
and either two or three of the other four parameters. Due to the strong degeneracies between parameters,
we found from an initial set of sparsely sampled fits
that the fitted shape over the radial range constrained by the data, along
with the fit quality, was similar regardless of the total number of varied parameters
as long as one of them was $\beta$.
Based on these test fits, we chose to vary $\textrm{P}_0$, $\alpha$, and $\beta$ in our final fits
because that combination produced the best fit quality among the three-parameter fits (with
$c_{500} = 1.4$ and $\gamma = 0.3$). Numerical values for all of the fitted parameters
are provided in Table~\ref{tab:gnfw_individual} of the Appendix.

Using our full sample of 40 galaxy clusters, we then searched for trends in the values
of the fitted parameters according to the parameterizations used by
\citet{Battaglia2012}, with
\begin{align}
  \log_{10}(\textrm{A}) =
    \log_{10}(\textrm{A}_0) + a_{\textrm{m}} \log_{10}(\textrm{M}_{500}/10^{15}) \nonumber \\
    + a_z \log_{10}(1+z)
  \label{eqn:gnfw_mz}
\end{align}
and the values of $a_{\textrm{m}}$ and $a_z$ describing the mass and redshift
dependence of the parameter. For each fitted gNFW parameter, we determined
the values of A$_0$, $a_{\textrm{m}}$, and $a_z$, along with the fractional
intrinsic cluster-to-cluster scatter in A, using a grid search.
The derived values are given in Table~\ref{tab:gnfw}.

\begin{deluxetable*}{ccccc}
\tablenum{3}
\tablecaption{gNFW Fit Parameters}
\tablewidth{0pt}
\tablehead{
  \colhead{Parameter} & \colhead{$\log_{10}($A$_0)$} & \colhead{a$_{\textrm{m}}$} &
  \colhead{a$_z$} & \colhead{$\sigma_{\textrm{int}}(\log_{10}(\textrm{A}))$}} 
\startdata
  \sidehead{Observational Sample}
  P$_{0}$ & $\phantom{-}0.74 \pm 0.17$ & $-0.27 \pm 0.45$ & $\phantom{-}2.10 \pm 1.29$ & $0.52 \pm 0.10$ \\
  $\alpha$ & $\phantom{-}0.12 \pm 0.11$ & $\phantom{-}0.12 \pm 0.29$ & $-0.41 \pm 0.83$ & $0.24 \pm 0.05$ \\
  $\beta$ & $\phantom{-}0.74 \pm 0.04$ & $\phantom{-}0.15 \pm 0.10$ & $\phantom{-}0.02 \pm 0.28$ & $0.10 \pm 0.02$ \\
  $\left[ \textrm{P}_0\alpha \right]_{\parallel}$ & $-0.63 \pm 0.20$ & $\phantom{-}0.30 \pm 0.53$ & $-2.09 \pm 1.52$ & $0.56 \pm 0.11$ \\
  $\left[ \textrm{P}_0\alpha \right]_{\perp}$ & $\phantom{-}0.41 \pm 0.03$ & $\phantom{-}0.00 \pm 0.08$ & $\phantom{-}0.48 \pm 0.24$ & $0.08 \pm 0.02$ \\
  \sidehead{\three}
  P$_{0}$ & $\phantom{-}0.77 \pm 0.16$ & $-0.04 \pm 0.59$ & $\phantom{-}1.70 \pm 1.14$ & $0.63 \pm 0.08$ \\
  $\alpha$ & $\phantom{-}0.12 \pm 0.10$ & $\phantom{-}0.03 \pm 0.35$ & $-0.24 \pm 0.68$ & $0.29 \pm 0.04$ \\
  $\beta$ & $\phantom{-}0.71 \pm 0.02$ & $\phantom{-}0.19 \pm 0.07$ & $\phantom{-}0.27 \pm 0.13$ & $0.07 \pm 0.01$ \\
  $\left[ \textrm{P}_0\alpha \right]_{\parallel}$ & $-0.66 \pm 0.19$ & $\phantom{-}0.51 \pm 0.69$ & $-1.65 \pm 1.32$ & $0.69 \pm 0.09$ \\
  $\left[ \textrm{P}_0\alpha \right]_{\perp}$ & $\phantom{-}0.42 \pm 0.02$ & $\phantom{-}0.01 \pm 0.08$ & $\phantom{-}0.47 \pm 0.16$ & $0.08 \pm 0.01$ \\
  \sidehead{\citet{Battaglia2012}}
  P$_{0}$ & $\phantom{-}1.16$ & $\phantom{-}0.27$ & $-1.50$ & \\
  $\alpha$ & $-0.05$ & $-0.05$ & $\phantom{-}0.50$ & \\
  $\beta$ & $\phantom{-}0.73$ & $\phantom{-}0.08$ & $\phantom{-}0.25$ & \\
  $\left[ \textrm{P}_0\alpha \right]_{\parallel}$ & $-1.08$ & $-0.26$ & $\phantom{-}1.57$ & \\
  $\left[ \textrm{P}_0\alpha \right]_{\perp}$ & $\phantom{-}0.43$ & $\phantom{-}0.06$ & $-0.15$ & \\
\enddata
\tablecomments{Generalized NFW fit parameters from our observational sample, \three, and \citet{Battaglia2012}.}
\label{tab:gnfw}
\end{deluxetable*}

\begin{figure}
  \centering
  \includegraphics[width=\columnwidth]{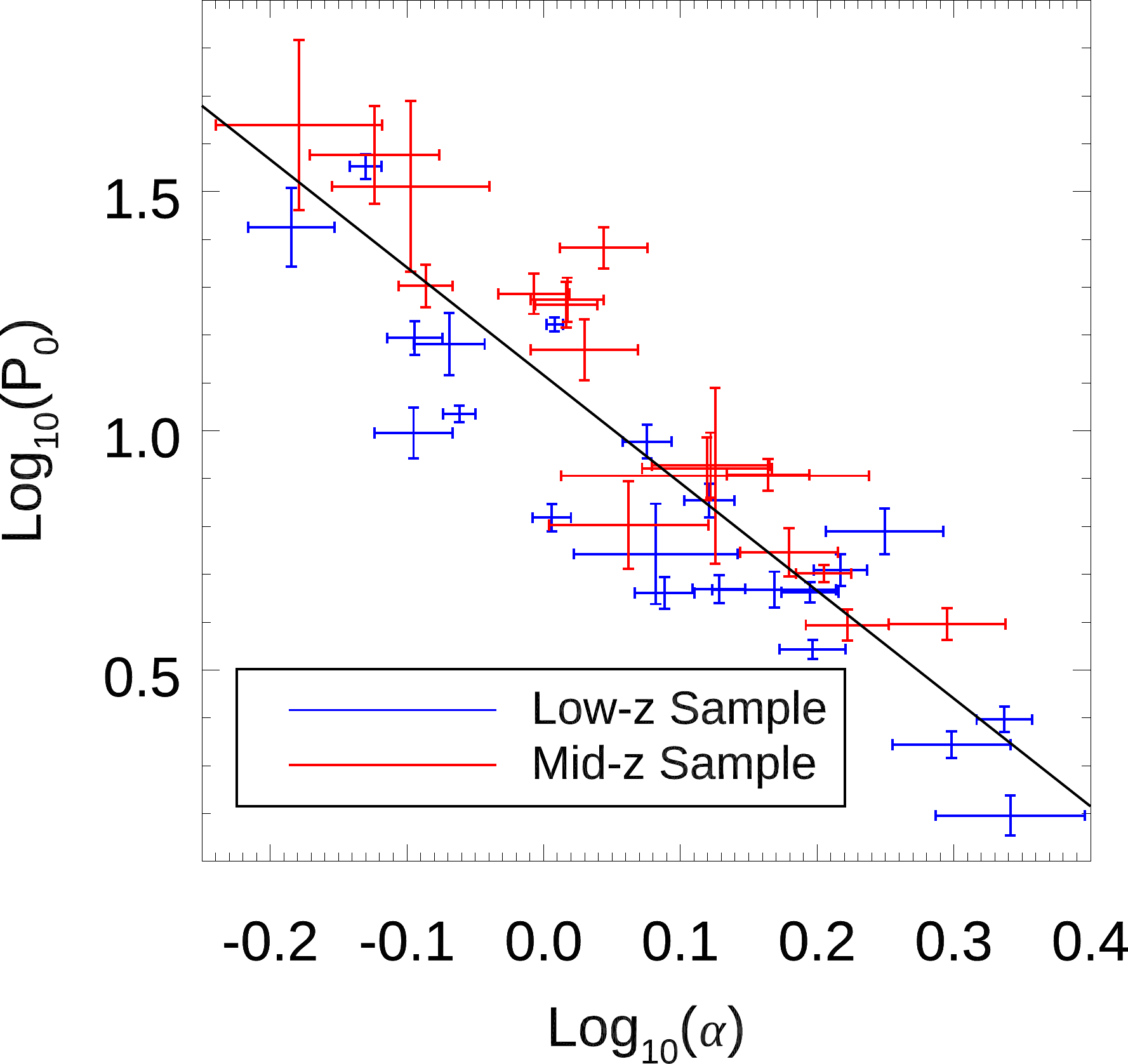}
  \caption{Fitted values of P$_0$ and $\alpha$
    for all of the galaxy clusters in our observational sample, illustrating
  the strong degeneracy between these two gNFW parameters.}
  \label{fig:p_alpha_degen}
\end{figure}

\begin{figure*}
  \centering
  \includegraphics[width=0.45\textwidth]{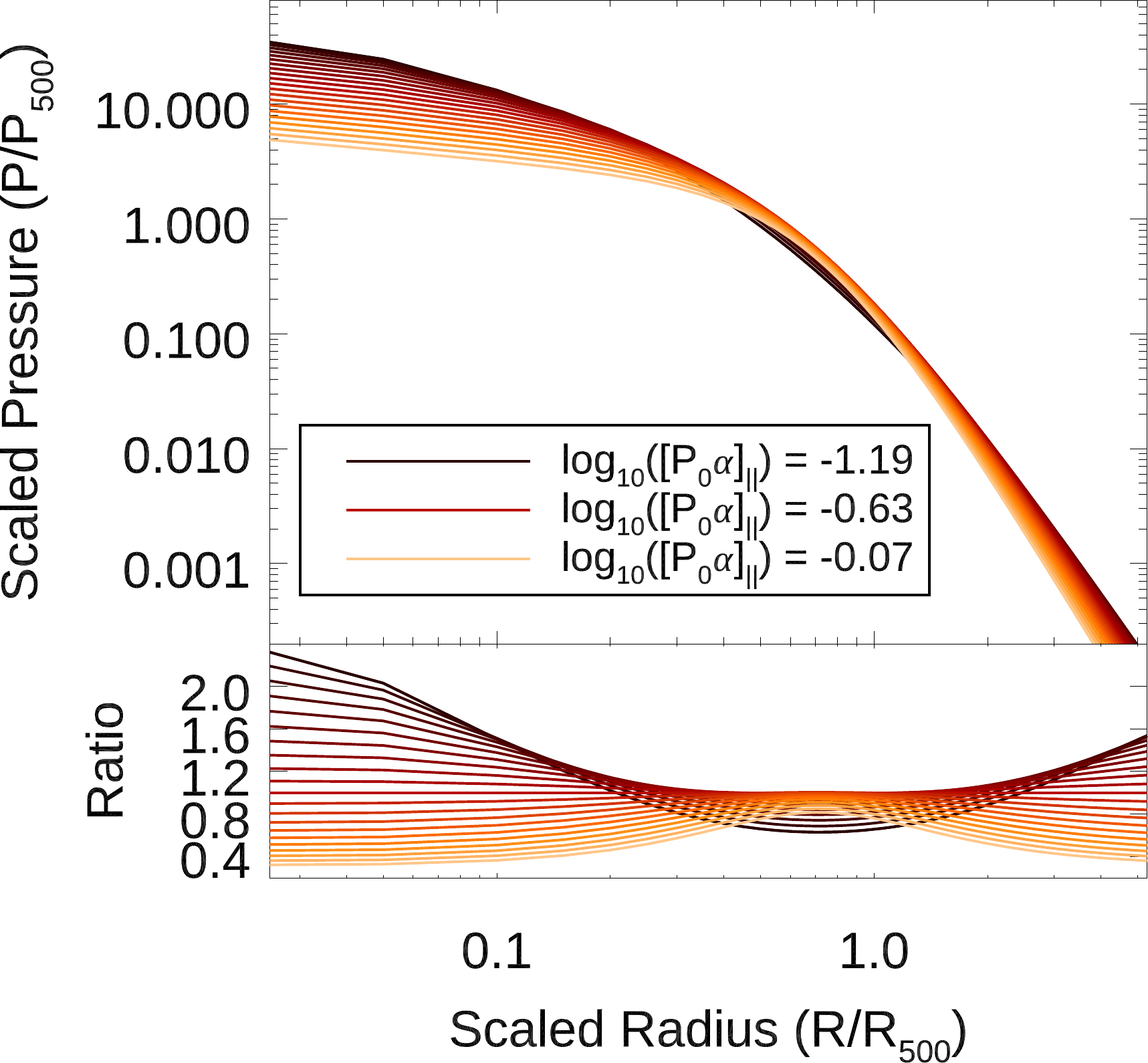}
  \hspace{0.03\textwidth}
  \includegraphics[width=0.45\textwidth]{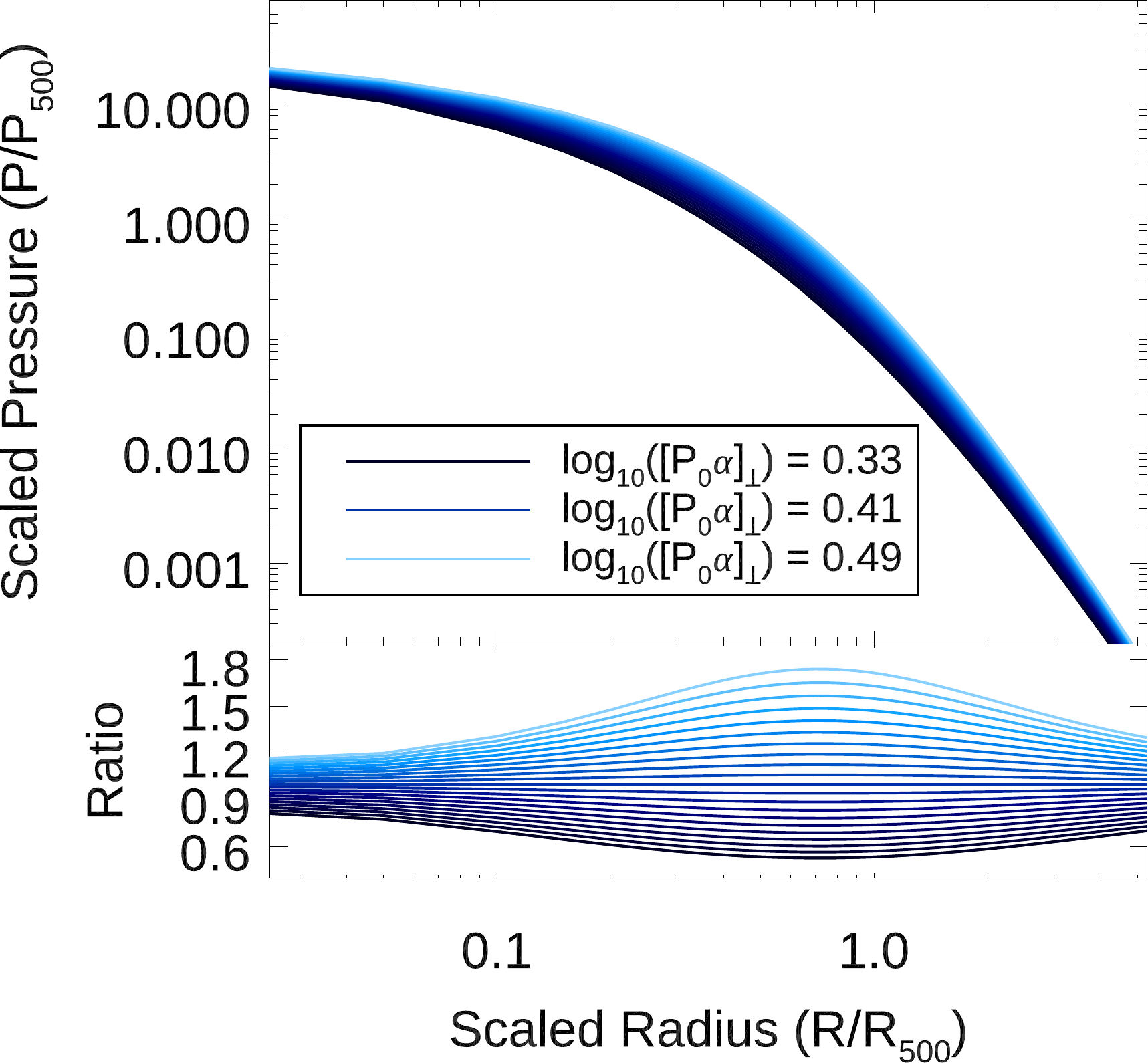}

  \vspace{12pt}
  \includegraphics[width=0.45\textwidth]{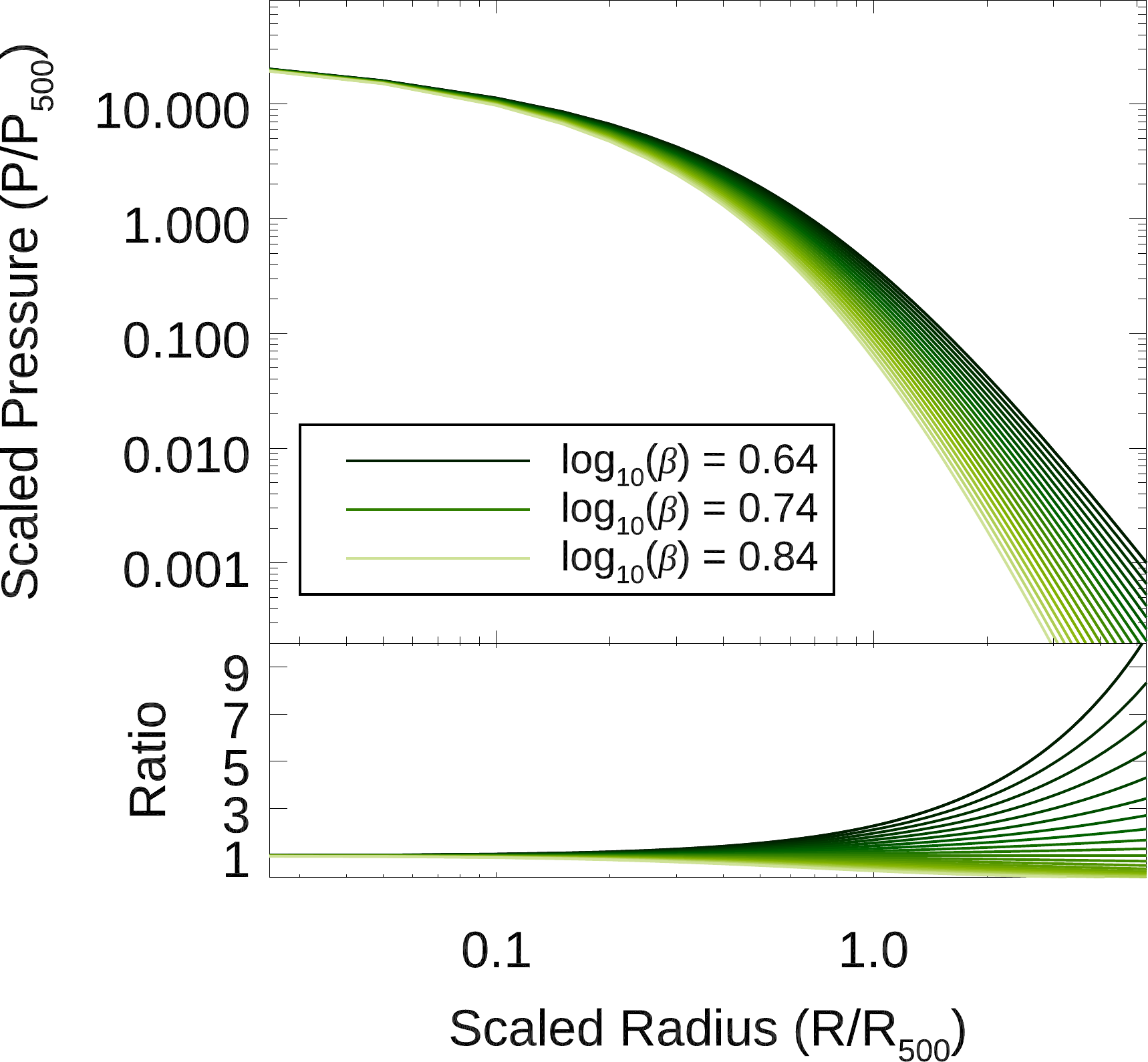}
  \caption{Generalized NFW scaled pressure profiles for the approximate range of values of
    $\left[ \textrm{P}_0\alpha \right]_{\parallel}$, $\left[ \textrm{P}_0\alpha \right]_{\perp}$,
    and $\beta$ consistent with the observed data (i.e., the range implied by the intrinsic scatter
    for each parameter given in Table~\ref{tab:gnfw}).
    The scaled pressure profile is primarily sensitive to $\left[ \textrm{P}_0\alpha \right]_{\parallel}$
    at small radii, $\left[ \textrm{P}_0\alpha \right]_{\perp}$ at intermediate radii, and
    $\beta$ at large radii. For reference, the median values of the three parameters
    are $-0.76$, 0.43, and 0.71 for the \lowz\ sample and $-0.99$, 0.50, and 0.75 for the
    \midz\ sample.}
  \label{fig:degen}
\end{figure*}

While we find that the parameters in Equation~\ref{eqn:gnfw_mz} are precisely measured for $\beta$,
the parameters associated with P$_0$ and $\alpha$ are relatively poorly constrained.
This is because, even when only varying three of the gNFW parameters, there is still
a significant degeneracy between P$_0$ and $\alpha$, as illustrated in Figure~\ref{fig:p_alpha_degen}.
To better explore this issue, we have therefore derived two new parameters
based on a linear fit to $\log_{10}(\textrm{P}_0)$ versus $\log_{10}(\alpha)$.
One is parallel to this linear fit, with
\begin{equation}
  \log_{10} (\left[ \textrm{P}_0\alpha \right]_{\parallel} ) = -0.91\log_{10}(\textrm{P}_0) + 0.41\log_{10}(\alpha),
\end{equation}
and the other is perpendicular, with
\begin{equation}
  \log_{10} (\left[ \textrm{P}_0\alpha \right]_{\perp} ) = 0.41\log_{10}(\textrm{P}_0) + 0.91\log_{10}(\alpha).
\end{equation}
Figure~\ref{fig:degen} shows how the overall gNFW profile shape depends on these these
parallel and perpendicular components, and illustrates that this decomposition of the
gNFW parameters is useful to interpreting the results. Specifically, the small radius
behavior is almost entirely governed by $\left[ \textrm{P}_0\alpha \right]_{\parallel}$,
the intermediate radius behavior is almost entirely governed by
$\left[ \textrm{P}_0\alpha \right]_{\perp}$, and the large radius behavior
is almost entirely governed by $\beta$. In Figure~\ref{fig:gnfw_pca} of the Appendix we
show plots of all the parameter pairs, and a an additional small degeneracy
between $\left[ \textrm{P}_0\alpha \right]_{\perp}$ and $\beta$ is evident. Unlike
the P$_0$--$\alpha$ degeneracy, accounting for this small degeneracy does not significantly
improve the scatter in the transformed parameters compared to the original parameters, and
so we do not include an additional parameter transformation in our analysis.

As expected, the results presented in Table~\ref{tab:gnfw} and Figure~\ref{fig:degen}
are consistent with what we found from the ensemble mean profiles
in Section~\ref{sec:mean}. In particular, both $\left[ \textrm{P}_0\alpha \right]_{\parallel}$
and $\left[ \textrm{P}_0\alpha \right]_{\perp}$ have measurable evolution
that results in on-average higher scaled pressure values at small and intermediate radii
with increasing redshift. In addition, $\beta$ has a measurable mass dependence
that indicates lower scaled pressure values with increasing mass. Thus, the general mass and
redshift trends that were
inferred in Section~\ref{sec:mean} based on various subsamples from \three\ are confirmed
in these gNFW fits. Furthermore,
these three parameters suggest a fractional intrinsic scatter in the scaled pressure that
is of order unity at small radii, approximately 50 per cent at intermediate radii,
and several times larger than unity at large radii.

For comparison, we also fit gNFW profiles to the set of simulated clusters selected
from \three, and performed an identical analysis to constrain the mass and redshift
dependence of the fitted parameters (see Table~\ref{tab:gnfw}). Again, as expected,
the results are largely consistent with what we find for the observational
sample, suggesting good agreement between the observations and simulations.
We note that, while the values of a$_z$ for $\beta$ are consistent given the measurement
uncertainties, \three\ find a non-zero value indicative of decreasing scaled pressure
with increasing redshift at large radii.

In addition, we also rescaled the fitted gNFW parameters from \citet{Battaglia2012}
for comparison. This involved generating a grid of pressure profiles in
mass and redshift based on the best-fit values given in Table~1 of that work
for the ``AGN Feedback $\Delta = 500$'' case using their Equation~10.
These profiles were then refit using our parameterization, with the
results given in Table~\ref{tab:gnfw}. Their value of $\beta$, including
its mass and redshift dependence, is consistent with what we find
in our analysis, suggesting good consistency in the scaled pressure values
at large radii. The same is generally true for $\left[ \textrm{P}_0\alpha \right]_{\perp}$,
where we find nearly identical values for A$_0$ and a$_m$. However,
in contrast to the evolution found in our analysis, \citet{Battaglia2012}
find a much weaker (and opposite) redshift dependence, indicating
minor differences in the scaled pressure at intermediate radii between
our analyses.
Finally, \citet{Battaglia2012} find a lower normalization
for $\left[ \textrm{P}_0\alpha \right]_{\parallel}$ and a positive rather
than negative value for the associated a$_z$. This indicates that
they find higher scaled pressures at small radii that decrease rather than increase
with redshift. However, as noted in Section~\ref{sec:mean}, our observed
trend with redshift appears to be largely due to selection effects with
regard to dynamical state, and so we caution that no general conclusions
should be drawn based on this comparison between our analyses
at small radii. 

\section{Discussion}
\label{sec:discussion}

The ensemble mean scaled pressure profiles we obtain for our \lowz\ and \midz\
samples are generally in good agreement with \three\ simulations
and previous observational studies. One exception is the profiles obtained
by previous observational studies at small to intermediate radii at \midz,
which are lower than those from our analysis. While the origin
of this difference is unclear, we speculate it may be due to the choice of
centering algorithm and/or the distribution
of dynamical states within the various samples used in these analyses. Similarly, we generally
find good agreement in the fractional intrinsic scatter about these
mean profiles, although our analysis suggests a higher scatter
at large radii compared to both \three\ and previous observational studies.
We also find generally good agreement for the mass and redshift
dependence of our measured profiles compared with the parametric
gNFW fits provided by \citet{Battaglia2012} based on their
simulations, where again the only notable discrepancy appears at smaller
radii where selection effects appear to play a dominant role.

We measure a significant difference in the mean scaled pressure profiles between
the \lowz\ and \midz\ samples, and we use \three\ simulations to better
understand these differences. At small radii, dynamical state has the largest
impact, with more relaxed systems having higher scaled pressure. While subdominant
to dynamical state, there are also modest indications of evolution playing
a role at small radii, with higher redshift systems having higher scaled pressure.
At intermediate radii, we find that evolution is the primary driver,
resulting in higher mean scaled pressure with increasing redshift. However,
dynamical state may also play a role here, given that cool core systems can
have higher pressures extending well beyond the core and the MACS X-ray
selection of the \midz\ sample has a significant bias towards such systems.
At large
radii, mass has the most influence on the mean scaled pressure, with lower
mass systems having higher scaled pressure. In addition, there are also some
indications of evolution at large radii, with higher redshift systems
having lower scaled pressure.
In order to facilitate a mapping of our measured mean scaled pressure profiles
to any combination of redshift and mass that is well constrained by
our study, corresponding to approximately
0.05~$\le z \le$~0.60 and $4 \times 10^{14}$~\msun~$\le \textrm{M}_{500} \le 30 \times 10^{14}$~\msun,
we provide a set of gNFW parameter values in Table~\ref{tab:gnfw}.

To better understand the physical processes responsible for the evolution
and mass dependence we find in our analysis, we consider the results from a
range of previous simulations. For instance, the pressure profile at large
radii is expected to steepen with increasing mass accretion rates due to
halo contraction during accretion and the further penetration of the accretion
shocks required to thermalize the new material
\citep{Battaglia2012-1, Diemer2014, Lau2015, Planelles2017,Aung2021, Diemer2022}.
Higher mass accretion rates should also reduce the overall thermalization fraction
of the ICM, and thus its pressure, particularly at large radii,
as suggested by recent measurements from \citet{Sereno2021}.
Given that the mass accretion
rate is enhanced in higher mass and/or higher redshift galaxy clusters
\citep[e.g.,][]{Diemer2014,Battaglia2015}, those systems are expected to
have lower scaled pressure at large radii.

\begin{figure}
  \centering
  \includegraphics[width=\columnwidth]{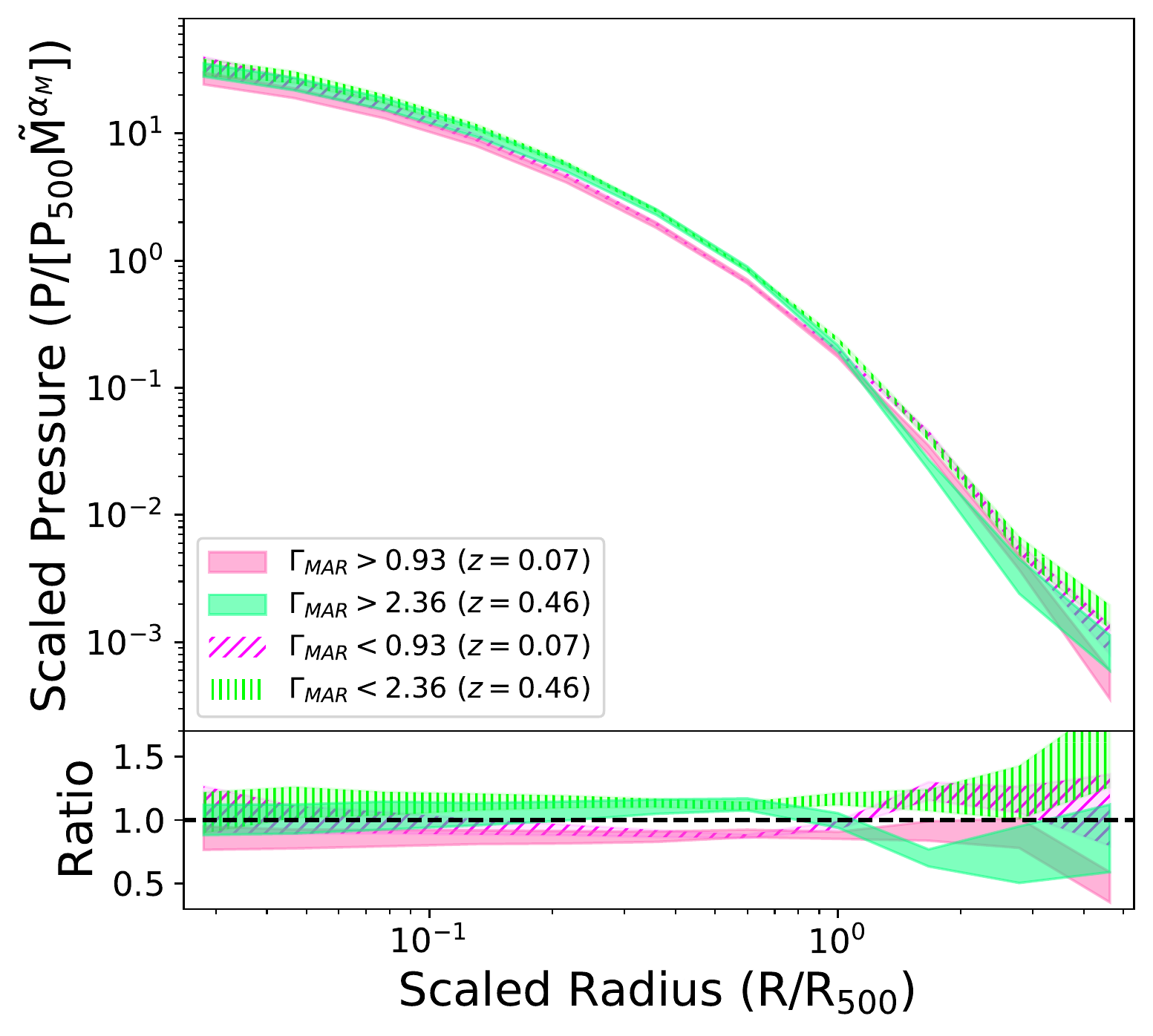}
  \caption{From the same \three\ samples as shown in Figure~\ref{fig:the300_evolved},
    the mean pressure profile for subsamples defined to be above and below
    the median $\Gamma_{\textrm{MAR}}$ at both $z=0.07$ and $z=0.46$.}
  \label{fig:MAR}
\end{figure}

To test this expectation,
we determined the mass accretion rate for all of simulated \three\ galaxy
clusters matched to the observed samples in Section~\ref{sec:three_comparison} and plotted
in Figure~\ref{fig:the300_evolved} according to the formalism of
\citet{Diemer2017} with
\begin{equation}
  \Gamma_{\textrm{MAR}} = \frac{\log_{10}(\textrm{M}_0) - \log_{10}(\textrm{M}_1)}
    {\log_{10}(1+z_1) - \log_{10}(1+z_0)},
\end{equation}
where M and $z$ are mass and redshift at epochs 0 and 1. We then determined
the mean pressure profile for samples selected to be above or below the median value
of $\Gamma_{\textrm{MAR}}$ at both $z=0.07$ and $z=0.46$, with the results shown
in Figure~\ref{fig:MAR}. We find the following: 1) within each redshift snapshot,
the higher $\Gamma_{\textrm{MAR}}$ sample has lower pressure and thus a steeper
profile beyond \rfive; 2) the on-average value of $\Gamma_{\textrm{MAR}}$ is a
factor of 2.5 higher in the $z=0.46$ sample than the $z=0.07$ sample, yet there is
no significant difference in the profiles outside of \rfive; 3) the higher
redshift (and thus higher $\Gamma_{\textrm{MAR}}$) samples do have higher pressure
at $\textrm{R} \lesssim \textrm{R}_{500}$, indicating a steeper profile in those
systems at intermediate radii. Thus, while we do find trends of steeper pressure
profiles at both intermediate and large radii in higher $\Gamma_{\textrm{MAR}}$
systems, there does not appear to be a simple mapping between $\Gamma_{\textrm{MAR}}$
and profile shape, with redshift and potentially other characteristics also playing
a role.

As another example, the relative impact of AGN feedback
is expected to be larger in low mass galaxy clusters,
resulting in more gas being ejected from the central regions
to larger radii during outbursts
\citep{Battaglia2010,McCarthy2011, Battaglia2012-1, LeBrun2015,
  Barnes2017, Truong2018, Henden2020}.
While this effect is most likely to be relevant in group-scale objects
with lower masses than the systems in our study, some simulations indicate
there may still be observable differences in low-to-moderate mass
galaxy clusters. Thus, our observed flattening of the scaled pressure profiles
in lower mass systems at both small and large radii may by caused in
part by AGN.
Other mechanisms, such as star formation, are also expected to have
a larger fractional impact on lower mass systems, producing
similar trends in the scaled pressure profile \citep{Battaglia2012-1, Henden2020}.
There are also indications of evolution in the mean scaled pressure profile
due to the increasing
prevalence of AGN at higher redshift, primarily resulting in an enhanced level of
thermal pressure at large radii \citep{Battaglia2012-1}.

Taken as a whole, our observed mass and redshift trends on the mean scaled pressure profile
are broadly consistent
with expectations from simulations.
We find that differences in mass accretion rate are a significant contributor
to these trends, although these differences are unable to fully explain the trends.
For instance, while the effects are generally expected to be minimal for
the high mass galaxy clusters typical of our study,
feedback from AGN and other relevant feedback processes also produce mass and redshift
trends in broad agreement with our observations.
  
Considering the intrinsic scatter rather than the mean scaled pressure profile,
we find no measurable difference between the \lowz\ and \midz\ samples.
The matched samples from \three\ also show consistent scatter at all radii,
further suggesting that the scatter is largely independent of population statistics.
While most results from simulations have focused more on the impact of
physical process on the mean shape rather than the scatter, \citet{Planelles2017}
examined the issue in some detail. For instance, they noted a significant
increase in scatter at small radii when AGN feedback was included, although
the scatter at larger radii was not sensitive to the included physics.
Furthermore, the increased clumpiness associated with higher mass
accretion rates (e.g., at higher redshift) was found to increase the
scatter. If such trends do exist, then they appear to be below
our measurement precision.

\begin{acknowledgments}
  Sayers and Mantz were supported by the National Aeronautics and Space
  Administration under Grant No. 80NSSC18K0920 issued through the ROSES
  2017 Astrophysics Data Analysis Program.
  Wan was supported by a Robert L. Blinkenberg Caltech Summer Undergraduate
  Research Fellowship and the Kavli Institute for Particle Astrophysics and
  Cosmology. Allen and Morris were supported by the U.S. Department of Energy
  under contract number DE-AC02-76SF00515.
\end{acknowledgments}

\vspace{5mm}
\facilities{CXO, \planck, Caltech Submillimeter Observatory, \rosat}

\software{MPFIT \citep{Markwardt2009},
          LRGS \citep{Mantz2016_LRGS}, 
          SZPACK \citep{Chluba2012,Chluba2013}
          }

\appendix

\section{Individual and Ensemble Mean Galaxy Cluster Scaled Pressure Profiles}

This appendix includes tabulated values for the individual galaxy cluster pressure profile deprojections and gNFW fits. It also includes tabulated values for the ensemble pressure profile deprojections for the observed and simulated data.

  \input{individual_cluster_deproj.tex}

  \input{individual_cluster_gnfw.tex}

  \input{lowz_20210312_latex_table.tex}

  \input{highz_20210312_latex_table.tex}

  \input{lowz_sim_lowz_dyn_selected_20220204_latex_table.tex}

  \input{highz_sim_highz_dyn_selected_20220204_latex_table.tex}

\section{Fitted gNFW Parameters}

This appendix includes plots of the fitted gNFW profiles for the observed galaxy clusters.

  \begin{figure*}
    \centering
    \includegraphics[width=0.3\textwidth]{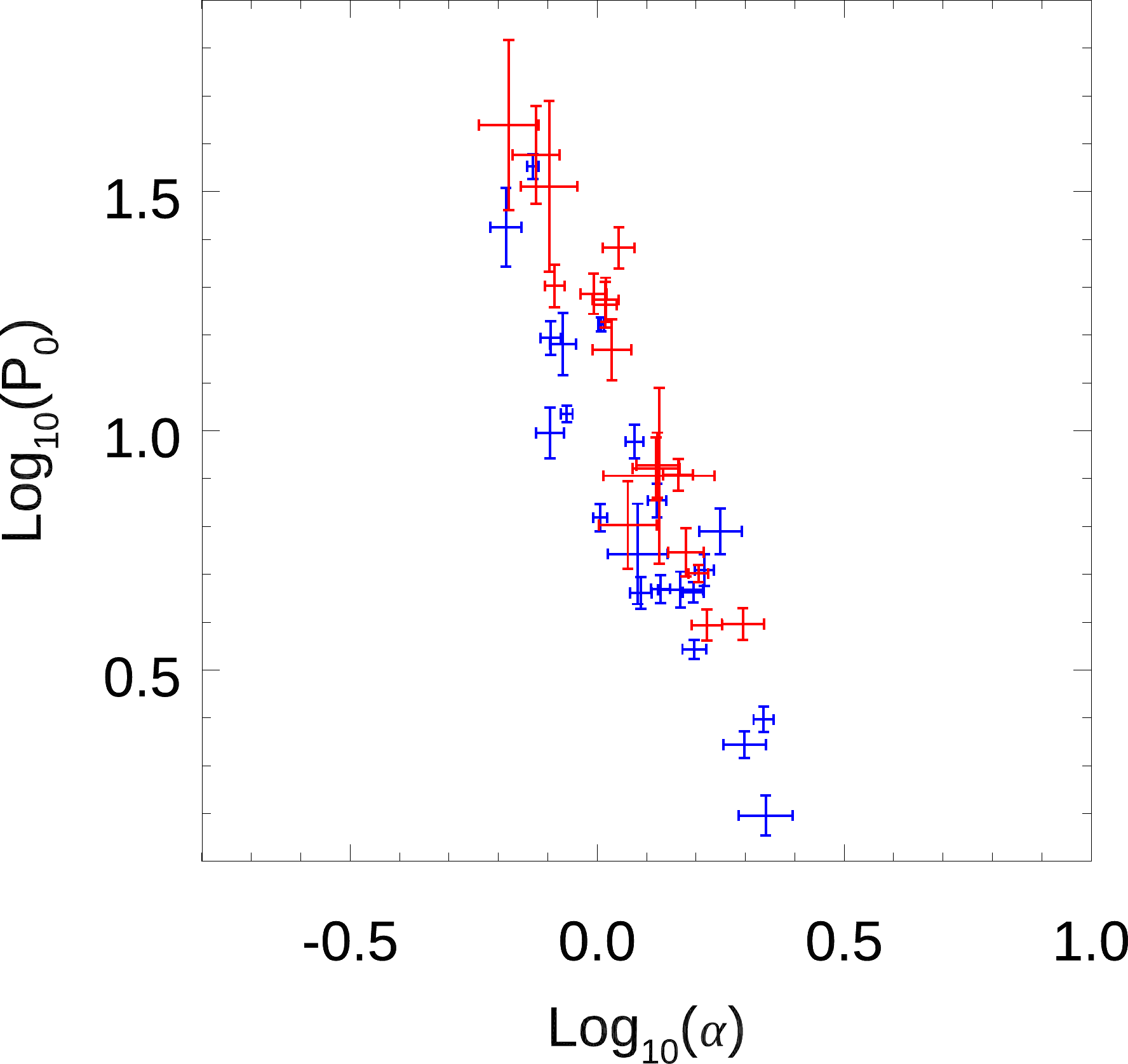}
    \hspace{0.03\textwidth}
    \includegraphics[width=0.3\textwidth]{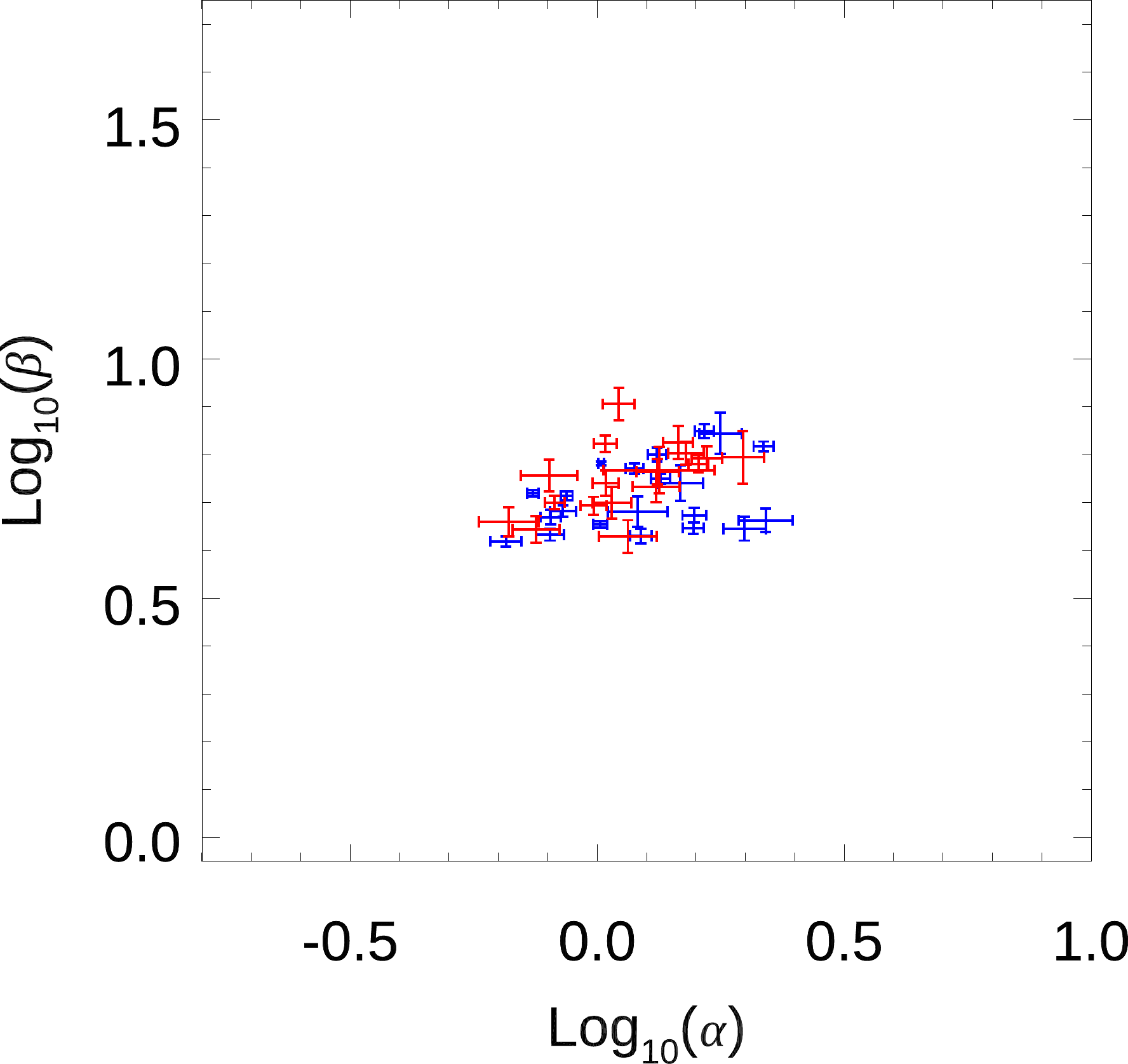}
    \hspace{0.03\textwidth}
    \includegraphics[width=0.3\textwidth]{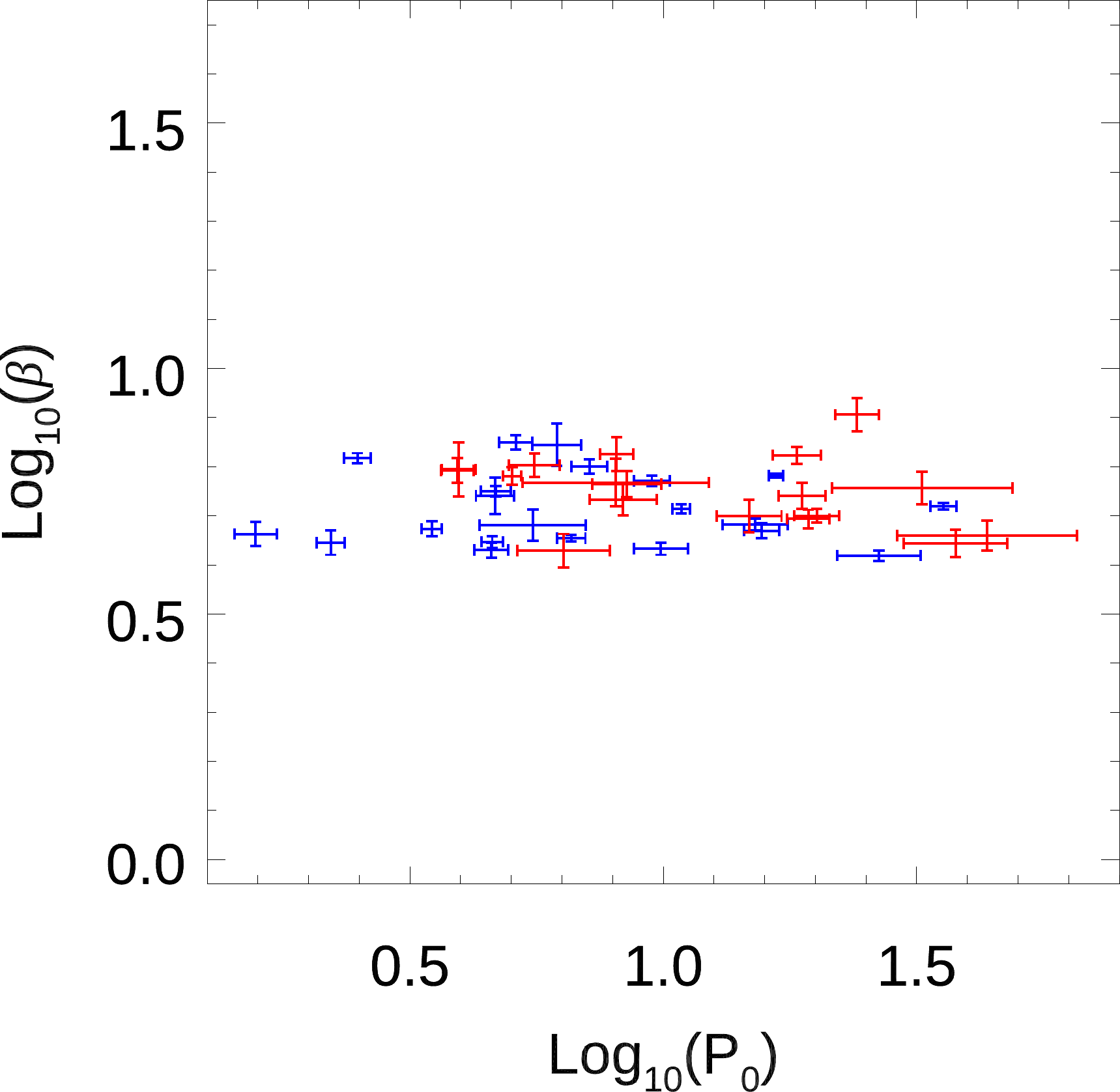}
    \includegraphics[width=0.3\textwidth]{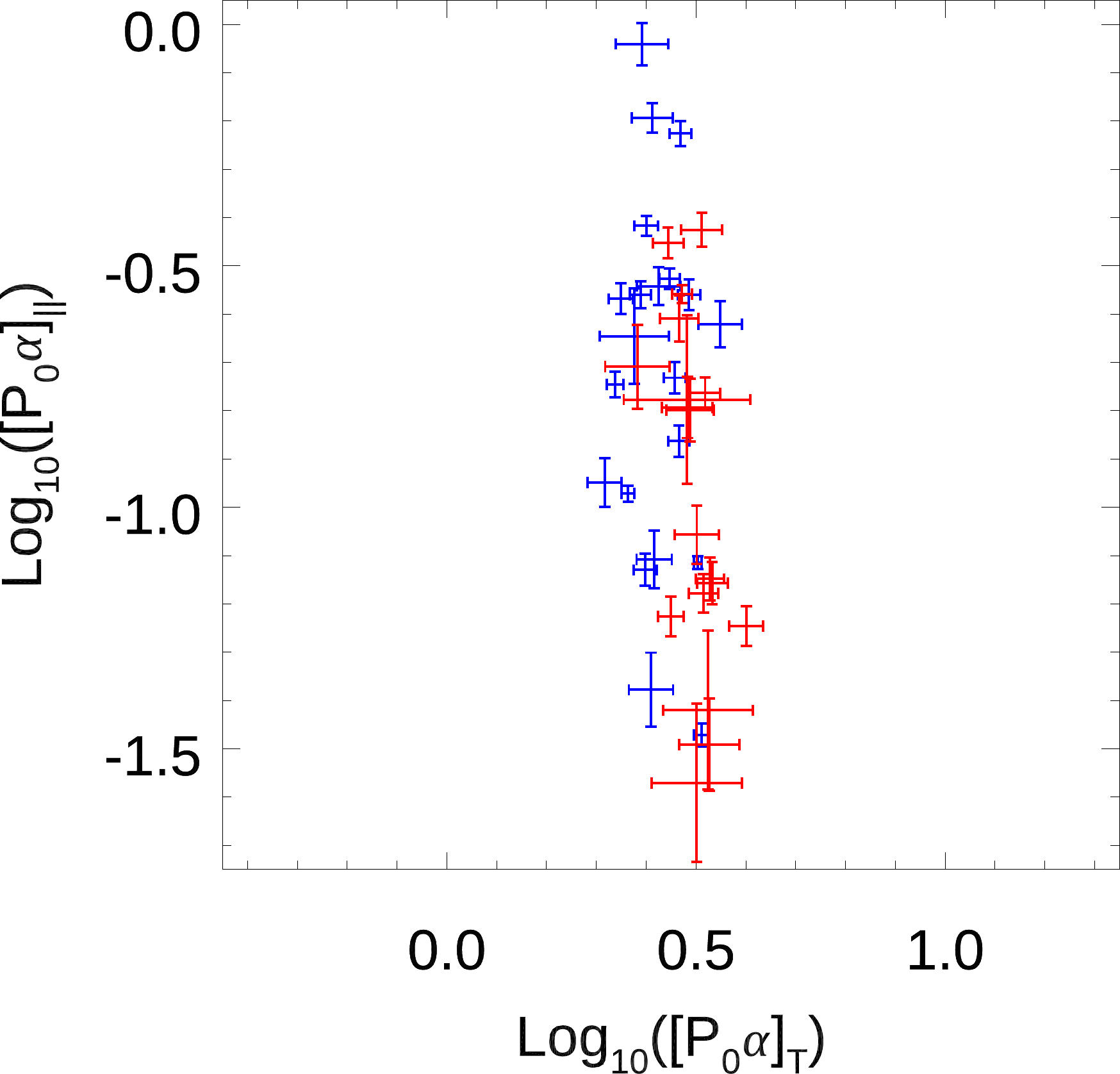}
    \hspace{0.03\textwidth}
    \includegraphics[width=0.3\textwidth]{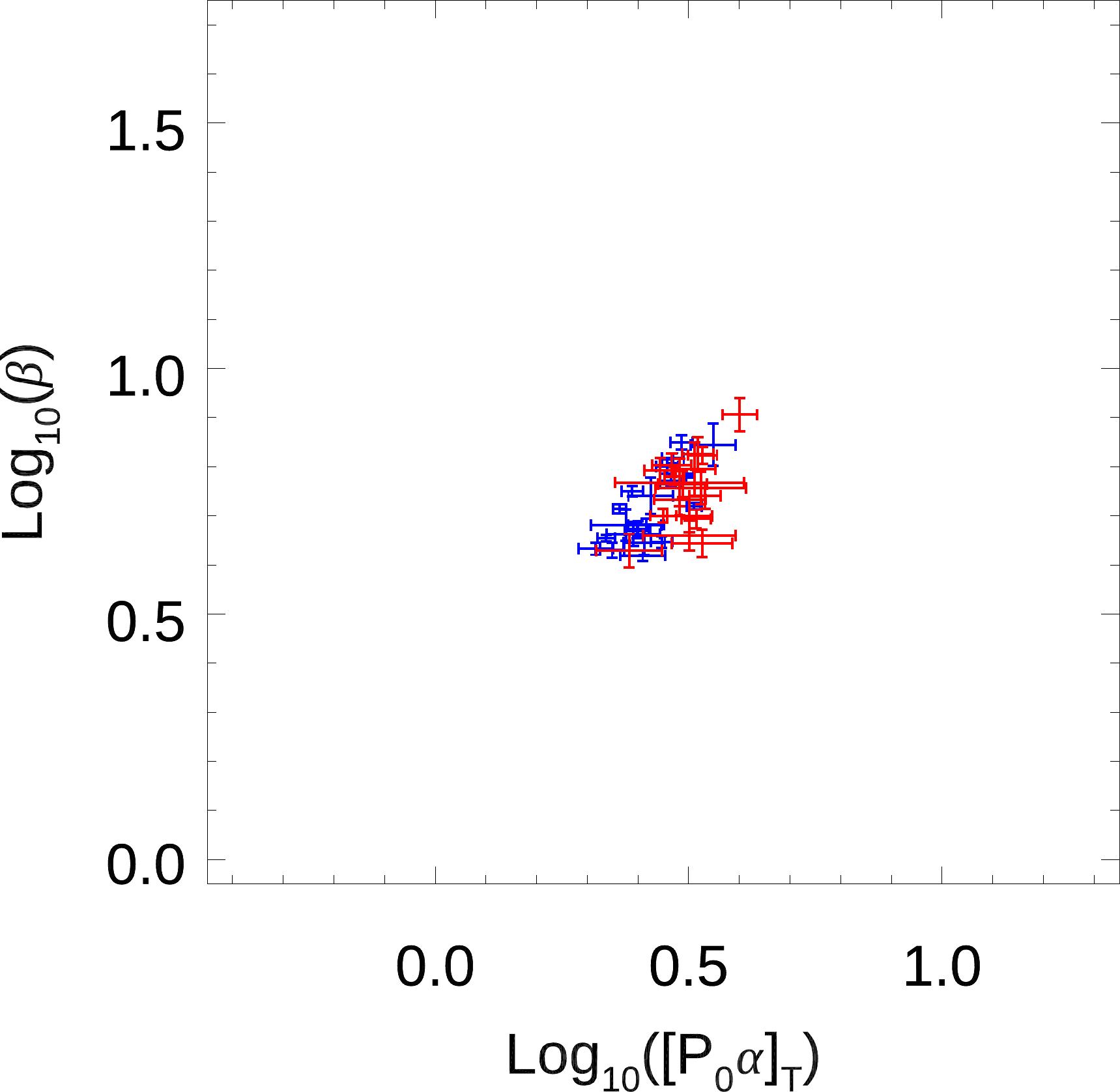}
    \hspace{0.03\textwidth}
    \includegraphics[width=0.3\textwidth]{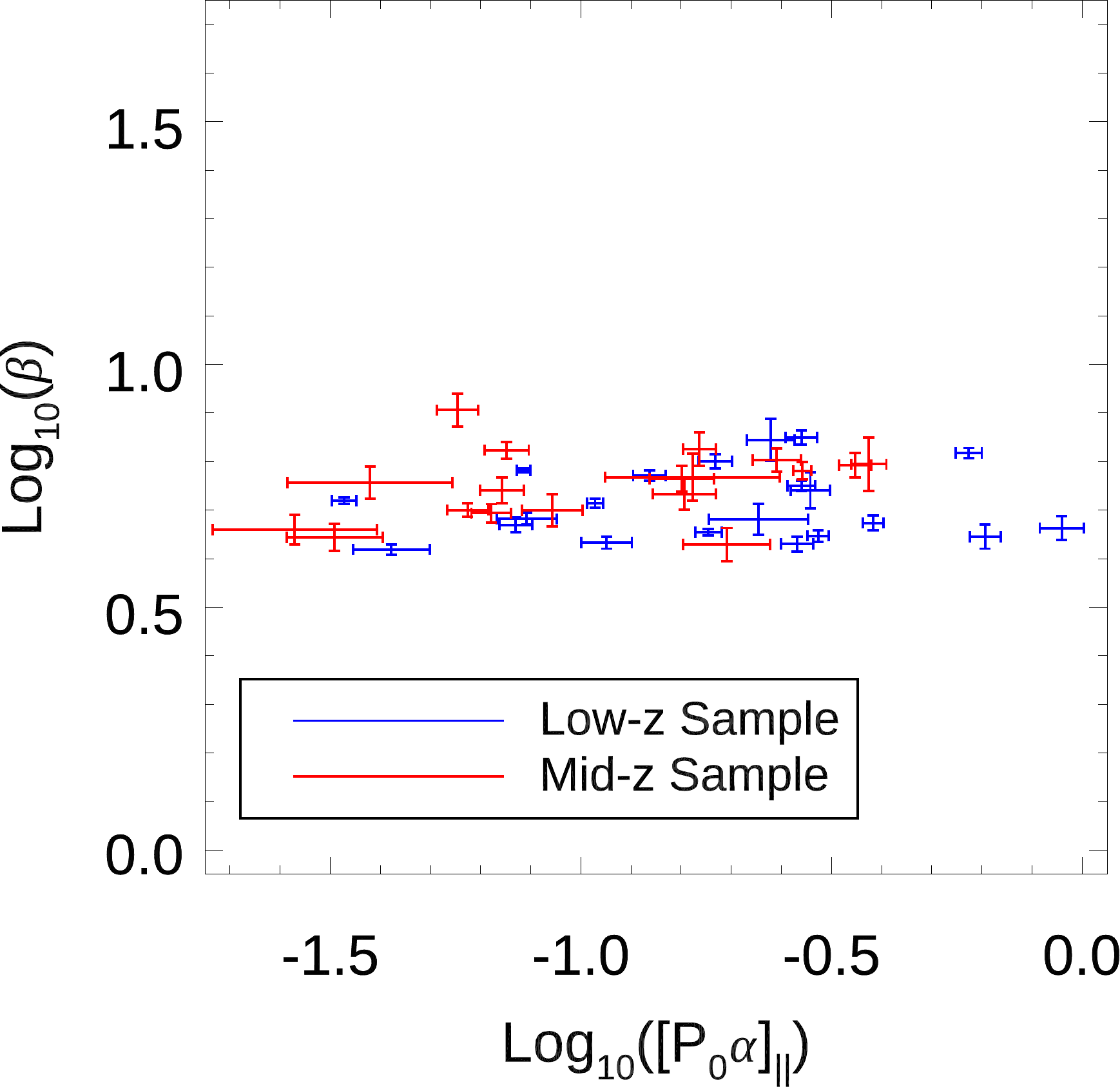}
    \caption{Fitted gNFW parameters for all the galaxy clusters in our \lowz\ (blue) and \midz\ (red) samples. The top
      row shows the original fitted parameters, with a large degeneracy between P$_0$ and $\alpha$.
      The bottom row shows the transformed parameters based on the equations described in Section~\ref{sec:gNFW}
      to remove the degeneracy. The axes span the same range in all plots.}
    \label{fig:gnfw_pca}
  \end{figure*}
  
\bibliography{references}{}
\bibliographystyle{aasjournal}

\end{document}

%% file: individual_cluster_deproj.tex
\begin{longrotatetable}
\begin{deluxetable*}{cccccccccccc}
  \tablewidth{0pt}
  \tablecaption{Scaled Pressure Deprojections}
  \tablehead{\colhead{R/R$_{500}$} & \colhead{0.028} & \colhead{0.047} & \colhead{0.078} & 
    \colhead{0.13} & \colhead{0.22} & \colhead{0.36} & \colhead{0.60} & \colhead{1.0} & 
    \colhead{1.7} & \colhead{2.8} & \colhead{4.6}}
  \startdata
    Abell 0754
 & 2.18E+01
 & 1.10E+01
 & 9.22E+00
 & 8.94E+00
 & 4.47E+00
 & 2.25E+00
 & 1.67E+00
 & 2.81E-01
 & 3.61E-02
 & 5.17E-03
 & $\le$4.5E-03
 \\
    Abell 0085
 & 1.62E+01
 & 1.57E+01
 & 9.80E+00
 & 5.77E+00
 & 3.62E+00
 & 1.56E+00
 & 8.19E-01
 & 2.54E-01
 & 6.32E-02
 & 7.07E-03
 & 2.88E-03
 \\
    Abell 3667
 & 1.28E+01
 & 1.01E+01
 & 7.67E+00
 & 5.21E+00
 & 3.01E+00
 & 1.28E+00
 & 6.33E-01
 & 1.86E-01
 & 3.74E-02
 & 3.93E-03
 & $\le$1.2E-03
 \\
    Abell 2256
 & 1.12E+01
 & 7.05E+00
 & 6.18E+00
 & 5.17E+00
 & 3.13E+00
 & 1.56E+00
 & 5.21E-01
 & 1.21E-01
 & 1.85E-02
 & 1.09E-03
 & 3.88E-04
 \\
    Abell 3158
 & ---
 & 1.20E+01
 & 1.04E+01
 & 6.90E+00
 & 3.70E+00
 & 1.70E+00
 & 7.74E-01
 & 2.30E-01
 & 6.61E-02
 & 4.47E-03
 & $\le$1.2E-03
 \\
    Abell 3266
 & 9.45E+00
 & 9.47E+00
 & 5.58E+00
 & 4.90E+00
 & 3.66E+00
 & 1.66E+00
 & 9.40E-01
 & 1.59E-01
 & 3.63E-02
 & 2.68E-03
 & 1.30E-03
 \\
    Abell 1795
 & 2.46E+01
 & 1.77E+01
 & 1.28E+01
 & 7.19E+00
 & 3.75E+00
 & 1.50E+00
 & 6.36E-01
 & 1.21E-01
 & 2.34E-02
 & 4.61E-03
 & 1.16E-02
 \\
    Abell 2065
 & 1.40E+01
 & 1.18E+01
 & 8.87E+00
 & 6.10E+00
 & 3.80E+00
 & 2.14E+00
 & 8.52E-01
 & 5.13E-02
 & 4.00E-02
 & $\le$6.5E-03
 & 8.61E-04
 \\
    Abell 3112
 & 3.80E+01
 & 2.67E+01
 & 1.71E+01
 & 1.04E+01
 & 5.09E+00
 & 2.24E+00
 & 6.50E-01
 & 7.55E-02
 & 6.08E-02
 & 9.71E-03
 & 2.31E-07
 \\
    Abell 2029
 & 3.81E+01
 & 2.52E+01
 & 1.56E+01
 & 8.97E+00
 & 3.98E+00
 & 1.61E+00
 & 5.64E-01
 & 3.64E-01
 & 4.13E-03
 & 4.03E-02
 & $\le$5.5E-04
 \\
    Abell 2255
 & ---
 & 3.63E+00
 & 4.94E+00
 & 4.80E+00
 & 2.44E+00
 & 1.83E+00
 & 8.03E-01
 & 2.10E-01
 & 5.94E-02
 & $\le$1.6E-03
 & $\le$2.8E-03
 \\
    Abell 1650
 & 2.63E+01
 & 2.19E+01
 & 1.49E+01
 & 8.41E+00
 & 4.76E+00
 & 1.87E+00
 & 6.85E-01
 & 2.19E-01
 & 1.30E-02
 & 1.15E-02
 & 8.92E-03
 \\
    Abell 1651
 & 2.71E+01
 & 2.02E+01
 & 1.35E+01
 & 9.79E+00
 & 4.71E+00
 & 1.98E+00
 & 7.41E-01
 & 1.48E-01
 & 4.64E-02
 & 7.58E-05
 & $\le$5.4E-03
 \\
    Abell 2420
 & 2.64E+01
 & 2.02E+01
 & 1.46E+01
 & 9.54E+00
 & 5.82E+00
 & 2.59E+00
 & 7.99E-01
 & 1.77E-01
 & 2.73E-02
 & 4.13E-03
 & 6.91E-04
 \\
    Abell 0478
 & 3.15E+01
 & 2.10E+01
 & 1.60E+01
 & 8.24E+00
 & 3.77E+00
 & 1.37E+00
 & 4.67E-01
 & 1.94E-01
 & 8.54E-03
 & 1.91E-02
 & $\le$4.2E-03
 \\
    Abell 2142
 & 1.74E+01
 & 1.19E+01
 & 8.48E+00
 & 5.88E+00
 & 3.12E+00
 & 1.38E+00
 & 6.52E-01
 & 2.18E-01
 & 2.22E-02
 & 3.33E-03
 & 4.14E-03
 \\
    Abell 3695
 & 1.20E+01
 & 2.09E+00
 & 5.58E+00
 & 4.67E+00
 & 2.87E+00
 & 1.39E+00
 & 4.95E-01
 & 2.00E-01
 & 3.76E-02
 & 9.77E-03
 & 6.90E-05
 \\
    Abell 3921
 & 1.03E+01
 & 8.98E+00
 & 9.76E+00
 & 5.45E+00
 & 3.62E+00
 & 1.85E+00
 & 6.46E-01
 & 1.05E-01
 & 7.18E-02
 & 1.15E-04
 & $\le$3.9E-03
 \\
    Abell 2244
 & 2.74E+01
 & 1.96E+01
 & 1.61E+01
 & 8.58E+00
 & 4.27E+00
 & 1.71E+00
 & 6.44E-01
 & 8.10E-02
 & 5.67E-02
 & 9.44E-03
 & 1.11E-03
 \\
    Abell 2426
 & 2.66E+01
 & 2.03E+01
 & 1.27E+01
 & 9.82E+00
 & 5.53E+00
 & 2.53E+00
 & 8.30E-01
 & 1.77E-01
 & 2.66E-02
 & 2.46E-03
 & $\le$9.1E-04
 \\
    Abell 3827
 & 2.42E+01
 & 1.90E+01
 & 1.60E+01
 & 8.93E+00
 & 4.99E+00
 & 2.59E+00
 & 7.55E-01
 & 1.80E-01
 & 8.23E-02
 & 3.74E-03
 & $\le$3.1E-03
 \\
    MACS J0416.1
 & 3.27E+01
 & 1.94E+01
 & 1.24E+01
 & 8.88E+00
 & 7.58E+00
 & 2.93E+00
 & 1.37E+00
 & 6.80E-02
 & $\le$1.8E-02
 & $\le$4.8E-03
 & $\le$4.0E-03
 \\
    MACS J2211.7
 & 7.63E+01
 & 4.24E+01
 & 3.05E+01
 & 1.34E+01
 & 6.07E+00
 & 2.56E+00
 & 7.97E-01
 & 7.26E-02
 & 1.16E-06
 & 1.50E-02
 & 1.67E-04
 \\
    MACS J0429.6
 & 6.74E+01
 & 3.12E+01
 & 2.83E+01
 & 1.14E+01
 & 8.76E+00
 & 3.01E+00
 & 1.33E+00
 & 3.78E-01
 & $\le$3.8E-02
 & $\le$6.7E-03
 & $\le$1.7E-02
 \\
    MACS J0451.9
 & 3.04E+01
 & 3.74E+01
 & 1.40E+01
 & 1.19E+01
 & 3.38E+00
 & 1.90E+00
 & 7.42E-01
 & 8.34E-02
 & 1.95E-05
 & 2.54E-02
 & 2.35E-02
 \\
    MACS J1206.2
 & 3.33E+01
 & 2.85E+01
 & 1.71E+01
 & 1.03E+01
 & 4.82E+00
 & 2.34E+00
 & 7.99E-01
 & 2.17E-01
 & 3.30E-04
 & 1.03E-02
 & 2.41E-04
 \\
    MACS J0417.5
 & 1.93E+01
 & 7.19E+00
 & 1.31E+01
 & 6.80E+00
 & 3.85E+00
 & 2.04E+00
 & 8.21E-01
 & 1.06E-01
 & $\le$1.1E-02
 & 1.90E-02
 & 7.08E-04
 \\
    MACS J0329.6
 & 6.07E+01
 & 5.85E+01
 & 2.91E+01
 & 1.73E+01
 & 8.35E+00
 & 2.32E+00
 & 1.25E+00
 & 3.46E-01
 & $\le$3.2E-02
 & $\le$1.7E-02
 & $\le$1.2E-02
 \\
    MACS J1347.5
 & 6.07E+01
 & 4.57E+01
 & 3.88E+01
 & 1.53E+01
 & 8.63E+00
 & 2.60E+00
 & 7.12E-01
 & 1.26E-01
 & 2.85E-05
 & $\le$4.4E-05
 & 4.23E-02
 \\
    MACS J1311.0
 & 5.09E+01
 & 4.91E+01
 & 3.48E+01
 & 1.68E+01
 & 9.16E+00
 & 4.17E+00
 & 1.28E+00
 & 4.85E-01
 & 1.60E-02
 & 5.53E-04
 & 3.80E-02
 \\
    MACS J2214.9
 & 2.77E+01
 & 2.23E+01
 & 1.73E+01
 & 1.08E+01
 & 5.43E+00
 & 2.21E+00
 & 9.76E-01
 & 1.79E-01
 & 1.44E-03
 & $\le$4.5E-03
 & $\le$3.2E-03
 \\
    MACS J0257.1
 & 3.96E+01
 & 2.46E+01
 & 1.64E+01
 & 8.75E+00
 & 6.52E+00
 & 2.39E+00
 & 1.40E+00
 & 1.41E-01
 & $\le$2.0E-02
 & 3.81E-04
 & 8.63E-03
 \\
    MACS J0911.2
 & 2.43E+01
 & 1.62E+01
 & 1.43E+01
 & 9.38E+00
 & 5.15E+00
 & 2.44E+00
 & 9.63E-01
 & 2.66E-01
 & 5.43E-02
 & 6.63E-03
 & 1.56E-03
 \\
    MACS J0454.1
 & 1.64E+01
 & 1.81E+01
 & 1.57E+01
 & 1.01E+01
 & 6.09E+00
 & 2.64E+00
 & 9.79E-01
 & 1.31E-01
 & $\le$1.3E-02
 & $\le$3.5E-03
 & 3.73E-02
 \\
    MACS J1423.8
 & 5.84E+01
 & 4.53E+01
 & 2.54E+01
 & 1.26E+01
 & 7.11E+00
 & 2.25E+00
 & 9.78E-01
 & 7.23E-02
 & $\le$4.1E-01
 & $\le$4.2E-03
 & $\le$8.5E-03
 \\
    MACS J1149.6
 & 1.49E+01
 & 1.22E+01
 & 8.91E+00
 & 5.93E+00
 & 3.37E+00
 & 1.53E+00
 & 5.22E-01
 & 1.24E-01
 & 2.09E-02
 & 2.53E-03
 & 2.13E-04
 \\
    MACS J0018.5
 & 2.06E+01
 & 1.62E+01
 & 1.06E+01
 & 6.93E+00
 & 3.93E+00
 & 2.15E+00
 & 7.39E-01
 & 1.85E-01
 & 1.20E-02
 & 5.53E-03
 & 2.85E-03
 \\
    MACS J0025.4
 & 5.16E+00
 & 6.78E+00
 & 8.17E+00
 & 5.04E+00
 & 4.93E+00
 & 2.54E+00
 & 9.89E-01
 & 1.13E-01
 & $\le$4.2E-04
 & $\le$5.2E-05
 & $\le$5.6E-06
 \\
    MACS J0647.8
 & 4.98E+01
 & 2.81E+01
 & 2.35E+01
 & 1.33E+01
 & 7.43E+00
 & 2.97E+00
 & 1.64E+00
 & 3.26E-01
 & $\le$1.8E-02
 & $\le$5.8E-03
 & $\le$5.3E-03
 \\
    MACS J2129.4
 & 3.11E+01
 & 4.16E+01
 & 2.17E+01
 & 1.61E+01
 & 5.02E+00
 & 1.87E+00
 & 1.14E+00
 & 5.19E-02
 & $\le$1.9E-02
 & $\le$5.6E-03
 & $\le$3.7E-03
 \\
  \enddata
  \tablecomments{Scaled pressure deprojections for all of the observed
    galaxy clusters. Due to strong degeneracies between the values, 
    error estimates are omitted from this table. Missing values at small 
    radii are interior to the radius of the innermost \chandra/\rosat\ X-ray 
    deprojection bin. Upper limits (at 68 per cent confidence) are given 
    for pressure values at large radii that are consistent with zero.}
  \label{tab:deproj_individual}
\end{deluxetable*}
\end{longrotatetable}

%% file: individual_cluster_gnfw.tex
\begin{deluxetable}{cccccc}
  \tablewidth{0pt}
  \tablecaption{gNFW Fit Parameters}
  \tablehead{\colhead{Cluster} & \colhead{P$_0$} & \colhead{$\alpha$} & \colhead{$\beta$} & 
    \colhead{$\gamma$} & \colhead{$c_{500}$}}
  \startdata
    Abell 0754 & \phn\phn4.60 & 1.57 & 4.43 & 0.30 & 1.40 \\
    Abell 0085 & \phn26.64 & 0.65 & 4.16 & 0.30 & 1.40 \\
    Abell 3667 & \phn\phn4.67 & 1.34 & 5.62 & 0.30 & 1.40 \\
    Abell 2256 & \phn\phn2.49 & 2.17 & 6.57 & 0.30 & 1.40 \\
    Abell 3158 & \phn\phn4.58 & 1.23 & 4.27 & 0.30 & 1.40 \\
    Abell 3266 & \phn\phn3.49 & 1.57 & 4.71 & 0.30 & 1.40 \\
    Abell 1795 & \phn\phn9.89 & 0.80 & 4.30 & 0.30 & 1.40 \\
    Abell 2065 & \phn\phn4.66 & 1.47 & 5.51 & 0.30 & 1.40 \\
    Abell 3112 & \phn10.85 & 0.87 & 5.18 & 0.30 & 1.40 \\
    Abell 2029 & \phn16.69 & 1.02 & 6.05 & 0.30 & 1.40 \\
    Abell 2255 & \phn\phn2.21 & 1.99 & 4.42 & 0.30 & 1.40 \\
    Abell 1650 & \phn15.18 & 0.85 & 4.81 & 0.30 & 1.40 \\
    Abell 1651 & \phn\phn7.15 & 1.32 & 6.32 & 0.30 & 1.40 \\
    Abell 2420 & \phn\phn6.17 & 1.78 & 6.99 & 0.30 & 1.40 \\
    Abell 0478 & \phn35.70 & 0.74 & 5.24 & 0.30 & 1.40 \\
    Abell 2142 & \phn\phn6.58 & 1.01 & 4.51 & 0.30 & 1.40 \\
    Abell 3695 & \phn\phn1.57 & 2.19 & 4.60 & 0.30 & 1.40 \\
    Abell 3921 & \phn\phn5.12 & 1.65 & 7.07 & 0.30 & 1.40 \\
    Abell 2244 & \phn15.64 & 0.80 & 4.67 & 0.30 & 1.40 \\
    Abell 2426 & \phn\phn5.53 & 1.21 & 4.80 & 0.30 & 1.40 \\
    Abell 3827 & \phn\phn9.50 & 1.19 & 5.91 & 0.30 & 1.40 \\
    MACS J0416.1 & \phn\phn8.33 & 1.32 & 5.41 & 0.30 & 1.40 \\
    MACS J2211.7 & \phn18.35 & 1.04 & 6.66 & 0.30 & 1.40 \\
    MACS J0429.6 & \phn14.77 & 1.07 & 5.01 & 0.30 & 1.40 \\
    MACS J0451.9 & 204.07 & 0.40 & 3.94 & 0.30 & 1.40 \\
    MACS J1206.2 & \phn20.08 & 0.82 & 5.01 & 0.30 & 1.40 \\
    MACS J0417.5 & \phn\phn5.03 & 1.60 & 6.04 & 0.30 & 1.40 \\
    MACS J0329.6 & \phn43.58 & 0.66 & 4.57 & 0.30 & 1.40 \\
    MACS J1347.5 & \phn24.12 & 1.11 & 8.05 & 0.30 & 1.40 \\
    MACS J1311.0 & \phn37.74 & 0.75 & 4.41 & 0.30 & 1.40 \\
    MACS J2214.9 & \phn\phn8.05 & 1.34 & 5.86 & 0.30 & 1.40 \\
    MACS J0257.1 & \phn\phn8.47 & 1.32 & 5.82 & 0.30 & 1.40 \\
    MACS J0911.2 & \phn\phn6.35 & 1.15 & 4.26 & 0.30 & 1.40 \\
    MACS J0454.1 & \phn\phn8.09 & 1.46 & 6.69 & 0.30 & 1.40 \\
    MACS J1423.8 & \phn19.33 & 0.98 & 4.94 & 0.30 & 1.40 \\
    MACS J1149.6 & \phn\phn3.93 & 1.67 & 6.20 & 0.30 & 1.40 \\
    MACS J0018.5 & \phn\phn5.57 & 1.51 & 6.36 & 0.30 & 1.40 \\
    MACS J0025.4 & \phn\phn3.95 & 1.97 & 6.23 & 0.30 & 1.40 \\
    MACS J0647.8 & \phn18.78 & 1.04 & 5.50 & 0.30 & 1.40 \\
    MACS J2129.4 & \phn32.42 & 0.80 & 5.71 & 0.30 & 1.40 \\
  \enddata
  \tablecomments{Generalized NFW parameters for all of the observed galaxy clusters.
    Due to the strong degeneracies between parameters, error estimates are 
    omitted from this table.}
  \label{tab:gnfw_individual}
\end{deluxetable}

%% file: lowz_20210312_latex_table.tex
\begin{rotatetable*} 
\begin{deluxetable*}{cccccccccccc} 
  \tablecolumns{12} 
  \tablewidth{0pt} 
  \tablecaption{\lowz\ Observational Sample} 
  \tablehead{\colhead{R/R$_{500}$} & \colhead{0.028} & \colhead{0.047} & \colhead{0.078} & 
    \colhead{0.13} & \colhead{0.22} & \colhead{0.36} & \colhead{0.60} & \colhead{1.0} &  
    \colhead{1.7} & \colhead{2.8} & \colhead{4.6}} 
  \startdata 
    \sidehead{Mean Pressure} 
     & 1.94E+01 & 1.47E+01 & 1.10E+01 & 7.16E+00 & 3.91E+00 & 1.79E+00 & 7.29E-01 & 1.79E-01 & 3.87E-02 & 6.92E-03 & 1.48E-03 \\ 
    \sidehead{Measurement Covariance Matrix} 
    0.028 & 1.50E+01 & 8.41E+00 & 4.72E+00 & 2.09E+00 & 6.89E-01 & 8.19E-02 & -5.97E-02 & 2.15E-02 & -1.35E-02 & 9.08E-03 & 5.91E-04 \\ 
    0.047 & --- & 5.92E+00 & 3.22E+00 & 1.33E+00 & 4.68E-01 & 6.29E-02 & -4.76E-02 & 1.05E-02 & -5.76E-03 & 5.09E-03 & 4.92E-04 \\ 
    0.078 & --- & --- & 2.00E+00 & 7.66E-01 & 2.55E-01 & 3.58E-02 & -3.84E-02 & 3.01E-03 & -2.03E-03 & 2.83E-03 & 2.02E-04 \\ 
    0.13 & --- & --- & --- & 3.90E-01 & 1.34E-01 & 3.22E-02 & 8.96E-04 & 3.27E-03 & -1.40E-03 & 1.05E-03 & -1.68E-05 \\ 
    0.22 & --- & --- & --- & --- & 6.41E-02 & 1.91E-02 & 2.08E-03 & -3.93E-05 & -4.05E-04 & 1.80E-04 & -1.32E-06 \\ 
    0.36 & --- & --- & --- & --- & --- & 1.25E-02 & 3.48E-03 & -3.91E-04 & 1.59E-04 & -7.23E-05 & -3.75E-05 \\ 
    0.6 & --- & --- & --- & --- & --- & --- & 7.29E-03 & 6.66E-04 & 4.71E-05 & -6.65E-05 & -1.04E-05 \\ 
    1.0 & --- & --- & --- & --- & --- & --- & --- & 8.69E-04 & -7.34E-05 & 6.75E-05 & -8.83E-07 \\ 
    1.7 & --- & --- & --- & --- & --- & --- & --- & --- & 5.67E-05 & -1.31E-05 & -2.68E-06 \\ 
    2.8 & --- & --- & --- & --- & --- & --- & --- & --- & --- & 1.28E-05 & -1.16E-07 \\ 
    4.6 & --- & --- & --- & --- & --- & --- & --- & --- & --- & --- & 1.19E-06 \\ 
    \sidehead{Intrinsic Scatter Covariance Matrix} 
    0.028 & 2.69E+02 & 1.50E+02 & 8.34E+01 & 3.74E+01 & 1.22E+01 & 1.42E+00 & -9.24E-01 & 3.83E-01 & -2.38E-01 & 1.60E-01 & 7.51E-03 \\ 
    0.047 & --- & 1.05E+02 & 5.69E+01 & 2.36E+01 & 8.25E+00 & 1.04E+00 & -7.50E-01 & 1.96E-01 & -1.01E-01 & 9.03E-02 & 6.91E-03 \\ 
    0.078 & --- & --- & 3.53E+01 & 1.36E+01 & 4.49E+00 & 6.03E-01 & -6.12E-01 & 6.07E-02 & -3.80E-02 & 5.02E-02 & 2.60E-03 \\ 
    0.13 & --- & --- & --- & 7.05E+00 & 2.40E+00 & 5.60E-01 & 2.45E-02 & 5.82E-02 & -2.42E-02 & 1.85E-02 & -6.24E-04 \\ 
    0.22 & --- & --- & --- & --- & 1.17E+00 & 3.37E-01 & 3.65E-02 & 1.72E-03 & -7.00E-03 & 3.27E-03 & -1.17E-04 \\ 
    0.36 & --- & --- & --- & --- & --- & 2.30E-01 & 5.99E-02 & -6.09E-03 & 2.64E-03 & -1.26E-03 & -6.42E-04 \\ 
    0.6 & --- & --- & --- & --- & --- & --- & 1.27E-01 & 1.07E-02 & 8.90E-04 & -1.11E-03 & -1.61E-04 \\ 
    1.0 & --- & --- & --- & --- & --- & --- & --- & 1.57E-02 & -1.25E-03 & 1.19E-03 & -1.48E-05 \\ 
    1.7 & --- & --- & --- & --- & --- & --- & --- & --- & 1.01E-03 & -2.30E-04 & -4.36E-05 \\ 
    2.8 & --- & --- & --- & --- & --- & --- & --- & --- & --- & 2.35E-04 & -2.86E-06 \\ 
    4.6 & --- & --- & --- & --- & --- & --- & --- & --- & --- & --- & 2.13E-05 \\ 
  \enddata 
\tablecomments{Ensemble mean pressure (in units of \pfive) and associated covariance (in units of \pfive$^2$).} 
\label{tab:deproj_lowz} 
\end{deluxetable*} 
\end{rotatetable*}

%% file: highz_20210312_latex_table.tex
\begin{rotatetable*} 
\begin{deluxetable*}{cccccccccccc} 
  \tablecolumns{12} 
  \tablewidth{0pt} 
  \tablecaption{\midz\ Observational Sample} 
  \tablehead{\colhead{R/R$_{500}$} & \colhead{0.028} & \colhead{0.047} & \colhead{0.078} & 
    \colhead{0.13} & \colhead{0.22} & \colhead{0.36} & \colhead{0.60} & \colhead{1.0} &  
    \colhead{1.7} & \colhead{2.8} & \colhead{4.6}} 
  \startdata 
    \sidehead{Mean Pressure} 
     & 3.68E+01 & 2.78E+01 & 1.93E+01 & 1.07E+01 & 5.76E+00 & 2.33E+00 & 9.48E-01 & 1.71E-01 & 1.59E-02 & 4.77E-03 & 7.65E-03 \\ 
    \sidehead{Measurement Covariance Matrix} 
    0.028 & 5.77E+01 & 3.23E+01 & 2.30E+01 & 6.63E+00 & 3.33E+00 & 4.68E-01 & 1.23E-01 & 1.73E-02 & -1.75E-02 & -2.56E-04 & 3.59E-03 \\ 
    0.047 & --- & 3.03E+01 & 1.57E+01 & 6.39E+00 & 1.87E+00 & 6.53E-02 & 4.68E-03 & -1.56E-04 & -1.33E-02 & -1.90E-03 & 9.09E-03 \\ 
    0.078 & --- & --- & 1.24E+01 & 3.85E+00 & 1.74E+00 & 2.79E-01 & 2.24E-02 & 1.63E-02 & -7.55E-03 & -1.25E-03 & 9.32E-03 \\ 
    0.13 & --- & --- & --- & 1.67E+00 & 4.69E-01 & 4.98E-02 & 1.85E-02 & 1.21E-03 & -3.03E-03 & -4.67E-04 & 2.53E-03 \\ 
    0.22 & --- & --- & --- & --- & 3.96E-01 & 6.69E-02 & 2.54E-02 & 6.25E-03 & -1.37E-03 & -1.03E-03 & 1.47E-03 \\ 
    0.36 & --- & --- & --- & --- & --- & 2.85E-02 & 4.67E-03 & 2.30E-03 & -1.66E-04 & -4.64E-05 & 3.90E-04 \\ 
    0.6 & --- & --- & --- & --- & --- & --- & 1.01E-02 & 8.81E-04 & -2.43E-04 & -1.18E-04 & -1.32E-04 \\ 
    1.0 & --- & --- & --- & --- & --- & --- & --- & 1.58E-03 & 4.21E-05 & -3.53E-05 & -4.83E-06 \\ 
    1.7 & --- & --- & --- & --- & --- & --- & --- & --- & 3.08E-05 & -9.41E-07 & -6.21E-06 \\ 
    2.8 & --- & --- & --- & --- & --- & --- & --- & --- & --- & 1.09E-05 & -1.42E-06 \\ 
    4.6 & --- & --- & --- & --- & --- & --- & --- & --- & --- & --- & 3.88E-05 \\ 
    \sidehead{Intrinsic Scatter Covariance Matrix} 
    0.028 & 8.94E+02 & 4.84E+02 & 3.53E+02 & 9.86E+01 & 5.03E+01 & 7.09E+00 & 1.80E+00 & 2.78E-01 & -2.51E-01 & -2.18E-03 & 4.78E-02 \\ 
    0.047 & --- & 4.56E+02 & 2.34E+02 & 9.62E+01 & 2.71E+01 & 1.08E+00 & 1.09E-01 & 1.42E-02 & -1.88E-01 & -2.45E-02 & 1.28E-01 \\ 
    0.078 & --- & --- & 1.91E+02 & 5.82E+01 & 2.65E+01 & 4.13E+00 & 3.49E-01 & 2.30E-01 & -1.05E-01 & -1.88E-02 & 1.29E-01 \\ 
    0.13 & --- & --- & --- & 2.53E+01 & 6.83E+00 & 7.39E-01 & 2.69E-01 & 2.03E-02 & -4.22E-02 & -6.51E-03 & 3.53E-02 \\ 
    0.22 & --- & --- & --- & --- & 6.10E+00 & 1.01E+00 & 3.64E-01 & 8.53E-02 & -1.93E-02 & -1.50E-02 & 2.05E-02 \\ 
    0.36 & --- & --- & --- & --- & --- & 4.32E-01 & 6.93E-02 & 3.26E-02 & -2.27E-03 & -6.85E-04 & 5.13E-03 \\ 
    0.6 & --- & --- & --- & --- & --- & --- & 1.53E-01 & 1.28E-02 & -3.38E-03 & -1.61E-03 & -1.80E-03 \\ 
    1.0 & --- & --- & --- & --- & --- & --- & --- & 2.40E-02 & 5.75E-04 & -4.99E-04 & -1.17E-04 \\ 
    1.7 & --- & --- & --- & --- & --- & --- & --- & --- & 4.62E-04 & -1.32E-05 & -7.63E-05 \\ 
    2.8 & --- & --- & --- & --- & --- & --- & --- & --- & --- & 1.64E-04 & -2.38E-05 \\ 
    4.6 & --- & --- & --- & --- & --- & --- & --- & --- & --- & --- & 5.76E-04 \\ 
  \enddata 
\tablecomments{Ensemble mean pressure (in units of \pfive) and associated covariance (in units of \pfive$^2$).} 
\label{tab:deproj_midz} 
\end{deluxetable*} 
\end{rotatetable*}

%% file: lowz_sim_lowz_dyn_selected_20220204_latex_table.tex
\begin{rotatetable*} 
\begin{deluxetable*}{cccccccccccc} 
  \tablecolumns{12} 
  \tablewidth{0pt} 
  \tablecaption{\lowz\ Matched Sample from \three} 
  \tablehead{\colhead{R/R$_{500}$} & \colhead{0.028} & \colhead{0.047} & \colhead{0.078} & 
    \colhead{0.13} & \colhead{0.22} & \colhead{0.36} & \colhead{0.60} & \colhead{1.0} &  
    \colhead{1.7} & \colhead{2.8} & \colhead{4.6}} 
  \startdata 
    \sidehead{Mean Pressure} 
     & 2.49E+01 & 1.94E+01 & 1.32E+01 & 7.92E+00 & 4.11E+00 & 1.81E+00 & 6.63E-01 & 1.87E-01 & 3.73E-02 & 4.62E-03 & 5.73E-04 \\ 
    \sidehead{Measurement Covariance Matrix} 
    0.028 & 1.73E+01 & 1.06E+01 & 5.89E+00 & 2.70E+00 & 9.79E-01 & 2.43E-01 & 7.64E-03 & -1.88E-02 & -1.13E-03 & 8.69E-04 & -3.59E-05 \\ 
    0.047 & --- & 7.03E+00 & 4.11E+00 & 1.93E+00 & 7.01E-01 & 1.75E-01 & 8.12E-03 & -1.12E-02 & -5.14E-04 & 5.51E-04 & -1.83E-05 \\ 
    0.078 & --- & --- & 2.54E+00 & 1.25E+00 & 4.75E-01 & 1.19E-01 & 7.60E-03 & -6.54E-03 & -5.00E-04 & 2.86E-04 & -1.13E-05 \\ 
    0.13 & --- & --- & --- & 6.80E-01 & 2.80E-01 & 7.12E-02 & 5.20E-03 & -3.29E-03 & -4.70E-04 & 6.92E-05 & -4.62E-06 \\ 
    0.22 & --- & --- & --- & --- & 1.33E-01 & 3.62E-02 & 3.16E-03 & -1.55E-03 & -3.74E-04 & -7.51E-06 & -2.24E-06 \\ 
    0.36 & --- & --- & --- & --- & --- & 1.21E-02 & 1.69E-03 & -4.66E-04 & -1.46E-04 & -6.64E-06 & -2.19E-07 \\ 
    0.6 & --- & --- & --- & --- & --- & --- & 7.57E-04 & -1.26E-05 & -3.55E-05 & -4.64E-06 & -8.80E-07 \\ 
    1.0 & --- & --- & --- & --- & --- & --- & --- & 7.96E-05 & 1.34E-05 & -7.01E-07 & -1.24E-07 \\ 
    1.7 & --- & --- & --- & --- & --- & --- & --- & --- & 1.07E-05 & 7.50E-07 & 3.58E-08 \\ 
    2.8 & --- & --- & --- & --- & --- & --- & --- & --- & --- & 4.15E-07 & 2.59E-08 \\ 
    4.6 & --- & --- & --- & --- & --- & --- & --- & --- & --- & --- & 1.32E-08 \\ 
    \sidehead{Intrinsic Scatter Covariance Matrix} 
    0.028 & 4.86E+02 & 2.96E+02 & 1.64E+02 & 7.46E+01 & 2.68E+01 & 6.64E+00 & 2.28E-01 & -5.14E-01 & -3.71E-02 & 2.32E-02 & -1.12E-03 \\ 
    0.047 & --- & 1.96E+02 & 1.14E+02 & 5.32E+01 & 1.91E+01 & 4.78E+00 & 2.38E-01 & -3.03E-01 & -1.78E-02 & 1.47E-02 & -5.20E-04 \\ 
    0.078 & --- & --- & 7.09E+01 & 3.48E+01 & 1.31E+01 & 3.30E+00 & 2.17E-01 & -1.77E-01 & -1.62E-02 & 7.49E-03 & -3.03E-04 \\ 
    0.13 & --- & --- & --- & 1.90E+01 & 7.86E+00 & 1.99E+00 & 1.47E-01 & -9.04E-02 & -1.45E-02 & 1.67E-03 & -1.18E-04 \\ 
    0.22 & --- & --- & --- & --- & 3.74E+00 & 1.02E+00 & 8.92E-02 & -4.24E-02 & -1.11E-02 & -2.55E-04 & -5.56E-05 \\ 
    0.36 & --- & --- & --- & --- & --- & 3.39E-01 & 4.73E-02 & -1.28E-02 & -4.14E-03 & -1.80E-04 & -1.27E-06 \\ 
    0.6 & --- & --- & --- & --- & --- & --- & 2.14E-02 & -3.08E-04 & -9.82E-04 & -1.22E-04 & -2.34E-05 \\ 
    1.0 & --- & --- & --- & --- & --- & --- & --- & 2.28E-03 & 3.79E-04 & -1.88E-05 & -3.71E-06 \\ 
    1.7 & --- & --- & --- & --- & --- & --- & --- & --- & 3.09E-04 & 2.10E-05 & 1.12E-06 \\ 
    2.8 & --- & --- & --- & --- & --- & --- & --- & --- & --- & 1.18E-05 & 7.21E-07 \\ 
    4.6 & --- & --- & --- & --- & --- & --- & --- & --- & --- & --- & 3.86E-07 \\ 
  \enddata 
\tablecomments{Ensemble mean pressure (in units of \pfive) and associated covariance (in units of \pfive$^2$).} 
\label{tab:deproj_lowz_sim} 
\end{deluxetable*} 
\end{rotatetable*}

%% file: highz_sim_highz_dyn_selected_20220204_latex_table.tex
\begin{rotatetable*} 
\begin{deluxetable*}{cccccccccccc} 
  \tablecolumns{12} 
  \tablewidth{0pt} 
  \tablecaption{\midz\ Matched Sample from \three} 
  \tablehead{\colhead{R/R$_{500}$} & \colhead{0.028} & \colhead{0.047} & \colhead{0.078} & 
    \colhead{0.13} & \colhead{0.22} & \colhead{0.36} & \colhead{0.60} & \colhead{1.0} &  
    \colhead{1.7} & \colhead{2.8} & \colhead{4.6}} 
  \startdata 
    \sidehead{Mean Pressure} 
     & 3.72E+01 & 2.89E+01 & 1.90E+01 & 1.11E+01 & 5.77E+00 & 2.47E+00 & 8.60E-01 & 2.14E-01 & 2.49E-02 & 2.54E-03 & 4.20E-04 \\ 
    \sidehead{Measurement Covariance Matrix} 
    0.028 & 2.80E+01 & 1.90E+01 & 8.89E+00 & 3.50E+00 & 1.20E+00 & 2.70E-01 & -2.10E-02 & -1.88E-02 & -2.24E-03 & -2.04E-04 & -8.92E-05 \\ 
    0.047 & --- & 1.40E+01 & 7.05E+00 & 2.82E+00 & 9.87E-01 & 2.12E-01 & -1.38E-02 & -1.35E-02 & -1.40E-03 & -1.26E-04 & -4.69E-05 \\ 
    0.078 & --- & --- & 4.04E+00 & 1.66E+00 & 6.26E-01 & 1.39E-01 & -6.70E-03 & -9.02E-03 & -8.74E-04 & -8.22E-05 & -2.24E-05 \\ 
    0.13 & --- & --- & --- & 7.34E-01 & 2.97E-01 & 7.06E-02 & -2.04E-03 & -3.97E-03 & -4.56E-04 & -3.43E-05 & -7.31E-06 \\ 
    0.22 & --- & --- & --- & --- & 1.44E-01 & 4.09E-02 & 8.89E-04 & -2.21E-03 & -2.79E-04 & -3.09E-05 & -1.66E-06 \\ 
    0.36 & --- & --- & --- & --- & --- & 1.57E-02 & 1.89E-03 & -5.98E-04 & -1.01E-04 & -1.41E-05 & -8.57E-08 \\ 
    0.6 & --- & --- & --- & --- & --- & --- & 1.30E-03 & 9.68E-05 & -9.03E-06 & -1.50E-06 & -1.67E-07 \\ 
    1.0 & --- & --- & --- & --- & --- & --- & --- & 1.82E-04 & 1.24E-05 & 8.23E-07 & -8.54E-08 \\ 
    1.7 & --- & --- & --- & --- & --- & --- & --- & --- & 3.06E-06 & 1.67E-07 & -6.76E-09 \\ 
    2.8 & --- & --- & --- & --- & --- & --- & --- & --- & --- & 1.11E-07 & 9.21E-09 \\ 
    4.6 & --- & --- & --- & --- & --- & --- & --- & --- & --- & --- & 4.33E-09 \\ 
    \sidehead{Intrinsic Scatter Covariance Matrix} 
    0.028 & 9.68E+02 & 6.59E+02 & 3.10E+02 & 1.22E+02 & 4.15E+01 & 9.40E+00 & -6.25E-01 & -6.53E-01 & -7.44E-02 & -6.81E-03 & -3.03E-03 \\ 
    0.047 & --- & 4.89E+02 & 2.47E+02 & 9.80E+01 & 3.43E+01 & 7.34E+00 & -4.15E-01 & -4.64E-01 & -4.58E-02 & -4.43E-03 & -1.64E-03 \\ 
    0.078 & --- & --- & 1.41E+02 & 5.76E+01 & 2.16E+01 & 4.74E+00 & -2.29E-01 & -3.02E-01 & -2.82E-02 & -2.84E-03 & -7.99E-04 \\ 
    0.13 & --- & --- & --- & 2.55E+01 & 1.02E+01 & 2.40E+00 & -7.09E-02 & -1.33E-01 & -1.48E-02 & -1.20E-03 & -2.66E-04 \\ 
    0.22 & --- & --- & --- & --- & 4.91E+00 & 1.38E+00 & 2.93E-02 & -7.29E-02 & -9.16E-03 & -1.05E-03 & -6.41E-05 \\ 
    0.36 & --- & --- & --- & --- & --- & 5.31E-01 & 6.38E-02 & -1.99E-02 & -3.41E-03 & -4.77E-04 & -4.41E-06 \\ 
    0.6 & --- & --- & --- & --- & --- & --- & 4.48E-02 & 3.29E-03 & -3.40E-04 & -5.30E-05 & -5.93E-06 \\ 
    1.0 & --- & --- & --- & --- & --- & --- & --- & 6.18E-03 & 4.00E-04 & 2.48E-05 & -2.94E-06 \\ 
    1.7 & --- & --- & --- & --- & --- & --- & --- & --- & 1.03E-04 & 5.77E-06 & -2.04E-07 \\ 
    2.8 & --- & --- & --- & --- & --- & --- & --- & --- & --- & 3.78E-06 & 3.12E-07 \\ 
    4.6 & --- & --- & --- & --- & --- & --- & --- & --- & --- & --- & 1.50E-07 \\ 
  \enddata 
\tablecomments{Ensemble mean pressure (in units of \pfive) and associated covariance (in units of \pfive$^2$).} 
\label{tab:deproj_midz_sim} 
\end{deluxetable*} 
\end{rotatetable*}